\documentclass[10pt,epsfig]{article} 

\usepackage{enumerate}
\usepackage{float}
\usepackage{subfig}
\usepackage{color}
\usepackage{slashed}
\usepackage{multirow}
\usepackage{cite}

\let\counterwithin\relax
\usepackage{chngcntr}
\usepackage{amssymb, amsmath,mathrsfs}

\usepackage{graphics}
\usepackage{graphicx}
\usepackage{epsf}
\usepackage{epsfig}
\usepackage{float}
\usepackage{makecell}
\usepackage{multirow}
\usepackage{color}
\usepackage{xcolor}

\usepackage[utf8]{inputenc}
\usepackage{amsmath}
\usepackage{amssymb}
\usepackage{subfig}
\usepackage[normalem]{ulem}

\usepackage[colorlinks=true,
            linkcolor=black,
            urlcolor=blue,
            citecolor=blue]{hyperref}

\long\def\rpl#1!!#2!!{\textcolor{red}{#1} \textcolor{blue}{#2}}
\def\baselinestretch{1.3}
\date{\today}

\newcommand{\beq}{\begin {equation}}  
\newcommand{\eeq}{\end   {equation}} 
\newcommand{\bea}{\begin {eqnarray}} 
\newcommand{\eea}{\end   {eqnarray}}  
\newcommand{\baa}{\begin {array}   } 
\newcommand{\eaa}{\end   {array}   }     
\newcommand{\bit}{\begin {itemize} }
\newcommand{\eit}{\end   {itemize} }
\newcommand{\be }{\begin {equation}} 
\newcommand{\ee }{\end   {equation}}
\newcommand{\nn }{\nonumber        }

\allowdisplaybreaks
\begin{document}

\begin{center}

{\Large \textbf {Determining the Shape of the Higgs Potential at Future Colliders}}\\[10mm]

Pankaj Agrawal$^{a,b}$\footnote{agrawal@iopb.res.in}, Debashis Saha$^{a,b}$\footnote{debasaha@iopb.res.in}, Ling-Xiao Xu$^{c}$\footnote{lingxiaoxu@pku.edu.cn}, ,
Jiang-Hao Yu$^{d, e}$\footnote{jhyu@itp.ac.cn, corresponding author}, C.-P. Yuan$^{f}$\footnote{yuan@pa.msu.edu}. \\[10mm]

\noindent 
$^a${\em \small Institute of Physics,  Sainik School Post, Bhubaneswar 751 005, India}                                                                   \\
$^b${\em \small Homi Bhabha National Institute, Training School Complex, Anushakti Nagar, Mumbai 400085, India}                                          \\
$^c${\em \small Department of Physics and State Key Laboratory of Nuclear Physics and Technology,  Peking University, Beijing 100871, China}             \\
$^d${\em \small CAS Key Laboratory of Theoretical Physics, Institute of Theoretical Physics,  Chinese Academy of Sciences, Beijing 100190, P. R. China}  \\
$^e${\em \small School of Physical Sciences, University of Chinese Academy of Sciences,  No.19A Yuquan Road, Beijing 100049, P.R. China}                 \\
$^f${\em \small Department of Physics and Astronomy, Michigan State University,  East Lansing, Michigan 48824, USA}                                       

\end{center}

\begin{abstract} 
	
Although the Higgs boson has been discovered, its self-couplings are poorly constrained. This leaves the nature of the Higgs boson undetermined. Motivated by different Higgs potential scenarios other than the Landau-Ginzburg type in the standard model, we systematically organize various new physics scenarios -- elementary Higgs, Nambu-Goldstone Higgs, Coleman-Weinberg Higgs, and Tadpole-induced Higgs, etc. We find that double-Higgs production at the 27 TeV high energy LHC can be used to discriminate different Higgs potential scenarios, while it is necessary to use triple-Higgs production at a future 100 TeV proton-proton collider to fully determine the shape of the Higgs potential. 

\end{abstract}

\newpage

\tableofcontents

\setcounter{footnote}{0}

\def\baselinestretch{1.5}
\counterwithin{equation}{section}

\newpage

\section{Introduction}
\label{sec:intro}

After a long wait of about half a century, in 2012, the Higgs boson was discovered at the Large Hadron Collider (LHC) by the CMS and ATLAS Collaborations. With the discovery of this missing piece, all the particles of the standard model (SM) have now been discovered. With the measured value of the Higgs boson mass, all the parameters in the SM are now known. The main goal of the LHC machine now is to measure the properties and interactions of the Higgs boson, as well as to look for signatures of possible new physics beyond the SM. Until now, direct searches for evidence of new physics (NP) have not yielded anything of significance. This has pushed the new physics scale to be around TeV. On the other hand, precision measurements on various SM processes provide us with an indirect way to probe new physics. The Higgs boson couplings to the gauge bosons and the SM fermions have been measured at the LHC through various production processes and decay modes. However, the Higgs self-couplings are not yet determined at the end of the Run-2 of the LHC~\cite{Sirunyan:2018two,Aaboud:2018sfw,Aaboud:2018knk,Aaboud:2018ksn,Aaboud:2018zhh,Aaboud:2018ewm,CMS:2018ccd,ATLAS:2018ch1}.

The self-couplings of the Higgs boson, including the trilinear and quartic Higgs couplings, are still mysteries. Experimentally, the trilinear and quartic Higgs couplings can be directly measured using double- and triple-Higgs production processes $pp\to hh$ and $pp\to hhh$, respectively, at hadron colliders. The ATLAS and CMS Collaborations have been looking for the $hh$ production signal in the data collected so far at the LHC. This data can put a very loose bound on the trilinear Higgs coupling. The $hhh$ production signal has not yet been investigated with the Run-2 data. It is quite challenging to measure the Higgs self-couplings at the LHC, and this provides a strong motivation for building future high energy colliders.

\begin{figure}[!htb]
\begin{center}
\includegraphics[scale=0.5]{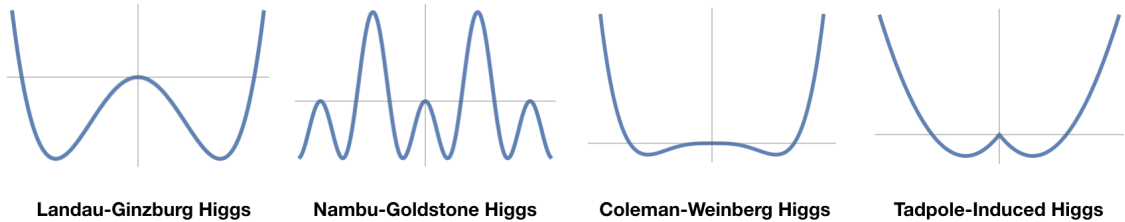}
\caption{The shapes of Higgs potential for various scenarios studied in this work.}
\label{shapePotential}
\end{center}
\end{figure}

Theoretically, there are still many unknowns about the Higgs boson, such as the nature of the Higgs boson, the origin of electroweak symmetry breaking (EWSB), the shape of the Higgs potential, and the strength of the electroweak phase transition, etc. All these questions can only be addressed after the Higgs self-couplings are determined. So far the Higgs self-couplings are not tightly constrained. The Higgs potential can be very different from the Landau-Ginzburg type in the SM. In this work, we systematically investigate various classes of new physics scenarios based on different types of Higgs potential. To be specific, we consider the following Higgs scenarios:
\begin{itemize}
\item Elementary Higgs boson, in which the Higgs boson is taken as an elementary scalar with rescaled self-couplings. The Higgs mass parameter is negative and thus triggers EWSB.
\item Nambu-Goldstone Higgs, in which the Higgs boson is taken as a pseudo Nambu-Goldstone (PNG) boson~\cite{Kaplan:1983fs,Kaplan:1983sm} emerging from strong dynamics at a high scale (see Refs.~\cite{Contino:2010rs,Panico:2015jxa,Bellazzini:2014yua} for comprehensive reviews).
\item Coleman-Weinberg (CW) Higgs, in which EWSB is triggered by renormalization group (RG) running effects~\cite{Hill:2014mqa,Helmboldt:2016mpi,Hashino:2015nxa} with classical scale invariance.
\item Tadpole-induced Higgs, in which EWSB is triggered by the Higgs tadpole~\cite{Galloway:2013dma,Chang:2014ida}, and the Higgs boson mass parameter is taken to be positive.
\end{itemize}
In general, the Higgs potentials could be organized according to their analytic structure. The key structure of the Higgs potential in each scenario is as follows:
\bea
	V(H) \simeq
	\begin{cases} 
	-m^2  H^\dagger H + \lambda ( H^\dagger H)^2+\frac{c_6 \lambda}{\Lambda^2} (H^\dagger H)^3, & \textrm{Elementary Higgs} \\
	-a \sin^2 ( \sqrt{ H^\dagger  H}/f) + b \sin^4 ( \sqrt{ H^\dagger  H}/f), & \textrm{Nambu-Goldstone Higgs} \\
	\lambda ( H^\dagger H)^2
	+ \epsilon ( H^\dagger H)^2 \log\frac{ H^\dagger H}{\mu^2},  & \textrm{Coleman-Weinberg Higgs} \\
	- \kappa^3 \sqrt{ H^\dagger  H} +  m^2   H^\dagger H, & \textrm{Tadpole-induced Higgs}  
	\end{cases}
\eea
where $f$ denotes the decay constant of the NG Higgs boson, and $\mu$ denotes the renormalization scale in case EWSB is triggered by radiative corrections, $m^2, \lambda, c_6, \Lambda, a, b, \epsilon, \kappa$ are dimensionful or dimensionless parameters in each new physics scenario. The shapes of the Higgs potential are schematically illustrated in Fig.~\ref{shapePotential}, respectively. In both the elementary and Nambu-Goldstone Higgs cases, the Higgs potential could be expanded in the powers of $H^\dagger H$, which could recover the Landau-Ginzburg effective theory description if a truncation on the series provides a good approximation. The decoupling limit of these two scenarios corresponds to the case when new physics sets in at a much higher energy scale than the EW scale. However, such kind of decoupling limit does not exist in either the Coleman-Weinberg Higgs or the Tadpole-induced Higgs scenario. In all the above cases, the trilinear and quartic Higgs couplings could be very different from those in the SM.

All the above mentioned scenarios can be described in an effective field theory (EFT) framework. One of the most popular EFT frameworks is the SMEFT~\cite{Buchmuller:1985jz,Grzadkowski:2010es,Giudice:2007fh}, which assumes new physics decouple at a high energy scale, and EW symmetry is in the unbroken phase. The SMEFT  is suitable for describing the elementary Higgs and the Nambu-Goldstone Higgs scenarios, when the Higgs non-linearity effect can be neglected~\cite{Li:2019ghf}. On the other hand, the Coleman-Weinberg Higgs and the Tadpole-induced Higgs scenarios cannot be described within the SMEFT framework due to the existence of non-decoupling effects. Hence, to compare all the four NP scenarios in one theory framework, we utilize the EFT framework in the broken phase of EW symmetry, which is known as the Higgs EFT~\cite{Appelquist:1980vg,Longhitano:1980iz,Koulovassilopoulos:1993pw,Contino:2010mh,Alonso:2012px,Buchalla:2013rka,Buchalla:2013eza}. Adopting the Higgs EFT framework, we summarize the general Higgs effective couplings in various scenarios, and parameterize the scaling behavior of multi-Higgs production cross sections at various high energy hadron colliders.

In this work, we study how to utilize the measurements of the $hh$ and $hhh$ production rates 
in hadron collision to discriminate the above mentioned scenarios. 
The $hh$ production process, via gluon-gluon fusion, has been extensively studied in the literature for measuring the trilinear Higgs boson coupling~\cite{Glover:1987nx,Baur:2003gp,Baur:2003gpa,Dolan:2012rv,Baglio:2012np,Papaefstathiou:2012qe,Goertz:2013kp,Barger:2013jfa,Barr:2013tda,Li:2015yia,deLima:2014dta,Alves:2017ued,Adhikary:2017jtu,Goncalves:2018yva,Borowka:2018pxx,Homiller:2018dgu,Kim:2018cxf,Heinrich:2019bkc,Cepeda:2019klc,Kim:2019wns} and the $t\bar{t}hh$ couplings in the EFT framework~\cite{Kanemura:2008ub,Contino:2012xk, Li:2019uyy}, and for probing various new physics models~\cite{Cao:2015oaa,Cao:2016zob,Chen:2014xra,Kim:2018uty,Huang:2017nnw,Basler:2018dac,Babu:2018uik,Azatov:2015oxa,Dawson:2015oha, Grober:2010yv,Gillioz:2012se,Grober:2016wmf}. In particular, probing the composite Higgs models via studying the $hh$ production process has been studied in~\cite{Grober:2010yv,Gillioz:2012se,Grober:2016wmf}. 
For completeness, we have reproduced some of the results shown in the literature, but with somewhat different emphasis on its analysis so as to compare the predictions on the $hh$ production rates from the above mentioned Higgs potential scenarios side-by-side.

We compare total cross sections, kinematical distributions, and various interference effects at the 14 TeV high luminosity LHC (HL-LHC), the 27 TeV high energy LHC (HE-LHC), and a future $100$ TeV proton-proton (pp) collider (which may be FCC-hh~\cite{Millet:2019oqe} or SppS~\cite{CEPC-SPPCStudyGroup:2015csa}) for various NP scenarios.  We estimate the $1\sigma$ uncertainty in the measurement of cross sections in various scenarios by scaling the SM signal-to-background studies available in the literature~\cite{Goncalves:2018yva}. We find that different scenarios of the Higgs potential can be distinguished by measuring the double-Higgs production cross sections at the HE-LHC and the 100 TeV pp collider. We also consider the possibility of constraining the trilinear Higgs coupling in these scenarios, assuming certain accuracies for the measured cross section. 

Next, we compare total cross sections and kinematical distributions for the process $gg \to hhh$ in various Higgs potential scenarios at the 100 TeV pp collider. 
This process was first studied in~\cite{Plehn:2005nk}, by scaling the SM Higgs boson self-couplings, for exploring the potential of the 100 TeV pp collider for measuring the quartic Higgs coupling. To reduce the backgrounds, the most promising signature has one of the Higgs bosons decaying into the rare decay channel of two photons, while
the other two Higgs bosons each decay into a pair of bottom jets~\cite{Chen:2015gva, Fuks:2015hna}.
In this work, we study various interference effects between the diagrams for $gg \to hhh$ production process in order to understand the dependence of these terms on various couplings, including quartic Higgs couplings. The dependence of the cross sections on the  quartic Higgs coupling is found to be weak. In the composite Higgs model, the presence of $t\bar{t}hhh$ coupling further complicates the situation. Assuming that the triple-Higgs production cross section can be measured to a certain accuracy at the 100 TeV pp collider, we could obtain the possible bounds on the strength of the quartic Higgs coupling.
 We estimate the $1\sigma$ uncertainty in the measurement of cross sections in various scenarios by scaling the SM signal-to-background studies available in the literature~\cite{Papaefstathiou:2015paa,Chen:2015gva,Fuks:2015hna,Fuks:2017zkg,Kilian:2017nio}. 
We find that the potential of the 100 TeV pp collider to discriminate various NP scenarios strongly depends on the tagging efficiency of multiple bottom jets in the data analysis. 
 
The paper is organized as follows. In Sections 2 and 3, we lay out the general framework of Higgs effective couplings and discuss various NP scenarios that could yield a different Higgs potential from the SM. In Section 4, we consider the theoretical constraints on the strength of Higgs boson self-couplings, by examining the conditions of tree-level partial wave unitarity and vacuum stability. In Section 5, we consider the $pp \to hh$ process for its potential to discriminate various Higgs potential scenarios. In Section 6, we examine the usefulness of the process $pp \to hhh$ for measuring the quartic Higgs coupling and for determining the shape of the Higgs potential. Our conclusion is presented in Section 7.

\section{Effective field theory frameworks}
\label{sec:effL}

In an EFT framework, new physics effect in the Higgs sector could be described using Higgs EFT and SMEFT in the broken and unbroken phase of electroweak symmetry, respectively. Higgs EFT could describe all the above-mentioned NP scenarios, while SMEFT is only suitable for describing NP models with decoupling behavior, such as the elementary Higgs scenario and the Nambu-Goldstone Higgs scenario with negligible Higgs non-linearity.

\subsection{Higgs EFT: Higgs in the Broken Phase}
\label{sec:heft}

In the broken phase of electroweak symmetry, it is convenient to use the Higgs EFT Lagrangian~\cite{Appelquist:1980vg,Longhitano:1980iz,Koulovassilopoulos:1993pw,Contino:2010mh,Alonso:2012px,Buchalla:2013rka,Buchalla:2013eza} to describe the interactions of the top quark, the Higgs boson, and the Goldstone bosons eaten by the massive gauge bosons $W^\pm$ and $Z$.\footnote{In this work, we focus on the effects of Higgs boson couplings in the double- and triple-Higgs production processes. Hence, we could take the gauge-less limit, {\it i.e.}, taking $g, g' \to 0$ limit.} Only the $U(1)_{EM}$ symmetry is manifest (or equivalently, the SM gauge symmetry $SU(2)_L\times U(1)_Y$ is non-linearly realized) in the broken phase.  Furthermore, the custodial symmetry $SU(2)_V$ should be respected and the Higgs boson $h$ is taken as a custodial singlet, when constructing the effective Lagrangian.  With the nonlinearly-realized symmetry $SU(2)_L\times SU(2)_R/SU(2)_V$, the leading Higgs EFT Lagrangian, in the limit of turning off gauge couplings, is~\cite{Appelquist:1980vg,Longhitano:1980iz,Koulovassilopoulos:1993pw,Contino:2010mh,Alonso:2012px,Buchalla:2013rka,Buchalla:2013eza}
\begin{align}
\mathcal{L}=\ &\frac{1}{2}(\partial_\mu h)^2-V(h)+\frac{v^2}{4}\text{Tr}[(\partial_\mu U)^\dagger \partial^\mu U]\left(1+2 a \frac{h}{v}+b\frac{h^2}{v^2}+\cdots\right)\nn\\
&-\frac{v}{\sqrt{2}}(\bar{t}_L,\bar{b}_L)U\left(1+c_1\frac{h}{v}+c_2\frac{h^2}{v^2}+c_3\frac{h^3}{v^3}+\cdots\right)\left(\baa{c} y_t t_R\\y_b b_R\eaa\right)+\text{h.c.}\ ,
\label{eff}
\end{align}
where $V(h)$ is the Higgs potential, $U$ is the Goldstone matrix of $SU(2)_L\times SU(2)_R/SU(2)_V$, and 
\bea
U=e^{\frac{iw^a \tau^a}{v}}. 
\label{eq:umatrix}
\eea
Here, $v$ ($= 246$ GeV) denotes the electroweak scale, $\tau^a$ are the Pauli matrices (with $a=1,2,3$), and $w^a$ are the Goldstone bosons eaten by $W^\pm, Z$ bosons. In general, the unknown coefficients $a, b, c_1, c_2$ are independent from each other. The SM corresponds to $a=b=c_1=1$ and other couplings equal to zero.
Note that the standard model gauge symmetry $SU(2)_L\times U(1)_Y$ is a 
subgroup of $SU(2)_L\times SU(2)_R$. After turning on the gauge coupling, 
one needs to replace the usual derivative with the gauge covariant derivative 
in the above equation, as $\partial_\mu\to D_\mu$, and the form in the unitary gauge can be easily obtained by setting $U\to 1$. For convenience, we will work with the above effective Lagrangian with the gauge couplings being turned off.

To be specific, the general Higgs potential is written as 
\begin{align}
V(h)&=\frac{1}{2}m_h^2 h^2+d_3\left(\frac{m_h^2}{2v}\right)h^3+d_4\left(\frac{m_h^2}{8v^2}\right)h^4+\cdots\nn\\
&\equiv\frac{1}{2}m_h^2 h^2+\frac{\lambda_3}{3!} h^3+\frac{\lambda_4}{4!} h^4+\cdots\ ,
\label{pot}
\end{align}
and the Goldstone matrix $U$ can be parameterized as 
\bea
U=\sqrt{1-\left(\frac{w_a \tau^a}{v}\right)^2}+i\frac{w_a \tau^a}{v}\ ,
\eea
where $d_{3,4}$ are the independent Higgs self-couplings. In the SM, $d_3=d_4=1$. It is easy to see that the normalization of the Goldstone matrix satisfies the condition $U^\dagger U=1$, thus the above parameterization for the Goldstone bosons is equivalent to the one in the exponential form, cf. Eq.~(\ref{eq:umatrix}). With this parameterization, the derivative-coupled interactions of the Goldstone bosons take a relatively simple form as
\bea
\text{Tr}[(\partial_\mu U)^\dagger\partial^\mu U]=\frac{2}{v^2}\partial_\mu w^a\partial^\mu w^a+\frac{2}{v^2}\frac{(w^a\partial_\mu w^a)^2}{v^2-w^2}\ ,
\eea
which would give rise to the usual kinetic terms for $w^a$ and their derivative-coupled interactions with the Higgs boson $h$.
In this work, we neglect the effects of heavy particles contributing to the contact interactions between gluons and the Higgs boson, $h^n G^{\hat{a}}_{\mu\nu}G^{\hat{a}\mu\nu}$, as these effective couplings vanish when the heavy particles decouple. For simplicity, we assume these particles are heavy enough and thus the $h^n G^{\hat{a}}_{\mu\nu}G^{\hat{a}\mu\nu}$ interactions can be safely neglected.

\subsection{SM EFT: Higgs in the Unbroken Phase}

Depending on the nature of the Higgs boson, SMEFT could be a good general framework to parameterize the Higgs couplings. In the scenarios of Coleman-Weinberg Higgs and Tadpole-induced Higgs, the SMEFT cannot be used because of the non-decoupling behavior of new particles. On the other hand, the elementary Higgs and Nambu-Goldstone Higgs scenarios could be well described in the SMEFT framework, because of the decoupling feature of these new physics models. In the following, we present the SMEFT framework and provide the correspondence between the SMEFT and the Higgs EFT defined above. 

If the new physics scale $\Lambda$ is much higher than the electroweak scale, and can be decoupled as $\Lambda \to \infty$, the SMEFT with higher-dimensional operators will be a useful framework to describe the effects of new physics at the weak scale. 
 The SM gauge symmetry $SU(2)_L\times U(1)_Y$ is manifested (or linearly-realized) in this case.
Neglecting lepton-number violating operator at the dimension $D=5$ (irrelevant to our study), the leading effective operators arise from dimension $D=6$. The non-redundant set of $D=6$ operators was summarized in Ref.~\cite{Grzadkowski:2010es}, known as the Warsaw basis. There are $53$ CP-even and $6$ CP-odd effective operators at the $D=6$ level. In this paper, we will focus on the CP-conserving case. By employing equations of motion, we can translate the $D=6$ operators in the Warsaw basis to the ones in the so-called strongly-interacting light Higgs (SILH) basis~\cite{Giudice:2007fh}. The Rosetta package~\cite{Falkowski:2015wza} can be used for translating between different bases. The main difference between these two bases are the operators involving fermionic currents (in the Warsaw basis) and the ones involving pure bosonic fields (in SILH basis); as 
$\sum_{\psi} Y_\psi O_{H\psi}\sim O_T, O_B$, and $O^\prime_{Hq}+O^\prime_{HL}\sim O_W$, where the sum is over all fermions with $Y_\psi$ denoting the corresponding hypercharge of $\psi$~\cite{Contino:2013kra}. When considering the $S$ parameter constraint, it is more convenient to consider $O_B$ and $O_W$ instead of $O_{H\psi}$ or $O^\prime_{Hq}, O^\prime_{HL}$, as the latter operators can induce vertex corrections and modify the Fermi constant. Furthermore, the operators such as $O_{WW}$ and $O_{BB}$ (in the Warsaw basis) can be reparameterized by the linear combinations of the operators $O_{W,B,HW,HB,\gamma}$ (in the SILH basis)~\cite{Contino:2013kra}. 

For discussing the production processes of multi-Higgs bosons via gluon-gluon fusion, we list the following relevant $D=6$ operators as
\begin{align}
\mathcal{L}_{D=6}=&\ \frac{c_H}{2\Lambda^2}\partial_\mu (H^\dagger H) \partial^\mu (H^\dagger H)-\frac{c_6}{\Lambda^2}\lambda (H^\dagger H)^3-\left(\frac{c_t}{\Lambda^2}y_t H^\dagger H \bar{Q}_L H^c t_R+\rm{h.c.}\right)\nn\\
&+\frac{\alpha_s}{4\pi} \frac{c_g}{\Lambda^2} H^\dagger H G^a_{\mu\nu} G^{a\mu\nu}+\frac{\alpha^\prime}{4\pi}\frac{c_\gamma}{\Lambda^2} H^\dagger H B_{\mu\nu} B^{\mu\nu}\ ,
\label{silh}
\end{align}
where $\lambda$ and $y_t$ are, respectively, the SM quartic Higgs coupling and the top 
Yukawa coupling. $\alpha_s=g_s^2/4\pi$, $\alpha^\prime=e^2/4\pi$, and $c_i (i=H, 6, t, g, \gamma)$ are unknown Wilson coefficients. 
It is worth pointing out that there is another operator $O_T=\frac{1}{2}\left (H^\dagger {\overleftrightarrow{D}_\mu} H\right)^2$ which violates custodial symmetry at tree level, thus we neglect it in the following discussion.
Further complication introduced by the flavor structure of the $D=6$ Yukawa term will not be explored in this paper.

\subsection{Relating SM EFT to Higgs EFT}

Since the Higgs EFT is formulated at the broken phase of the electroweak symmetry, it is a more general description than the SM EFT. Hence, we could identify the SM EFT Wilson coefficients with the 
Higgs EFT coefficients.

With appropriate field redefinitions taken into account~\cite{Goertz:2014qta}, we can match the Higgs-Goldstone couplings, Higgs-top couplings, and Higgs self-couplings defined in Eqs.~(\ref{eff}) and~(\ref{pot}) to the Wilson coefficients in Eq.~(\ref{silh}), as
\begin{align}
a&=1-c_H \frac{v^2}{2\Lambda^2}+\mathcal{O}(\frac{1}{\Lambda^4})\ ,\\
b&=1-c_H \frac{2v^2}{\Lambda^2}+\mathcal{O}(\frac{1}{\Lambda^4})\ ,
\end{align}
\begin{align}
c_1&=1-c_H \frac{v^2}{2\Lambda^2}+c_t\frac{v^2}{\Lambda^2}+\mathcal{O}(\frac{1}{\Lambda^4})\ ,\\
c_2&=c_t \frac{3 v^2}{2 \Lambda^2}-c_H \frac{v^2}{2\Lambda^2}+\mathcal{O}(\frac{1}{\Lambda^4})\ ,\label{eq:wilsoncoef_c2}\\
c_3&=c_t \frac{v^2}{2 \Lambda^2}-c_H \frac{v^2}{6\Lambda^2}+\mathcal{O}(\frac{1}{\Lambda^4})\ ,
\end{align}
\begin{align}
d_3&=1+c_6\frac{v^2}{\Lambda^2}-c_H\frac{3v^2}{2\Lambda^2}+\mathcal{O}(\frac{1}{\Lambda^4})\ ,\label{eq:wilsoncoef1} \\
d_4&=1+c_6\frac{6 v^2}{\Lambda^2}-c_H\frac{25 v^2}{3\Lambda^2}+\mathcal{O}(\frac{1}{\Lambda^4})\ .
\label{eq:wilsoncoef2}
\end{align}

As we will see later, different Higgs couplings in the Higgs EFT are usually correlated for a given NP model. For simplicity, we again assume that the heavy particles decouple and the coefficients  $c_g$ and $c_\gamma$ vanish, though in general these two effective operators can be induced by heavy particles with nontrivial color or electric charges circulating in loops.

\section{Various Higgs Scenarios}
\label{sec:model}

In contrast to the model-independent discussions presented in the last section, we explicitly derive the Higgs effective couplings in some specific NP scenarios, {\it i.e.}, the elementary Higgs, Nambu-Goldstone Higgs, Coleman-Weinberg Higgs, and Tadpole-induced Higgs. To identify the Higgs boson's nature through Higgs self-interactions, we will derive the trilinear and quartic Higgs couplings for each scenario. Since different Higgs couplings are usually correlated for a specific NP model, we will also present the relevant $hVV (V=W^\pm,Z)$,  $ht\bar{t}$ and $hht\bar{t}$ couplings, when necessary. 

\subsection{Elementary Higgs Boson}

When the Higgs boson is an elementary scalar, we include in the Ginzburg-Landau potential, as in the SM, the dimension-six operator $(H^\dagger H)^3$ to effectively describe the new physics contributions parameterized in the SMEFT, cf. Eq.~(\ref{silh}).
In the NP models with scalar extensions, such as the singlet extension, the two Higgs doublet model, the real and complex triplets and the quadruplet models~\cite{Corbett:2017ieo, Dawson:2017vgm, Belusca-Maito:2016dqe}, the $(H^\dagger H)^3$ operator can be induced, which has been classified in Ref.~\cite{Corbett:2017ieo} based on group theory construction. 
Similarly, integrating out new heavy fermions and gauge bosons at the one-loop level 
could also induce the $(H^\dagger H)^3$ operator.

To be specific, 
the Ginzburg-Landau potential considered in this work is
\bea
V=-\mu^2 H^\dagger H+\lambda(H^\dagger H)^2+\frac{c_6}{\Lambda^2}\lambda (H^\dagger H)^3\ ,
\eea
where the Higgs doublet is $H=1/\sqrt{2}\ (0,v+h)^T$ in the unitary gauge, the Higgs boson mass term is $m_h^2=2\lambda v^2 (1+\frac{3 c_6 v^2}{2\Lambda^2})$, and the electroweak scale $v$ is obtained by solving
\bea
\mu^2=\lambda v^2 (1+\frac{3}{4}\frac{c_6v^2}{\Lambda^2}).
\eea
In the SMEFT description, the trilinear and quartic Higgs couplings are
\begin{align}
d_3&=1+c_6\frac{v^2}{\Lambda^2}-c_H\frac{3v^2}{2\Lambda^2}+\mathcal{O}(\frac{1}{\Lambda^4})\ ,\\
d_4&=1+c_6\frac{6 v^2}{\Lambda^2}-c_H\frac{25 v^2}{3\Lambda^2}+\mathcal{O}(\frac{1}{\Lambda^4})\ .
\end{align}

Here $c_H$, cf. Eq.~(\ref{silh}), modifies the kinetic term of the Higgs field, which universally shift the Higgs couplings  to electroweak gauge bosons. Thus, the coefficient $c_H$ is highly constrained by the measurement of the couplings of Higgs boson to weak gauge bosons. 
The coefficient $c_t$ is constrained by the measurements of $t\bar{t}h$ and Higgs production cross section via gluon-gluon fusion process.  
To probe the Higgs self-couplings, we assume in this work that the operator $(H^\dagger H)^3$ makes the most significant NP contribution and the other operators can be safely neglected.

\subsection{Nambu-Goldstone Higgs Boson}
\label{sec:pngb}

The Higgs boson can be a pseudo Nambu-Goldstone boson~\cite{Kaplan:1983fs,Kaplan:1983sm}, arising from strong dynamics at the TeV scale. The pseudo Nambu-Goldstone Higgs corresponds to one of the broken generators for some spontaneously broken global symmetry $\mathcal{G}/\mathcal{H}$, based on which all the operators, consistent with Higgs nonlinearity, can be systematically constructed~\cite{Coleman:1969sm,Callan:1969sn}.

With its PNG nature, the general Higgs potential is approximately
\bea
V(h) = -A f^4 \sin^2\left(\frac{h}{f}\right) + B f^4 \sin^4\left(\frac{h}{f}\right)+\cdots.
\eea
with higher order terms being neglected, where $A$ and $B$ are the two coefficients whose values are determined by the specific dynamics responsible for generating the Higgs potential, and $4\pi f$ denotes the NP scale. With the above notation, the coefficients $A$ and $B$ are positive. One can further define a ratio between the electroweak scale $v$ and the NP scale $f$ to denote the Higgs nonlinearity in this scenario. 
To be specific, the minimization condition of the Higgs potential gives\footnote{It is nontrivial to realize a small $\xi$ (less than about $0.1$), as required by precision measurements of Higgs couplings and electroweak precision data. See, {\it e.g.,} Ref.~\cite{Yu:2016bku,Yu:2016swa,Xu:2018ofw,Xu:2019xuo} for recent attempts to achieve this goal. It is also found experimentally challenging to extract out small $\xi$ values from measuring the Higgs couplings at the LHC~\cite{Cao:2018cms}. Note that the parameter $\xi$ is positive for compact cosets, while it is negative for non-compact cosets~\cite{Alonso:2016btr}. In this work, we only focus on models with compact cosets, as EWSB is difficult to be triggered in the models based on non-compact cosets.}
\bea
\xi\equiv\frac{v^2}{f^2} = \sin^2\left(\frac{\langle h \rangle}{f}\right) = \frac{A}{2B}.
\eea
By expanding the Higgs potential in the powers of $h$ after EWSB, we have
\begin{align}
V(h)= B f^2 \sin^2\left(\frac{2\langle h \rangle}{f}\right) h^2 
+ B f \sin\left(\frac{4\langle h \rangle}{f}\right) h^3 
+ B\left(-\frac 16 + \frac 76 \cos\left(\frac{4\langle h \rangle}{f}\right) \right) h^4+\cdots\ ,\nn\\
\end{align}
The Higgs boson mass is given by
\bea
 m_h^2 = 2 B f^2 \sin^2\left(\frac{2\langle h \rangle}{f}\right),
\eea
and the trilinear and quartic Higgs couplings are, respectively,
\begin{align}
d_3&=\frac{B f \sin\left(\frac{4\langle h \rangle}{f}\right)}{\left(\frac{m_h^2}{2v}\right)}=\frac{1-2\xi}{\sqrt{1-\xi}}\ ,\\
d_4&=\frac{\frac16 B \left(-1 + 7 \cos\left(\frac{4\langle h \rangle}{f}\right) \right)}{\left(\frac{m_h^2}{8v^2}\right)}=\frac{28\xi^2-28\xi+3}{3-3\xi}\ ,
\end{align}
where the ratio of $d_3$ and $d_4$ is obviously not one, depending on the parameter $\xi$.

Due to the Higgs nonlinear effects associated with its nature as a PNG, 
the Higgs couplings in the top sector (the $h\bar{t}t$, $hh\bar{t}t$, $hhh\bar{t}t$ couplings) and the Higgs couplings with electroweak gauge bosons can deviate from the SM values. Regarding the Higgs couplings in the top sector, the $h\bar{t}t$ and $hh\bar{t}t$, $hhh\bar{t}t$ couplings depend on the representation in which the top quark is embedded. 
As the two benchmarks, we consider the minimal composite Higgs model (MCH or MCHM)~\cite{Agashe:2004rs,Contino:2006qr}, where both the left-handed $t_L$ and right-handed $t_R$ are embedded in the fundamental representation $5$ of the global $SO(5)$ symmetry; and the composite twin Higgs model (CTH or CTHM)~\cite{Geller:2014kta,Barbieri:2015lqa,Low:2015nqa}, where the left-handed $t_L$ is embedded in the fundamental representation $8$, while the right-handed $t_R$ is a singlet of the global $SO(8)$ symmetry. The Higgs couplings in these two models are systematically derived in Ref.~\cite{Li:2019ghf} and collected in Table~\ref{tab:coupling}.

\subsection{Coleman-Weinberg Higgs Boson} 

Another theoretically attractive scenario is the Coleman-Weinberg Higgs, where the Higgs potential at the classical level is assumed to be scale invariant, i.e. only the quartic Higgs term is present at 
the tree level~\cite{Hill:2014mqa,Helmboldt:2016mpi,Hashino:2015nxa}. However, with quantum corrections, the Higgs mass term is usually generated at the one-loop level through the Coleman-Weinberg mechanism~\cite{Coleman:1973jx}. To be specific, the Higgs self-couplings are essentially determined by the $\beta$-function of the quartic Higgs coupling $\lambda$, and the electroweak scale $v$ is generated at quantum level~\cite{Hill:2014mqa}. The $\beta$-function of the quartic Higgs coupling $\beta_{\lambda}$ is positive-definite, and accordingly the running quartic coupling at the EW scale $\lambda (v)$ is negative~\cite{Hill:2014mqa}, which corresponds to the minimum of the Higgs potential.

The general Coleman-Weinberg Higgs boson $h$ has the following potential
\bea
V(h) = A h^4 + B h^4 \log\frac{h^2}{\Lambda_{\rm GW}^2},
\eea
where
\bea
	A = \sum_i n_i \frac{m_i^4}{64\pi^2 v^4}\left( \log\frac{m_i^2}{v^2} - c_i\right), \quad
	B = \sum_i n_i \frac{m_i^4}{64\pi^2 v^4}.
\eea
Here, the masses $m_i$\footnote{The specific types of the particles running in the loop is irrelevant at the one-loop order.} denote the masses of the particles circulating in the loop, which are defined in the vacuum background, $n_i$ denotes the internal degrees of freedom, and $c_i$ is the renormalization-scheme dependent constant\footnote{For example, in the $\overline{\text{MS}}$ scheme, $c_i=\frac{5}{6}$ for gauge bosons, while $c_i=\frac{3}{2}$ for scalars and fermions.}.  
The parameter $B$ is directly related to the $\beta$-function of the quartic Higgs coupling $\beta_{\lambda}$.
The minimization condition $\frac{\text{d}V(\langle h\rangle)}{\text{d}\langle h\rangle}=0$ gives~\cite{Gildener:1976ih} 
\bea
v=\langle h \rangle = \Lambda_{\rm GW} \exp\left[-\frac14 - \frac{A}{2B}\right],
\eea
which leads to a relation between $A$ and $B$. At this minimum, the running quartic coupling at the EW scale $\lambda (v)$ is negative. 
Since $v$ is determined from the dimensionless parameters, this is one specific realization of the dimensional transmutation mechanism.

After expanding the above Higgs potential in the powers of $h$, after EWSB, we obtain
\bea
V(h) \simeq  4 B  \langle h \rangle^2  h^2 + \frac{20}{3} B \langle h \rangle  h^3  +  \frac{11}{3} B   h^4+\cdots.
\eea
Here, all the Higgs self-couplings are related to the parameter $B$ (or equivalently $\beta_{\lambda}$). 
Note that the higher order terms, such as $h^5$, are neglected here. 
Therefore, the Higgs mass is
\bea
 m_h^2 = 8 B  \langle h \rangle^2
\eea
and the trilinear and quartic Higgs couplings are, respectively,
\begin{align}
d_3&=\frac{\frac{20}{3} B \langle h \rangle}{\left(\frac{m_h^2}{2\langle h\rangle}\right)}=\frac{5}{3}\ ,\\
d_4&=\frac{ \frac{11}{3} B}{\left(\frac{m_h^2}{8\langle h\rangle^2}\right)}=\frac{11}{3}\ .
\end{align}
We note that the trilinear and quartic Higgs couplings are fixed at the one-loop order. Small corrections to the above relations of $d_3$ and $d_4$ would appear only at the two-loop or higher orders~\cite{Hill:2014mqa,Helmboldt:2016mpi}.

\subsection{Tadpole-Induced Higgs Boson} 
 
Another interesting scenario is the Tadpole-induced Higgs. 
Because of the existence of a non-zero tadpole term, the electroweak symmetry is spontaneously broken.
 As a result, the Higgs self-couplings, both the trilinear and quartic Higgs couplings, are largely suppressed with respect to the SM prediction. 
In such models, an additional source of electroweak symmetry breaking other than the SM Higgs mechanism is needed. One specific realization of this class of models is the bosonic technicolor model~\cite{Simmons:1988fu,Carone:1993xc}. In the typical technicolor models~\cite{Susskind:1978ms, Weinberg:1975gm}, only the condensate of technifermions $\langle\bar{Q}_iQ_j\rangle \sim \Lambda_{\rm tech}^3$ triggers EWSB, and thus it predicts no Higgs boson. However, this has been ruled out due to the discovery of the Higgs boson at the LHC. On the other hand, in the bosonic technicolor model, an elementary Higgs boson is also there to trigger EWSB with vacuum expectation value (VEV) $v_h$ such that
\bea
v^2 \equiv v_h^2 + f^2,
\eea
where $f \equiv  \Lambda_{\rm tech}$. 
As both scalars can contribute to the $W^\pm$ and $Z$ boson masses, the scale $f$ should be suppressed with respect to the electroweak scale $v$,
so that the $hVV$ (with $V=W^\pm, Z$) couplings can be close to the SM predictions, as required by the experimental findings. This leads to $v\simeq v_h \gg f$.

At the low energy, the bosonic technicolor condensate could be parameterized as another effective scalar doublet field with the same quantum numbers as of the Higgs doublet. 
For convenience, let us name this auxiliary doublet as $\Sigma$, and the $\Sigma$ field is interpreted as the condensate of technifermions $\Sigma\sim\langle\bar{Q}_iQ_j\rangle/\Lambda_{\rm tech}^2$.
The simplified Lagrangian~\cite{Galloway:2013dma,Chang:2014ida} for the Tadpole-induced Higgs scenario is 
\bea
\mathcal{L}=(D^\mu H)^\dagger(D_\mu H)+(D^\mu\Sigma)^\dagger(D_\mu\Sigma)-V(H,\Sigma)
\eea
where
\bea
V(H,\Sigma)=m_h^2H^\dagger H-\left(\epsilon \Sigma^\dagger H+\textrm{h.c.}\right) - m_\Sigma^2 \Sigma^\dagger \Sigma + \lambda_S \left(\Sigma^\dagger \Sigma\right)^2.
\label{eq:induce_Lag}
\eea
Note that the mass term of the Higgs doublet $H$ is positive, so EWSB is not triggered by the $m_h^2H^\dagger H$ term as in the SM. In order for the Tadpole-induced mechanism to be dominant, the quartic term $\lambda_H (H^\dagger H)^2$ should be sub-dominant (thus negligible) in the above Higgs potential.
The vacuum structure is then parameterized as
\bea
\Sigma=\frac{1}{\sqrt{2}}\left(\baa{c} 0\\f\eaa\right), \ \ \ H=\frac{1}{\sqrt{2}}\left(\baa{c} 0\\v_h\eaa\right) \, ,
\eea
where the VEV $f$ is obtained from the sector with auxiliary doublet alone, and $v_h$ is obtained from the  $m_h^2H^\dagger H$ term and the mixing term between the two scalar sectors, such that
\bea
v_h=\frac{\epsilon f}{m_h^2}.
\eea

\begin{figure}
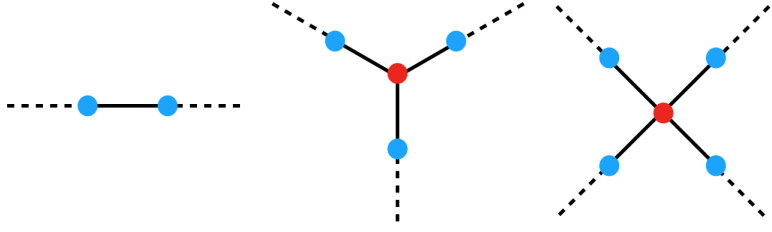

\begin{center}
\includegraphics[scale=0.45]{plots/induced_EWSB1.pdf}
\includegraphics[scale=0.45]{plots/induced_EWSB2.pdf}
\includegraphics[scale=0.45]{plots/induced_EWSB3.pdf}
\caption{The trilinear and quartic Higgs couplings obtained in the Tadpole-induced Higgs scenario. The blue dot denotes the mixing between the two doublets $H$ and $\Sigma$, while the red dot denotes the self-couplings of the auxiliary doublet $\Sigma$.}
\label{induced_EWSB}
\end{center}
\end{figure}

More interestingly, the self-couplings of the Higgs boson are highly suppressed in this class of models. Let us assume that the Higgs particle in the auxiliary scalar sector is heavy enough ($v\ll m_{\Sigma}$) such that one can integrate out the auxiliary scalar field and derive the tree-level effective potential for the Higgs boson. Because of the self-interactions of the auxiliary scalar and the mixing between the auxiliary field and the Higgs boson, trilinear and quartic Higgs couplings are induced, as shown in Fig.~\ref{induced_EWSB}.
To be specific, we have the tree-level effective Higgs potential as
\begin{align}
V(H)=\frac{1}{2}m_h^2 H^\dagger H-\epsilon f \sqrt{H^\dagger H}+&\left(\frac{\epsilon^2}{m_{\Sigma}} \left(\sqrt{H^\dagger H}\right)^2+\left(\frac{\epsilon}{m_{\Sigma}^2}\right)^3\frac{m^2_{\Sigma}}{f}\ \left(\sqrt{H^\dagger H}\right)^3\right.\nn\\
&+\left.\left(\frac{\epsilon}{m_{\Sigma}^2}\right)^4\frac{m^2_{\Sigma}}{4f^2}\ \left(\sqrt{H^\dagger H}\right)^4+\cdots\right)&\, ,
\end{align}
where, $H^\dagger H=(h+v_h)^2/2$ 
in the unitary gauge.
When $m_{\Sigma}$ is sufficiently large,  which in turn requires the self-interactions of the auxiliary scalar field to be strong, all the self-coupling terms of the Higgs field $h$ are suppressed. Hence, after performing a shift of ($h\to h+v_h$), after EWSB, to remove the tadpole term in Eq.~(\ref{eq:induce_Lag}), we find the self-couplings of the physical Higgs boson yield $d_3\simeq d_4\simeq 0$. 
However, if the Higgs potential $V(H,\Sigma)$ contains a quartic term $\lambda_H (H^\dagger H)^2$, the Higgs self-couplings can yield non-zero $d_3$ and $d_4$ values. In this work, we simply assume that the quartic term $\lambda_H (H^\dagger H)^2$ vanishes, cf. Eq.~(\ref{eq:induce_Lag}).

\subsection{Summary on Higgs Couplings}

We collect all the relevant Higgs couplings in Table~\ref{tab:coupling} for different NP scenarios -- the elementary Higgs (both the SM and the SMEFT with the operator $O_6$), Nambu-Goldstone Higgs (MCH and CTH models), Coleman-Weinberg Higgs and Tadpole-induced Higgs. As we will see, these couplings are important for deriving theoretical constraints, from the partial wave unitarity, the tree-level vacuum stability, and the study of the phenomenology of the double-Higgs production $gg\to hh$ and the triple-Higgs production $gg\to hhh$ at the LHC and the future hadron colliders.

\begin{table}
 \begin{center}
  \begin{tabular}{c|c|c|c|c|c|c|c}
   \hline\hline
 \  & $a$ & $b$ & $c_1$ & $c_2$ & $c_3$ & $d_3$ & $d_4$  \\[8pt]
  relevant couplings & $hVV$ & $hhVV$ & $h\bar{t}t$ & $hh\bar{t}t$ & $hhh\bar{t}t$ & $hhh$ & $hhhh$ \\
   \hline  
  SM  & $1$ & $1$ & $1$ & $0$ & $0$ & $1$ & $1$ \\ 
  \hline
  SMEFT (with $O_6$) & $1$ & $1$ & $1$ & $0$ & $0$ & $1+c_6\frac{v^2}{\Lambda^2}$ & $1+c_6\frac{6 v^2}{\Lambda^2}$ \\  
   \hline
  $\text{MCH}_{5+5}$ & $ 1-\frac{\xi}{2}$ & $ 1-2\xi$ & $ 1-\frac{3}{2}\xi$ & $ -2\xi$ & $ -\frac{2}{3}\xi$ &$ 1-\frac{3}{2}\xi$ & $ 1-\frac{25}{3}\xi$ \\
   \hline
   $\text{CTH}_{8+1}$ & $ 1-\frac{\xi}{2}$ & $ 1-2\xi$ & $ 1-\frac{1}{2}\xi$ & $ -\frac{1}{2}\xi$ & $ -\frac{1}{6}\xi$ & $ 1-\frac{3}{2}\xi$ & $ 1-\frac{25}{3}\xi$ \\
   \hline
   CW Higgs (doublet) & $1$ & $1$ & $1$ & $0$ & $0$ & $\frac{5}{3}(1.75)$ & $\frac{11}{3}(4.43)$ \\
   \hline
   CW Higgs (singlets) & $1$ & $1$ & $1$ & $0$ & $0$ & $\frac{5}{3}(1.91)$ & $\frac{11}{3}(4.10)$ \\
   \hline
   Tadpole-induced Higgs & $\simeq 1$ & $\simeq 1$ & $\simeq 1$ & $0$ & $0$ & $\simeq 0$ & $\simeq 0$ \\
   \hline\hline
  \end{tabular}
 \end{center}
  \caption{Higgs couplings, defined in Eqs.~(\ref{eff}) and~(\ref{pot}), for the SM and various NP scenarios. For the Coleman-Weinberg (CW) Higgs scenario, we also present  in the parenthesis the Higgs self-couplings up to the two-loop order, predicted in the two of the simplest conformal extensions of the scalar sector: SM Higgs doublet with another doublet~\cite{Hill:2014mqa}, and SM Higgs doublet with two additional singlets~\cite{Helmboldt:2016mpi}.}
  \label{tab:coupling}
\end{table}

Below, we summarize the specific assumptions made in each class of NP models, which yield the Higgs couplings listed in Table~\ref{tab:coupling}. 
\bit
\item In the SMEFT scenario, we only include the $O_6$ operator, for simplicity, since almost all the other operators are (and will be further) constrained by the precision measurements of the Higgs boson couplings to gauge bosons or fermions.
\item In the Nambu-Goldstone Higgs scenario, the Higgs self-couplings depend on the Higgs nonlinearity parameter $\xi$, whose value has been constrained by the precision $hVV$ coupling measurements. It is found that $\xi<0.1$ at the 1$\sigma$ level. To be concrete, we restrict ourselves to two specific benchmark models, $\text{MCH}_{5+5}$ and $\text{CTH}_{8+1}$. For consistency, we have included the Higgs nonlinear effects in deriving all the Higgs boson couplings.
Here, we neglect the contribution of the composite states to Higgs couplings, by assuming that all the composite particles are heavy enough. The effects of composite particles in Higgs couplings have been systematically discussed in Ref.~\cite{Li:2019ghf}. 
\item In the Coleman-Weinberg Higgs scenario, we simply take all the other Higgs couplings, except the Higgs self-couplings $d_3$ and $d_4$, to be identical to the SM values. This is the case when the extra scalar particles do not mix with the Higgs boson after the EWSB. It is found that the Higgs self-couplings $d_3=\frac{5}{3}$ and $d_4=\frac{11}{3}$, at the one-loop order. For completeness, their values at the two-loop order~\cite{Hill:2014mqa,Helmboldt:2016mpi} are also included in Table~\ref{tab:coupling}.
\item In the Tadpole-induced Higgs scenario, we approximate $d_3=d_4\simeq 0$, as they can be highly suppressed, though their exact values would depend on the self-couplings of the auxiliary scalar field. We also simply neglect the mixing between the auxiliary doublet and the Higgs doublet, as it is required by the result of the precision $hVV$ coupling measurement.
\eit

\section{Theoretical Constraints on Higgs Self-couplings}
\label{sec:unitarity}

\subsection{Tree-level Perturbative Unitarity}

In this section, we aim to obtain the unitarity constraints on Higgs couplings defined after the EWSB, especially, on the trilinear and quartic Higgs couplings. We adopt the method of the coupled-channel analysis to obtain the optimal bound~\cite{Lee:1977eg,Chanowitz:1978mv}, since the most restrictive limit would come from the largest eigenvalue of the matrix for all the coupled scattering processes. For constraining the trilinear and quartic Higgs couplings, we, therefore, consider the electric-neutral channels for the scatterings between the top quark ($t$), longitudinal $W^\pm$ and $Z$, and the Higgs boson at the energy $\sqrt{s}\gg m_t, m_W, m_Z, m_h$. According to the Goldstone equivalence theorem~\cite{Lee:1977eg}, the longitudinal $W^\pm$ and $Z$ are equivalent to the Goldstone bosons ($w^a$) when $\sqrt{s}\gg m_W, m_Z$. 

To be specific, the following coupled $2\to 2$ scattering processes at the tree level are considered
\begin{align}
&t^{\lambda_1}\bar{t}^{\lambda_2}\to t^{\lambda_3}\bar{t}^{\lambda_4},\ \ t^{\lambda_1}\bar{t}^{\lambda_2}\to w^b w^b,\ \ t^{\lambda_1}\bar{t}^{\lambda_2}\to hh,\nn\\
&w^a w^a\to t^{\lambda_3}\bar{t}^{\lambda_4},\ \ w^a w^a\to w^b w^b,\ \ w^a w^a\to h h,\nn\\
&hh\to t^{\lambda_3}\bar{t}^{\lambda_4}, \ \ hh \to w^b w^b, \ \ hh\to hh,
\end{align}
where $\lambda_{1,2,3,4}=\pm$ denote the helicity of the initial-state and final-state top and anti-top quarks, while $a=1,2,3$ and $b=1,2,3$ are the flavor indices for the initial and final state Goldstone bosons, respectively. It is worth noticing that the scattering process $w^a w^a\to w^b w^b$ occurs only when $a\ne b$.

In the isospin basis, the $2\to 2$ matrix element $\mathcal{M}_{if}(\sqrt{s}, \cos\theta)$ can be decomposed into partial waves ($a_j$) as
\bea
\mathcal{M}_{if}(\sqrt{s}, \cos\theta)=32 \pi \sum_{j=0}^{\infty}\frac{2j+1}{2}\ a_j(\sqrt{s})\ P_j(\cos\theta) \, ,
\eea
where $P_j(\cos\theta)$ are the orthogonal Legendre polynomials. Therefore, partial wave amplitudes are
\bea
a_j(\sqrt{s})=\frac{1}{32\pi}\int_0^\pi\text{d}\theta\sin\theta\ P_j(\cos\theta)\ \mathcal{M}_{if}(\sqrt{s}, \cos\theta)\ ,
\eea
which are bounded at the tree level as
\bea
|\text{Re}(a_j)|<\frac{1}{2}, \
\eea
to satisfy partial wave unitarity. For the coupled channels listed above, the $s$-wave ($j=0$) scattering matrix at high energies, $\sqrt{s}\gg m_t, m_W, m_Z, m_h$, is explicitly
\bea
a_0(\sqrt{s})=\frac{3}{16\pi}\frac{m_t}{v^2}
\left(\baa{cccc} 
-(c_1^2+1)m_t & 0 & (1-ac_1)\sqrt{\frac{s}{3}} & -2c_2\sqrt{\frac{s}{3}}\\ 
0 & -(c_1^2+1)m_t & (-1+ac_1)\sqrt{\frac{s}{3}} & 2c_2\sqrt{\frac{s}{3}}\\
(1-ac_1)\sqrt{\frac{s}{3}} & (-1+ac_1)\sqrt{\frac{s}{3}} & \frac{s}{3m_t}(1-a^2) & -\frac{s}{3m_t}(b-a^2)\\
-2c_2\sqrt{\frac{s}{3}} & 2c_2\sqrt{\frac{s}{3}} & -\frac{s}{3m_t}(b-a^2) & -d_4\frac{m_h^2}{m_t}
\eaa\right)
\label{scattering}
\eea
in the basis
\bea
\left\{t^{+}\bar{t}^{+},\ t^{-}\bar{t}^{-},\ \frac{1}{\sqrt{2}}w^a w^a,\ \frac{1}{\sqrt{2}}hh\right\}.
\eea
Here the factor $\frac{1}{\sqrt{2}}$ is due to identical particles in the initial and final states. Note that the states $t^{+}\bar{t}^{-}$ and $t^{-}\bar{t}^{+}$ do not contribute to the $s$-wave scatterings. 
For any given NP model, we can always diagonalize the scattering matrix in Eq.~(\ref{scattering}) numerically.

{\em Elementary Higgs, CW Higgs, Tadpole-induced Higgs in $2\to 2$ scatterings: }
The $s$-wave unitarity bounds on $d_3$ and $d_4$, obtained from the above $2\to 2$ processes, are quite loose if the $hV_LV_L$ couplings ($V=W^\pm,Z$) equal to the SM predictions. This corresponds to the elementary Higgs, Coleman-Weinberg Higgs and Tadpole-induced Higgs.
Moreover, many channels would further decouple when the $t\bar{t}hh$ contact interaction vanishes, and in this case we can solve the $s$-wave unitarity constraints on $d_4$ analytically. This leads to the result 
\bea
\lim_{\sqrt{s}\to\infty}|a_0(\sqrt{s})|=\frac{|d_4|}{32\pi}\frac{3 m_h^2}{v^2}<\frac{1}{2}.
\eea
Roughly, $|d_4|<16\pi$. The coefficient $d_3$ is only moderately bounded as $|d_3-1|<5$~\cite{DiLuzio:2017tfn}.
Alternatively, this bound on $d_4$ can be translated into the bound on the Wilson coefficient $c_6/\Lambda^2$ for the case of the SMEFT, which yields $|c_6|<\left(\frac{16\pi v^2}{3 m_h^2}-1\right)\frac{\Lambda^2}{6v^2}$. 

\begin{figure}[h]
\begin{centering}
\includegraphics[width=0.8\linewidth]{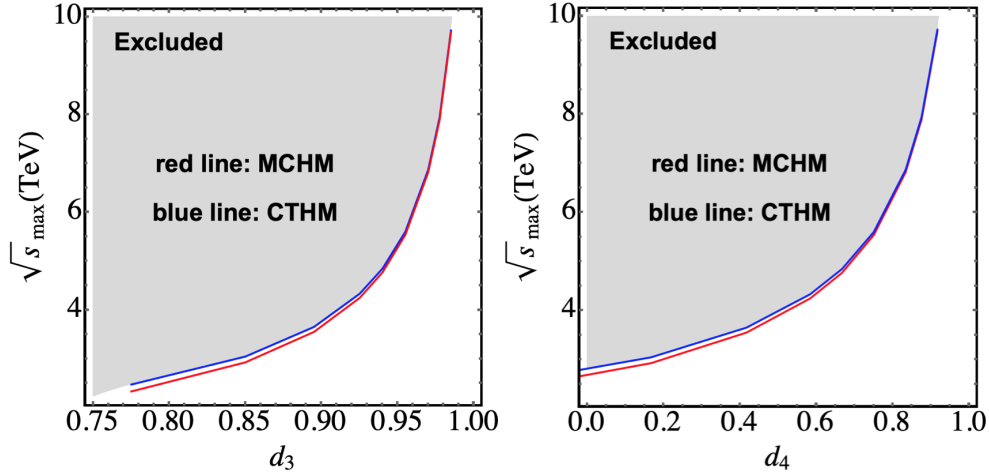}
\caption{Unitarity constraints on the trilinear (left) and quartic (right) Higgs couplings for the Nambu-Goldstone Higgs scenario. The vertical axis denotes the unitarity-violating scale while the horizontal axis denotes the Higgs self-interactions. The red line denotes the MCH, while the blue line denotes the CTH. The shaded region is excluded by unitarity.}
 \label{fig:unitarity}
 \end{centering}
\end{figure}

{\em Nambu-Goldstone Higgs in $2\to 2$ scatterings:}
When $hV_LV_L$ couplings ($V=W^\pm,Z$) deviate from the SM values, the $s$-wave unitarity bound from $2\to 2$ scatterings could be quite stringent\footnote{The unitarity bound mainly results from the deviation of Higgs-Goldstone (eaten by EW gauge bosons in the unitary gauge) couplings. One can explicitly check the eigenvectors after diagonalizing the scattering matrix and find that the $w^a w^a \to w^b w^b$ ($a\ne b$) channel contributes the most to the eigenstate that violates the $s$-wave unitarity.}. 
This applies to the case of the Nambu-Goldstone Higgs scenario due to the Higgs nonlinearity. The unitarity violating scale is found to be $\sqrt{s}\simeq 3 \ \rm{TeV}$, if the nonlinearity parameter $\xi\simeq 0.1$, which yields $d_3\simeq 1- \frac{3}{2}\xi\simeq 0.85$. However, this unitarity bound could be weakened with appropriate composite resonances in the bosonic sector~\cite{Contino:2011np}. In this work, we neglect the effects from those composite resonance states, by assuming that they are all very heavy. 
In Fig.~\ref{fig:unitarity}, we recast the unitarity constraints on Higgs self-couplings $d_3$ and $d_4$, with $\xi$ varying in the range $0.01<\xi<0.15$, for the NG Higgs scenario.

{\em CW Higgs and Tadpole-induced Higgs beyond $2\to 2$ scatterings:}
It is interesting to notice that a relatively stronger unitarity bound on the Higgs self-couplings can be obtained from $2\to n\ (n>2)$ processes, if the Higgs potential is non-analytical~\cite{Falkowski:2019tft,Chang:2019vez}. It corresponds to a pole in the Higgs potential, when $H^\dagger H\to 0$. This applies to the scenarios of Coleman-Weinberg Higgs and Tadpole-induced Higgs.
The non-analytical Higgs potential would correspond to the non-decoupling behaviour, such that the universal unitarity violating scale $4\pi v\sim 3$ TeV is obtained, regardless of how much $d_3$ and $d_4$ deviate from the SM values~\cite{Falkowski:2019tft,Chang:2019vez}. Schematically, for the high dimensional operator~\cite{Chang:2019vez}
\bea
\mathcal{L}_{int}=\frac{\lambda_n}{n_1!\cdots n_r!}\phi_1^{n_1}\phi_2^{n_2}\cdots\phi_r^{n_r}, 
\eea
the $2\to n\ (n>2)$ scattering process only matters when $\lambda_n$ is an order-one coefficient ($\lambda_n\sim \mathcal{O}(1)$), and the unitarity condition requires the energy is bounded roughly as $E<(1/\lambda_n)^{1/n}$~\cite{Chang:2019vez}. The stringent unitarity bound would come in the large $n$ limit.
Physically, $\lambda_n\sim \mathcal{O}(1)$ is only possible in non-decoupling theories, because there is no large scale that is responsible for suppressing this coefficient $\lambda_n$. On the other hand, one could expect that the coefficient $\lambda_n$ is highly suppressed by the cutoff scale in the NP models with decoupling behavior, then the unitarity bound from the $2\to n\ (n>2)$ process will be very loose. Thus, it is sufficient to consider the conventional $2\to 2$ scatterings for the NP models with decoupling behavior.

Based on the above discussion, we summarize the tree-level partial wave unitarity bound in Table~\ref{tab:unitarity} for each new physics scenario.
\begin{table}
 \begin{center}
  \begin{tabular}{|c|c|}
   \hline
  scenarios & unitarity constraints \\
   \hline
  SMEFT & $0<c_6<1584$ for $\Lambda=3$ TeV \\
  \hline
  NG Higgs  & $\sqrt{s}<4$ TeV for $\xi=0.05$  \\
  \hline
  CW Higgs & $\sqrt{s}<4\pi v\sim 3$ TeV \\
  \hline
  Tadpole Higgs  & $\sqrt{s}< 4\pi v\sim 3$ TeV \\
  \hline  
  \end{tabular}
 \end{center}
  \caption{Tree-level unitarity constraints from the scatterings of the Higgs boson, top quark, and longitudinal electroweak gauge bosons. For the SMEFT and NG Higgs scenarios, the bound is obtained from analyzing the $2\to 2$ scattering partial-wave amplitudes. For CW Higgs and Tadpole-induced Higgs scenarios, the unitarity violating scale is roughly $4\pi v\sim 3$ TeV due to their non-decoupling nature of the theories~\cite{Falkowski:2019tft}, which can be estimated from analyzing the $2\to n\ (n>2)$ scattering partial-wave amplitudes~\cite{Chang:2019vez}. Note that we require $c_6$ to be positive in the SMEFT, since the Higgs potential should be bounded from below.}
  \label{tab:unitarity}
\end{table}

\subsection{Tree-level Vacuum Stability}

\begin{figure}[h]
\begin{center}
\includegraphics[width=0.5\linewidth]{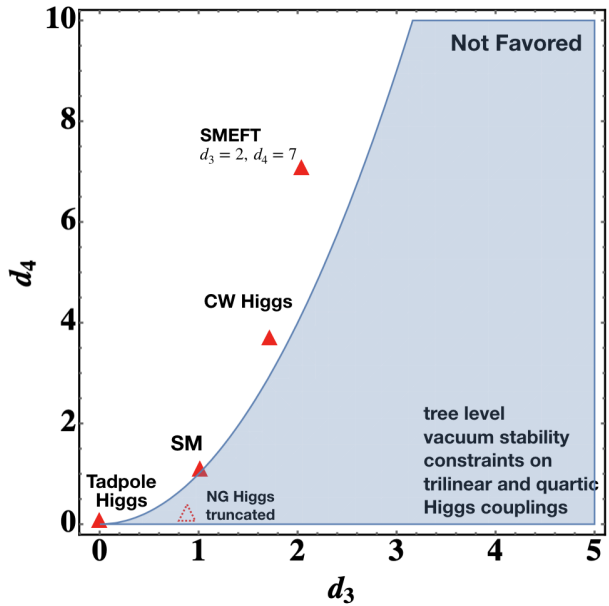}
\caption{Tree-level vacuum stability constraints on Higgs self-couplings $d_3$ and $d_4$. The shaded region is not favored if the EW vacuum is required to be the global minimum. We note that higher powers of Higgs 
self-couplings are relevant for stabilizing the EW vacuum for the NG Higgs scenario. Thus it is only an artifact that the NG Higgs scenario is in the shaded region, resulting from the truncation of the full Higgs potential in deriving the coefficient $d_4$. }  
 \label{fig:stability}
 \end{center}
\end{figure}

Even though the unitarity bound on the Higgs self-couplings is not very tight, the trilinear Higgs coupling $d_3$ cannot be arbitrary large if the EW vacuum is required to be the global minimum.
Based on the Higgs potential in Eq.~(\ref{pot}), this requirement is
\bea
 \frac{\text{d} V(h)}{\text{d} h} = m_h^2 h + d_3 \frac{3m_h^2}{2v} h^2+ d_4 \frac{m_h^2}{2v^2} h^3= 0.
\eea
When $9(d_3)^2-8 d_4$ is positive or zero, the roots of the above equation are explicitly 
\bea
h_1=0\ ;\quad\quad h_2=v \frac{\sqrt{9 (d_3)^2-8 d_4}-3 d_3}{2 d_4}\ ;\quad\quad h_3=v \frac{-\sqrt{9 (d_3)^2-8 d_4}-3 d_3}{2 d_4}.
\eea
In this case, $h_1=0$ corresponds to the EW vacuum, $h_3$ corresponds to another local minimum of the Higgs potential, while $h_2$ corresponds to the local maximum. Tree-level vacuum stability requires the EW local minimum to be the global minimum, i.e. $V(h_1)<V(h_3)$. When $9(d_3)^2-8 d_4$ is negative, only one solution, {\it i.e.,} $h=0$, exists for $\text{d} V(h)/ \text{d} h=0$, which corresponds to the only minimum of the Higgs potential.
As a result, we obtain the tree-level vacuum stability bound on $d_3$ and $d_4$, as shown in Fig.~\ref{fig:stability}\footnote{Given a NP model, $d_3$ and $d_4$ are usually correlated. Thus,  we only focus on the region where both $d_3$ and $d_4$ are positive, rather than treating them as independent parameters.}.
Consistent with Ref.~\cite{Falkowski:2019tft}, the conservative bound on the  trilinear Higgs coupling is obtained as $0<\Delta_3\equiv d_3-1<2$. Certainly, this bound on $d_3$ can be slightly relaxed in case when $d_4$ is much larger than the SM value. As we see in Fig.~\ref{fig:stability}, when $|\Delta_3|> 2$, $d_4$ is required to be more than $10$ times of the SM value.
For comparison, we also mark several benchmark points of various Higgs scenarios discussed in this work.

\section{Double-Higgs Production: Model Discrimination}

In this section, we utilize the double-Higgs production cross section measurements to discriminate different 
Higgs scenarios, since these scenarios predict very different strength of trilinear Higgs boson couplings. At a high energy hadron collider, 
the gluon-gluon fusion channel is the dominant production mechanism for the double-Higgs boson production. 
As mentioned in the introduction, this process has been widely considered in the literature for validating the SM cross section, measuring
 the trilinear Higgs coupling~\cite{Glover:1987nx,Baur:2003gp,Baur:2003gpa,Dolan:2012rv,Baglio:2012np,Papaefstathiou:2012qe,Goertz:2013kp,Barger:2013jfa,Barr:2013tda,Li:2015yia,deLima:2014dta,Alves:2017ued,Adhikary:2017jtu,Goncalves:2018yva,Borowka:2018pxx,Homiller:2018dgu,Kim:2018cxf,Heinrich:2019bkc,Cepeda:2019klc,Kim:2019wns} and the $t\bar{t}hh$ coupling~\cite{Contino:2012xk, Li:2019uyy}, and for probing various NP models~\cite{Cao:2015oaa,Cao:2016zob,Chen:2014xra,Kim:2018uty,Huang:2017nnw,Basler:2018dac,Babu:2018uik,Azatov:2015oxa,Gillioz:2012se,Dawson:2015oha}.
It remains to be established whether this process can be observed at the $5 \sigma$ level at the LHC.

The ATLAS and CMS Collaborations have been looking for the $hh$ signal in the data collected so far at the LHC and have accordingly set upper limits on its production cross section ~\cite{Sirunyan:2018two,Aaboud:2018sfw,Aaboud:2018knk,Aaboud:2018ksn,Aaboud:2018zhh,Aaboud:2018ewm}. Both collaborations have also examined the prospects of detecting $hh$ signal at the high-luminosity LHC (HL-LHC) and the high-energy LHC (HE-LHC)~\cite{CMS:2018ccd,ATLAS:2018ch1}. At the HL-LHC, without (with) the systematic uncertainty, the signal can be measured at $31\%$ ($40\%$) accuracy relative to the standard model prediction with the confidence level of $3.5 \;\sigma$ ($3 \; \sigma$), and the trilinear Higgs coupling can be constrained in the range $-0.1< \frac{\lambda}{\lambda_{SM}}< 2.7\ \rm{and}\ 5.5< \frac{\lambda}{\lambda_{SM}}< 6.9$ ($-0.4< \frac{\lambda}{\lambda_{SM}} < 7.3$). At the HE-LHC ($27$ TeV with $15\; \rm{ab}^{-1}$ of integrated luminosity), the signal can be measured at the confidence level of $7.1 \; \sigma$ and $11 \; \sigma$, without the systematic uncertainty, in the $b\bar{b}\gamma\gamma$ and $b\bar{b}\tau\tau$ channels, respectively~\cite{ATLAS:2018ch1}.  
A number of the above studies have performed detailed background analysis with optimized cut-based analysis or with
multivariate techniques. In this paper, we do not intend to perform any detailed signal-to-background analysis. Instead, we utilize the NP cross section, after some basic kinematic cuts, to calculate the confidence level of observing the double-Higgs production, as predicted in the specific NP scenario, 
by assuming the same background rates as reported in the literature for detecting the SM double-Higgs production. 
 As stated above,
we mostly focus on the double-Higgs production at the 27 TeV HE-LHC and the future 100 TeV pp collider, 
and explore the possibility of distinguishing various scenarios and extracting the unknown Higgs couplings, 
especially, the trilinear Higgs coupling.

\subsection{Cross Section and Distributions}

\begin{figure}[h]
\centering
\includegraphics[scale=0.8]{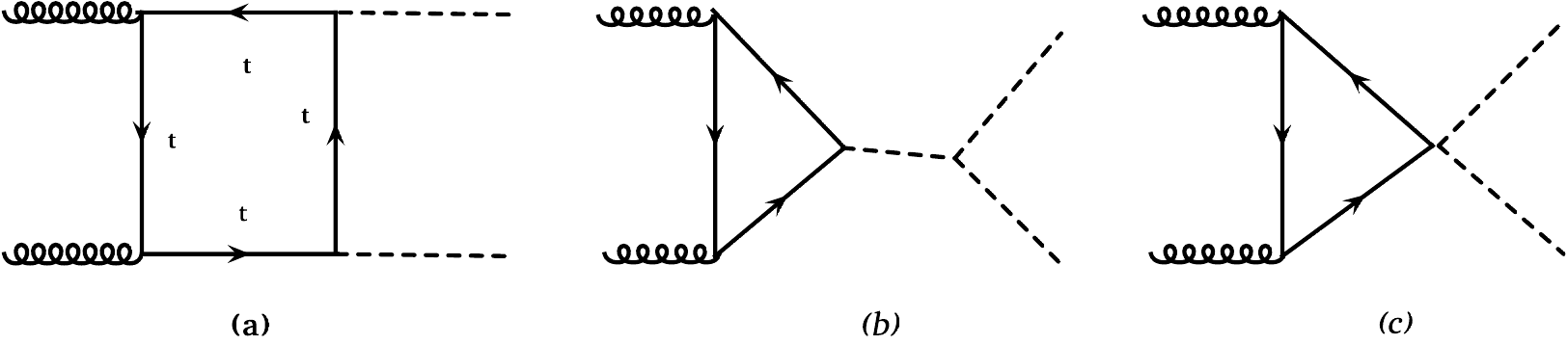}
\caption{Representative Feynman diagrams for the $pp \to hh$ production. The third diagram appears only if the $t {\bar t} hh$ coupling is non-vanishing.}  
 \label{fig:feyn-HH-bx-tr}
\end{figure}

With the Higgs effective couplings listed in Table~\ref{tab:coupling}, the total cross section for the 
 double-Higgs production at hadron colliders can be written as
\begin{eqnarray}
\sigma = c_1^4\,\sigma_{b}^{_{\rm SM}} + c_1^2d_3^2\,\sigma_{t}^{_{\rm SM}}+ c_1^3d_3\,\sigma_{bt}^{_{\rm SM}} + c_2^2\,\sigma_{t\bar{t}hh}+ c_1^2c_2\,\sigma_{b,\,t\bar{t}hh} + c_1d_3c_2\,\sigma_{t,\,t\bar{t}hh},\
\label{eqn:interfernce_HH_composite_intrmd}
\end{eqnarray}
where  $\sigma_b^{_{\rm SM}}$ ($\sigma_t^{_{\rm SM}}$) denotes the SM cross section with only the box (triangle) contribution,  $\sigma_{bt}^{_{\rm SM}}$ the interference between the box  and triangle contribution, 
$\sigma_{t\bar{t}hh}$ the new triangle contribution with non-vanishing SM-like $t {\bar t} hh$ coupling, and $\sigma_{b,\, t\bar{t}hh}$ ($\sigma_{t,\,t\bar{t}hh }$) the interference between new triangle and the box (triangle) contribution. 
A representative set of Feynman diagrams, including the triangle and box diagrams, is given in Fig.~\ref{fig:feyn-HH-bx-tr} for illustration. 


Our methodology of computing the $pp \to hh$ cross section has been discussed in Refs.~\cite{Agrawal:1998ch,Agrawal:2012as}, which we follow below.
We use the leading-order CTEQ parton distribution functions, CT14LLO~\cite{Dulat:2015mca}, with the renormalization and factorization scales chosen as $\sqrt{\hat{s}}$. 
The numerical result for each SM-like cross section, defined in Eq.~(\ref{eqn:interfernce_HH_composite_intrmd}), at the 14, 27, and 100 TeV proton-proton colliders, respectively, is listed in Table~\ref{table:HHformfactor}. To suppress the large QCD backgrounds, we apply a hard cut on the transverse momentum ($p_T$) of the Higgs boson, as discussed in next subsection. Thus Table~\ref{table:HHformfactor} also includes the results on the SM-like cross sections with a cut $p_T^h > 70$ GeV at the 14, 27, and 100 TeV proton-proton colliders, respectively. No further kinematic cuts are applied here, as we are not doing detailed signal-to-background study. 
Although higher order QCD corrections  up to NNLO in the SM~\cite{Grober:2015cwa, deFlorian:2017qfk,Grober:2017gut,Buchalla:2018yce} and the EFT framework~\cite{Dawson:1998py,Grigo:2013rya,Borowka:2016ehy,Borowka:2016ypz,deFlorian:2013jea, Grazzini:2018bsd} are known, e.g. the $K$-factor  is about 2.3 (1.7) for the 14 (100) TeV colliders~\cite{deFlorian:2013jea}. We do not include the $K$-factors in our numerical results presented in Table~\ref{table:HHformfactor}. 
According to the Table, the SM cross section without cuts (with $ p_T^h > 70\ \rm{GeV}$ cut) is 17.2 (15.4) fb at the 14 TeV collider, and it is 73.6 (66.2) fb at the 27 TeV collider, about 5 (4) times larger. 
The SM cross sections at 100 TeV collider are 830.1 fb and 756.8 fb, respectively, which are about 50 times larger than the results at the 14 TeV.


\begin{table}[h]
\begin{center}
\begin{tabular}{|c|c|c|c|c|c|c|c|c|}
\hline
Collider & $p_T^h$ & $\sigma_{b}^{_{\rm SM}}$ & $\sigma_{t}^{_{\rm SM}}$ & $\sigma_{bt}^{_{\rm SM}}$ & $\sigma_{b,\,t\bar{t}hh}$  & $\sigma_{t,\,t\bar{t}hh}$  & $\sigma_{t\bar{t}hh}$  \\
\hline
$14$ TeV & no cut &  36.1 &  4.9 & -23.8  &  -147.0 &  48.9 & 175.8   \\
\cline{2-8}                                                       
&$p_T^h > 70 $ GeV &  29.6 &  2.9 & -17.1 &  -122.4 &  36.3   & 151.9   \\
\hline                                                       
\hline
$27$ TeV& no cut & 149.2 &  18.9 &  -94.5  &  -618.9 &  197.92  &  777.0\\
 \cline{2-8}                                                       
 &$p_T^h> 70 $ GeV & 124.1 &  11.6 &  -69.6 &   -524.5 &  151.1 &  684.5 \\
\hline
\hline
{$100$ TeV}& no cut &  1607.6 &  184.3 &  -961.8 &  -6872 &  2077.3 &  9356 \\
 \cline{2-8}                                                       
 &$p_T^h> 70 $ GeV &  1370 &  118.8 &  -732 &  -5970 &  1645 &  8464 \\
\hline
\end{tabular}
\end{center}
\vspace{-0.4cm}
\caption{The SM-like cross section defined in  Eq.~(\ref{eqn:interfernce_HH_composite_intrmd}), without cuts and with $ p_T^h > 70\ \rm{GeV}$ cut, at the 14 TeV, 27 TeV, and 100 TeV proton-proton colliders, respectively.  }
\label{table:HHformfactor}
\end{table}

As seen from Table~\ref{table:HHformfactor}, there are also some interesting patterns
for the interference between different Feynman diagrams. These could help us understand how the cross sections and differential  distributions depend on various effective Higgs couplings. 
This will be discussed in more detail in next subsection. 

Combining the effective Higgs couplings in Table~\ref{tab:coupling} and the SM-like cross section in Table~\ref{table:HHformfactor},  we obtain the total cross sections in different models using the Eq.~(\ref{eqn:interfernce_HH_composite_intrmd}). Take the 27 TeV collider for example. The total cross sections in the Tadpole-induced Higgs model and the Coleman-Weinberg model are $149.2$ fb and $124.1$ fb, respectively, while with $ p_T^h > 70\ \rm{GeV}$ cut, they are $44.2$ fb and $40.3$ fb, respectively. 
For the MCH and CTH models, if we take the benchmark value $\xi=0.05$, the total cross sections without cuts  are  $97.7$ fb and $79.9$ fb, while with $ p_T^h > 70\ \rm{GeV}$, they are  $87.2$ fb and $71.5$ fb, respectively. 

\begin{figure}[!htb] 
\begin{center}
\includegraphics[angle=0,width=0.48\linewidth]{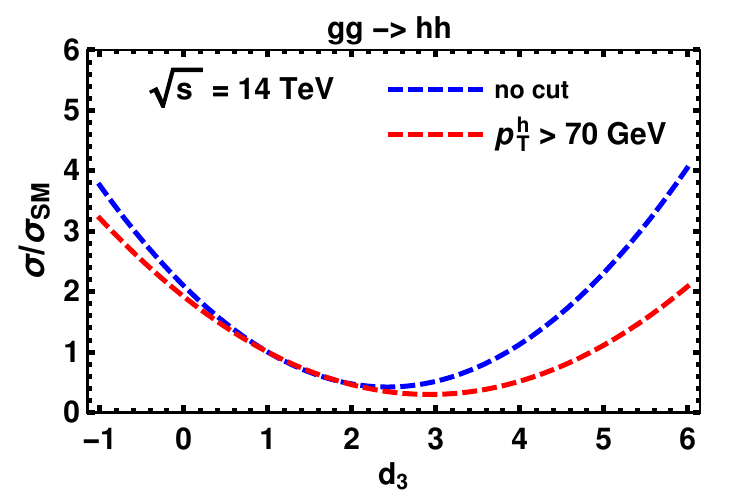}
\includegraphics[angle=0,width=0.48\linewidth]{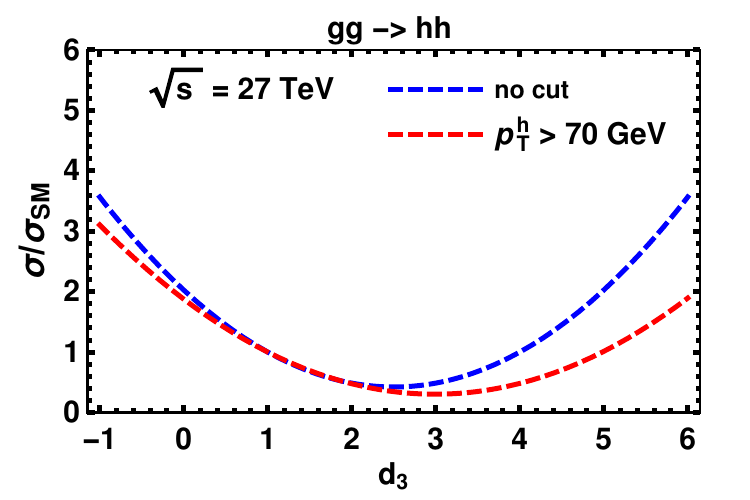}\\
\includegraphics[angle=0,width=0.48\linewidth]{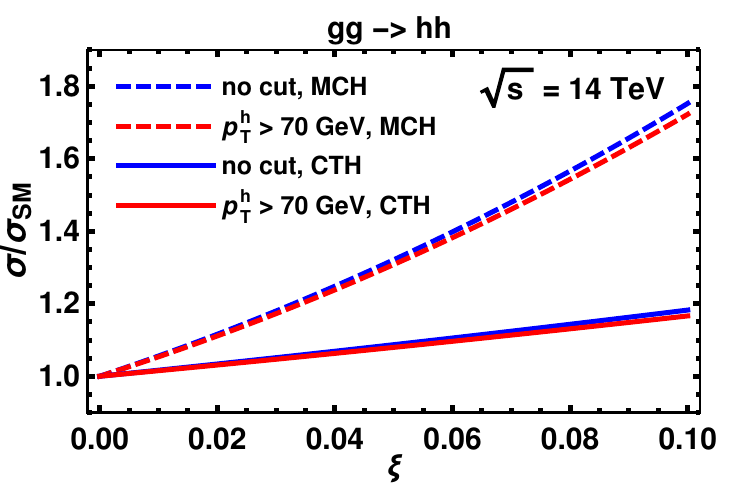}
\includegraphics[angle=0,width=0.48\linewidth]{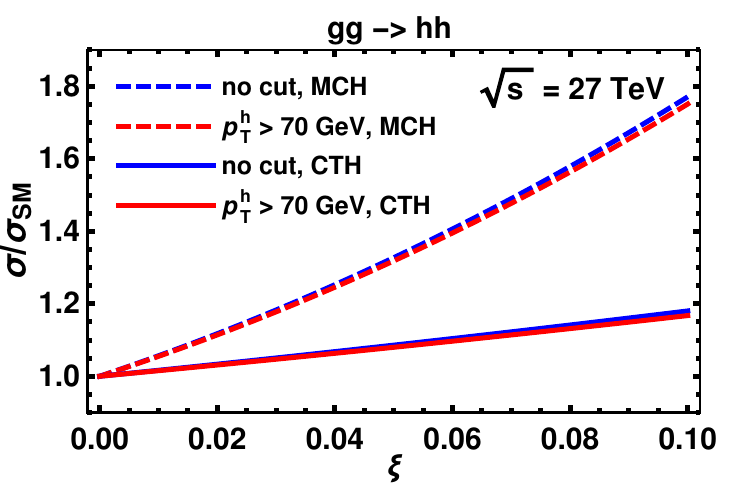}
\end{center}
\caption{Variation of the ratio of the new-physics cross section to that of the SM for $hh$ production, at the $14$ TeV HL-LHC and the $27$ TeV HE-LHC, respectively, as a function of  the trilinear Higgs coupling $d_3$ in the elementary Higgs, Coleman-Weinberg Higgs and Tadpole-induced Higgs scenarios (upper row), and as a function of the parameter $\xi$ in the Nambu-Goldstone Higgs scenario (lower row).}

 \label{fig:cs-HH-14-27TeV}
\end{figure}

\begin{figure}[!htb]
\centering
\includegraphics[angle=0,width=0.48\linewidth]{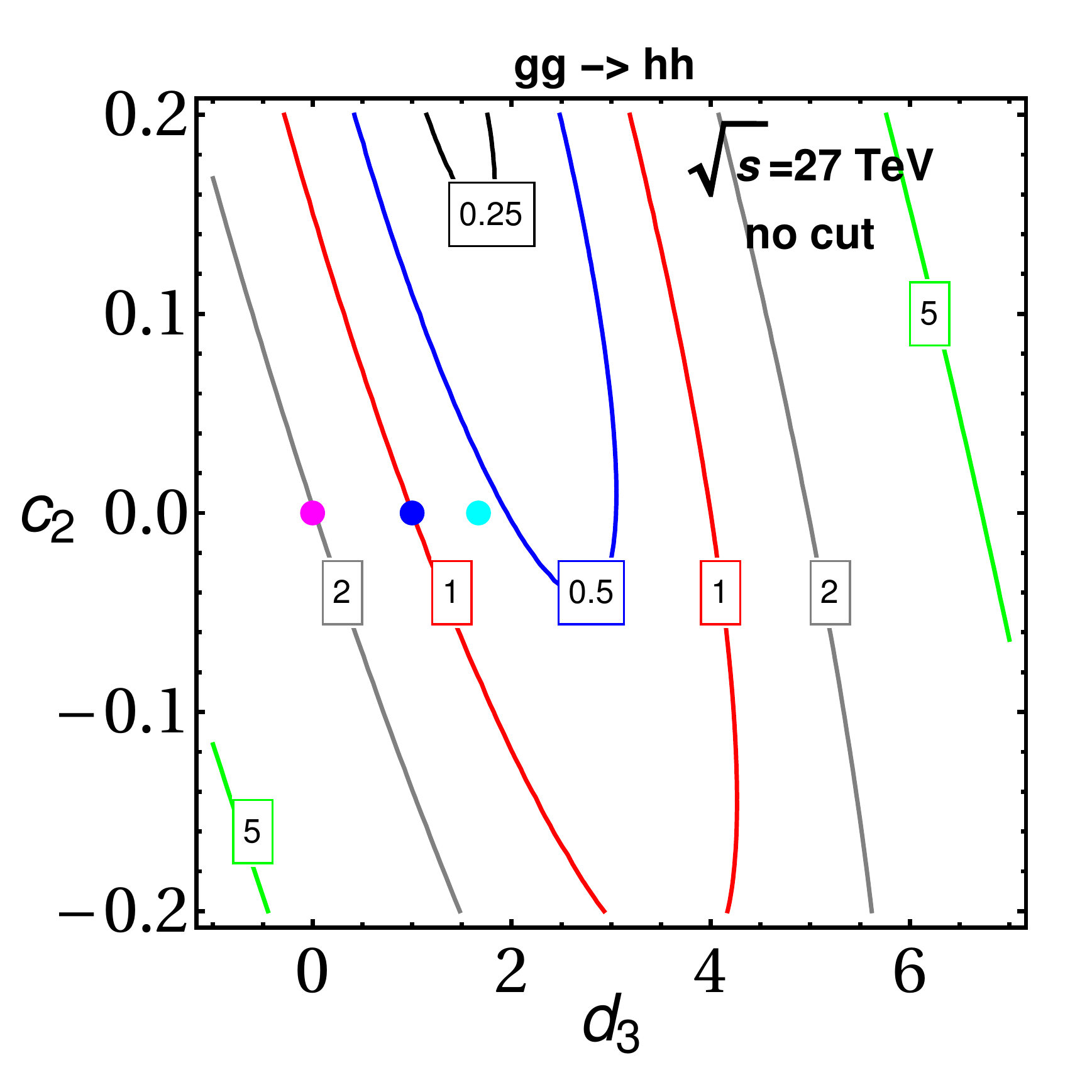}
\includegraphics[angle=0,width=0.48\linewidth]{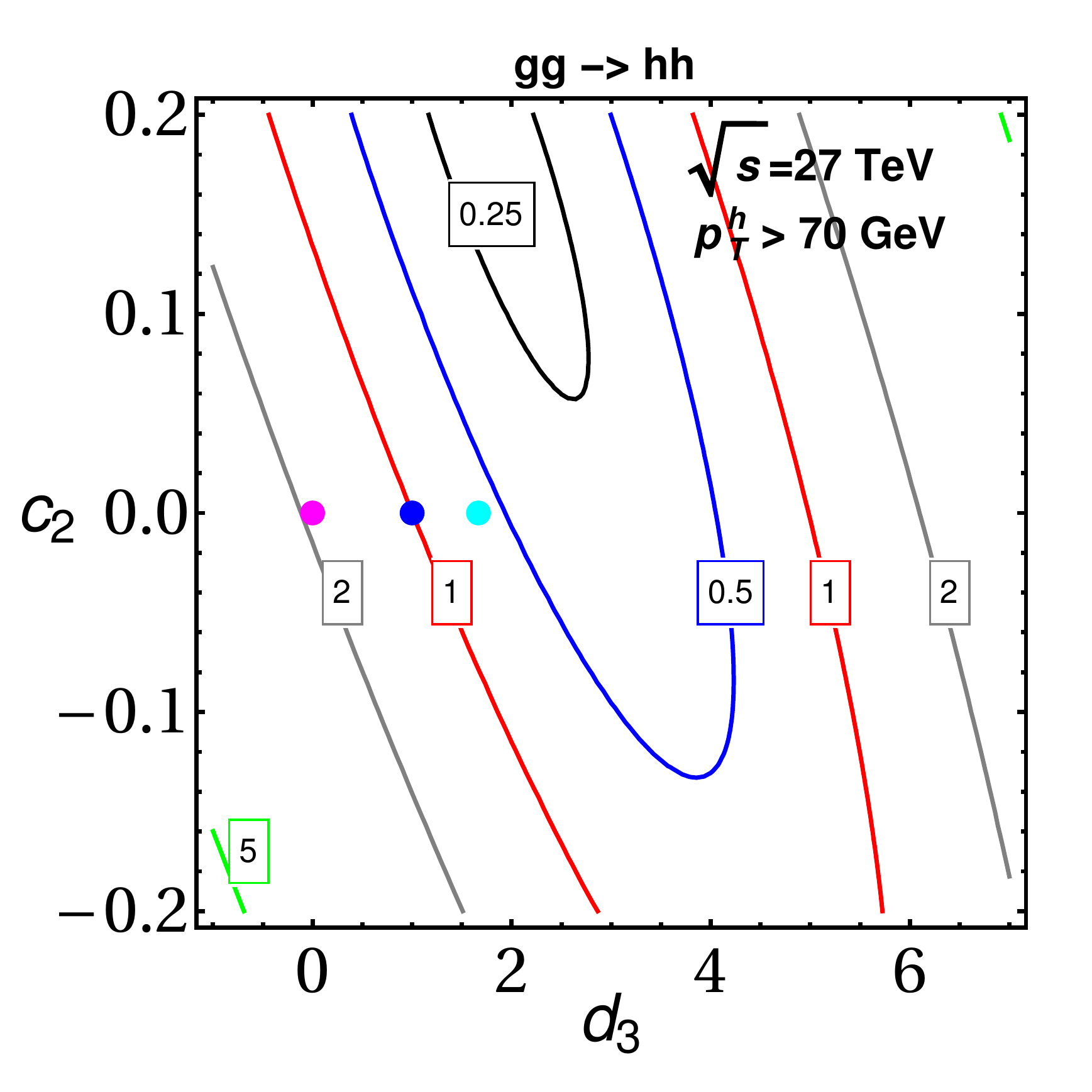}
\caption{Cross section ratio ${\sigma}/{\sigma_{SM}}$ as a function of $c_2$ and $d_3$: without any cut (left), and with the kinematic cut $p_T^h>$70 GeV (right). The SM cross sections without cuts and with cut, at the 27 TeV HE-LHC collider, are 73.6 fb and 66.2 fb, respectively. The total cross sections in the Tadpole-induced Higgs and Coleman-Weinberg scenarios are 149.2 fb and 124.1 fb, respectively, while with $p_T^h>$70 GeV, they are 44.2 fb and 40.3 fb, respectively. The magenta, blue, and cyan dots denote the ${\sigma}/{\sigma_{SM}}$ ratios in the Tadpole-induced Higgs, the SM, and Coleman-Weinberg scenarios, respectively.}
\label{fig:c2-d3_HH}
\end{figure}

In Fig.~\ref{fig:cs-HH-14-27TeV},
for illustration, we display the ratio of the new-physics cross sections to the SM cross section in various 
Higgs potential scenarios at the $14$ TeV LHC and the $27$ TeV HE-LHC, respectively. 
In the top row of Fig.~\ref{fig:cs-HH-14-27TeV},
the ratio of the cross sections exhibits the quadratic dependence of $d_3$ with minimum around $d_3 = 2 \sim 3$. This behavior will be explained below. 
The bottom row of the figure shows the ratio as a function of the parameter $\xi$ in the Nambu-Goldstone
Higgs scenario. In the case of the CTH model, the cross section ratio slowly increases as the $\xi$ increases, and this behavior does not change much when the $p_T^h> 70 $ GeV cut is imposed. 
This ratio could also be presented in the model-independent way using the general Higgs couplings. 
In Fig.~\ref{fig:c2-d3_HH}, this ratio is plotted in a two dimensional ($c_2$, $d_3$) contour with other parameters taken as the SM-like values.
In this figure, the values in the SM Higgs, Coleman-Weinberg Higgs, and 
Tadpole-induced Higgs scenarios are marked. The behavior of the cross section ratio in these models can be understood based on the interference patterns, as to be discussed
in the next section.

 \begin{figure}[!htb]
 \begin{center}
 \includegraphics[angle=0,width=0.45\linewidth]{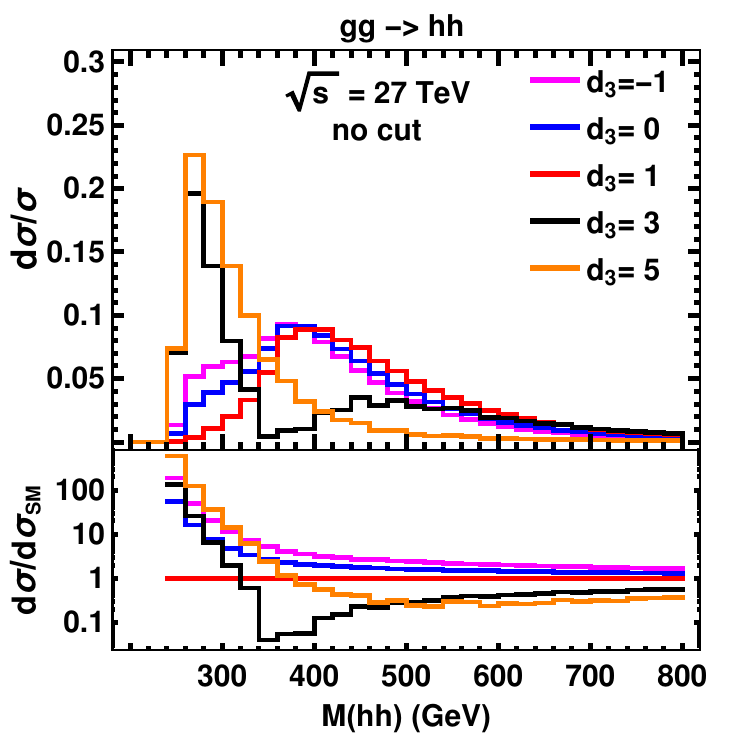}
 \includegraphics[angle=0,width=0.45\linewidth]{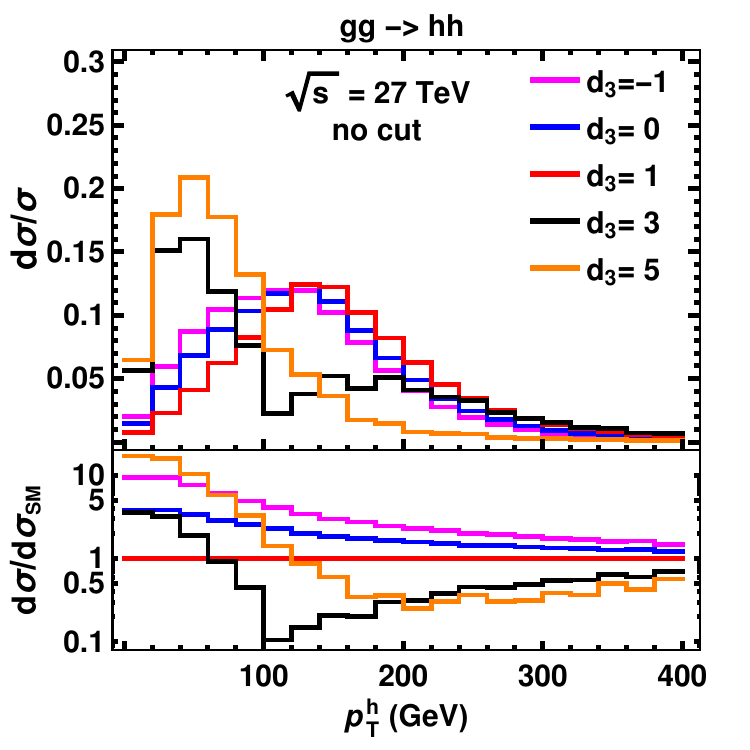}\\
 \includegraphics[angle=0,width=0.45\linewidth]{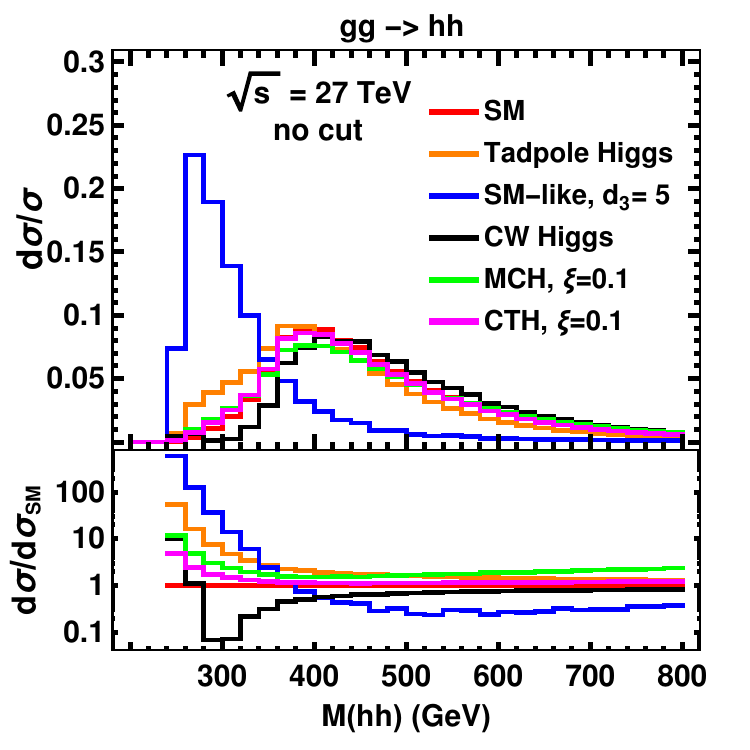}
 \includegraphics[angle=0,width=0.45\linewidth]{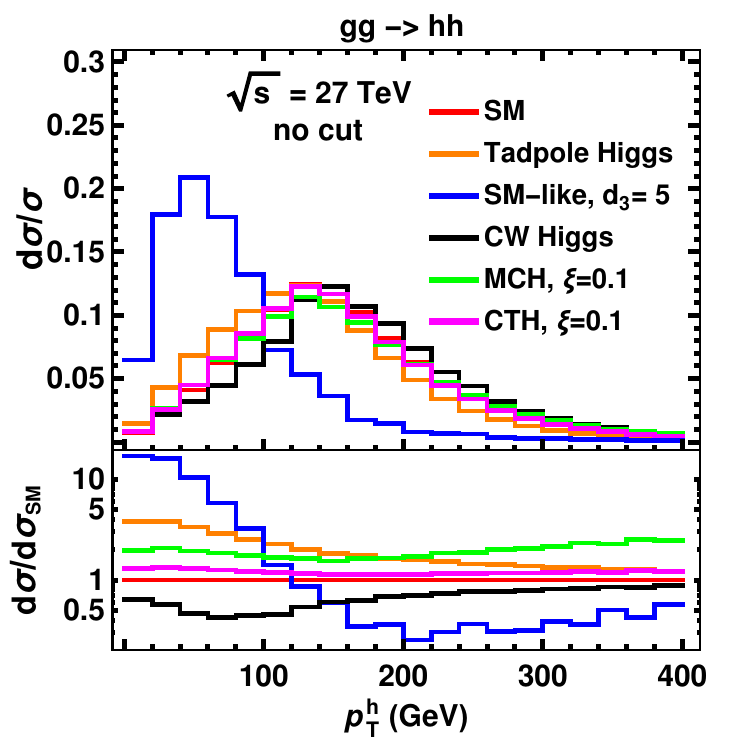}
\end{center}
\caption{Various normalized distributions on the di-Higgs invariant mass $M(hh)$ (left) and the Higgs $p_T$ (right) at the 27 TeV HL-LHC with different $d_3$ couplings (upper) and various Higgs potential models (lower). 
The case of $d_3=3$ shows an interesting feature, due to the competition between the triangle and box diagram contributions, as explained in the text.} 
\label{fig:dist-HH}
\end{figure}

In Fig.~\ref{fig:dist-HH}, we display the normalized di-Higgs invariant mass $M(hh)$ (left) and $p_T^h$ (right) distributions 
at the 14 TeV LHC and the 27 TeV HE-LHC, respectively. These distributions play an important role in determining suitable kinematic cuts to reduce the SM backgrounds.
The upper row of Fig.~\ref{fig:dist-HH} shows the normalized $M(hh)$ distribution with a range of values of $d_3$.  In the case of $d_3 = 3$, there is an interesting two-peak structure in the normalized $M(hh)$ distribution, arising from the competition between the triangle and box diagram contributions. We will come back to this in the next subsection, cf.  Fig.~\ref{fig:dist-HH-shat-d13}.

\subsection{Interference Effects}

As shown in Fig.~\ref{fig:feyn-HH-bx-tr}, the trilinear Higgs coupling is only presented in the triangle diagrams. However, as the box and triangle diagrams interfere, the contribution of the trilinear Higgs coupling to the cross section also depends on the box diagram contribution. Furthermore, their interference effect is destructive. Since some of the new Higgs potential scenarios would allow large deviations of the Higgs couplings from the SM value, the total cross section and various distributions could change significantly. 
Moreover\footnote{In the elementary Higgs scenario, $t\bar{t}hh$ can also be induced via integrating out heavy particles, as shown in Eq.~(\ref{eq:wilsoncoef_c2}). Here, for simplicity, we take the $t\bar{t}h$ coupling and the $hVV$ to be the SM ones, which eliminates the $t\bar{t}hh$ coupling because the $t\bar{t}hh$  and $t\bar{t}h$ are correlated in this model. }, the Nambu-Goldstone Higgs scenario also predicts non-zero $t\bar{t}hh$ coupling due to the Higgs non-linearity, and there is correlation among the  $hVV$, $t\bar{t}hh$ and $t\bar{t}h$ couplings~\cite{Li:2019ghf} for non-zero $\xi$. Because of this new $t\bar{t}hh$ interaction, two new triangle diagrams appear. These diagrams interfere with the SM triangle diagram destructively, and with the box diagram constructively. This behavior happens as the $t\bar{t}hh$ coupling has a negative sign relative to $t\bar{t}h$ coupling in this scenario, as shown in Table.~\ref{tab:coupling}. 
In Table \ref{table:HHformfactor}, we observe that taking the SM couplings, e.g., $d_3 = 1$, the pure box contribution is large, while the pure triangle contribution is small, and furthermore, the interference contribution is large but negative, i.e. destructive. This leads to a small total cross section for the $pp\to hh$ process in the SM.

\subsubsection{Interference effects without $t\bar{t}hh$}

Let us first consider scenarios without the $t\bar{t}hh$ vertex. They are the elementary Higgs, Coleman-Weinberg Higgs, and 
Tadpole-induced Higgs scenarios. As one can see from Eq.~(\ref{eqn:interfernce_HH_composite_intrmd}), the pure triangle contribution depends quadratically on $d_3$, whereas the interference term depends linearly on it. However, the pure box contribution does not depend on $d_3$. For the negative $d_3$, the cross section keeps on increasing with increasing  magnitude of $d_3$, cf. Fig.~\ref{fig:cs-HH-14-27TeV}, as both $\sigma_{t}^{_{\rm SM}}$ and $\sigma_{bt}^{_{\rm SM}}$ contributions increase. For positive $d_3$, however, the cross section first decreases and then keeps on increasing after reaching some threshold value of $d_3$, as first $\sigma_{bt}^{_{\rm SM}}$ dominates which decreases the cross section, then $\sigma_{t}^{_{\rm SM}}$ dominates which increases the cross section. This explains the feature found in the upper row of Fig.~\ref{fig:cs-HH-14-27TeV}.

 \begin{figure}[!htb] 
 \begin{center}
 \includegraphics[angle=0,width=0.45\linewidth,height=0.4\linewidth]{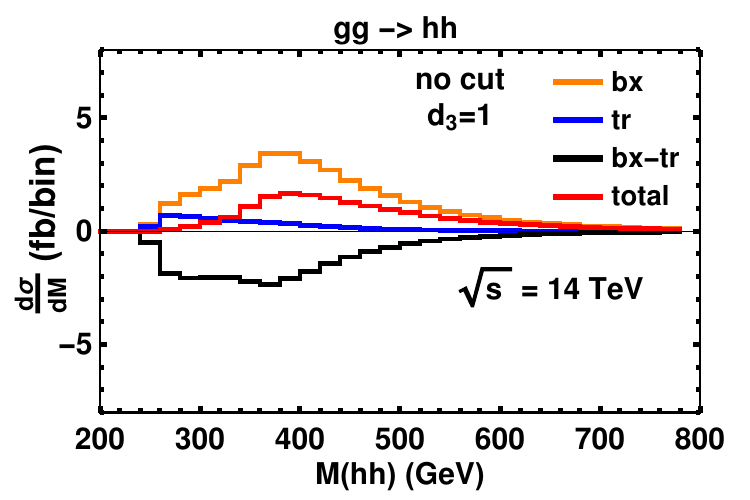}
 \includegraphics[angle=0,width=0.45\linewidth,height=0.4\linewidth]{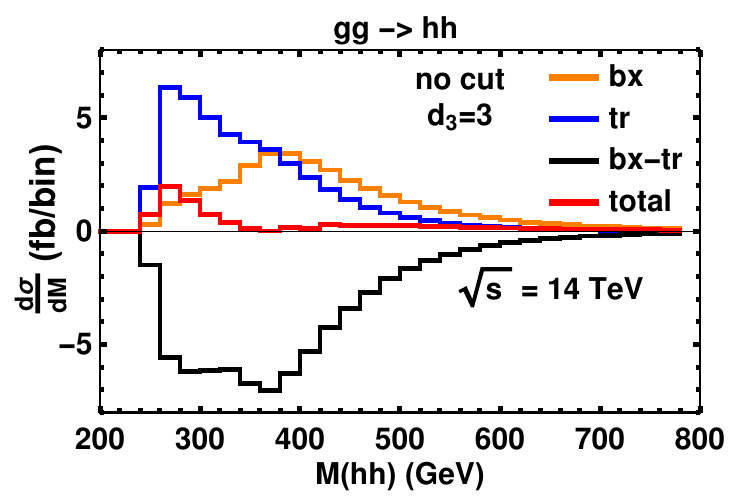}
 \end{center}
 \caption{Contribution of various classes of Feynman diagrams and their interference effects to the $M(hh)$ distribution of
  $hh$ production, for $d_3=1$ (left) and $d_3 = 3$ (right). When we increase the value of $d_3$ from 1 to 3, the triangle diagrams contribution and the negative interference term
  get scaled by $9$ and $3$, respectively. However, as the box-diagram contribution ``bx" does not depend on $d_3$, it remains the same. The peak of the total distribution gets shifted to the left with  the increase in $d_3$,  as the triangle diagram, being an s-channel process, contributes significantly near the threshold of $hh$ production.
  } \label{fig:dist-HH-shat-d13}
 \end{figure}

To understand the feature in Fig.~\ref{fig:dist-HH}, let us examine the contribution to the $M(hh)$ distribution by decomposing each class of Feynman diagrams and their interference effect. As shown in Fig.~\ref{fig:dist-HH-shat-d13}, the triangle diagram mostly contributes near the Higgs pair threshold, while the box diagram mainly contributes at the threshold of the top quark  pair system.
As the $d_3$ increases, the contribution to the $M(hh)$ and $p_T^h$ distributions from the triangle diagram keeps increasing and eventually exceeds the box diagram when $d_3$ becomes very large. 
When $d_3=3$, both the triangle and box diagrams are sizable. Together with their interference effect, they result in the two peaks in the $M(hh)$ and $p_T^h$ distributions, as shown in Fig.~\ref{fig:dist-HH}. 
Moreover, as we increase the minimum cut on the $p_T$ variable of the Higgs boson, which further suppresses the large QCD background, the contribution from the pure triangle diagrams decreases more than the interference and the pure box contributions relatively, as shown in Table~\ref{table:HHformfactor}. 
For the SM case, with the $p_T^h>70$ GeV cut on the Higgs boson, the pure triangle contribution decreases by a factor of around $1.7$, the magnitude of the interference term by $1.4$, and the one of the pure box term by $1.2$.
This explains why, in Fig.~\ref{fig:cs-HH-14-27TeV}, the minimum of the curves, in which the pure triangle contribution starts to dominate over the interference term, shifts to the right with the increase of the $p_T^h$ cut. 
For the SM case, since the triangle contribution is small, the reduction on the total cross section is not that steep with the increase in the minimum $p_T^h$ cut, e.g., the total contribution decreases by a factor of 1.1 only with the $p_T^h>70$ GeV cut. 
However, for larger positive $d_3$ values, the pure triangle contribution does not dominate over the negative interference before applying any cuts. 
Thus, even though the cross section without cuts is large, imposing certain minimum $p_T^h$ cut would lead to a larger reduction in the cross section. 
For instance, for $d_3=10$, the total cross section is $288.9$ fb with no $p_T$ cut; it reduces to $150.8$ fb when a $p_T>70$ GeV cut is applied, i.e., a reduction by a factor of $1.9$. 
The cross section for any $d_3$, before and after cuts, can easily be obtained using Table~\ref{table:HHformfactor}.

At the $14$ TeV HL-LHC, the double-Higgs production cross section is not statistically large. Thus, in the case of the most promising final state signature  `$bb \gamma \gamma$', there are only few tens of events with 3 ab$^{-1}$ luminosity, which could only put very loose constraints on $d_3$.  
Nevertheless, the cross sections at the $27$ TeV HE-LHC are about $5-6$ times larger than that at the HL-LHC. Therefore, even for the final state `$bb \gamma \gamma$', one could have significant constraint on $d_3$, and the $d_3$ value can be determined within around $20\%$~\cite{ATLAS:2018ch1}. Therefore, we expect at the $27$ TeV HE-LHC, it is possible to distinguish different Higgs potential scenarios which do not contain the $t {\bar t} hh$  vertex.

\subsubsection{Interference effects with $t\bar{t}hh$}

In the Nambu-Goldstone Higgs scenario, e.g. the MCH and CTH models, in addition to the appearance of a new $t\bar{t}hh$ vertex, the existing vertices, such as the $t\bar{t}h$ and $hhh$ couplings, also get modified from the SM ones, as displayed in Table~\ref{tab:coupling}. In Ref.~\cite{Li:2019ghf}, a global fit on the MCH and CTH parameter space was performed by using the available 
data from the LHC Run-2 data. The $95\%$ CL limit on $\xi$ is obtained to be $\xi<0.1$ for the MCH model. 
 Thus in this study, we will choose the value of $\xi$ up to $0.1$.

\begin{figure}[!htb]
\vspace{0cm} 
\begin{center}
\includegraphics[angle=0,width=0.4\linewidth]{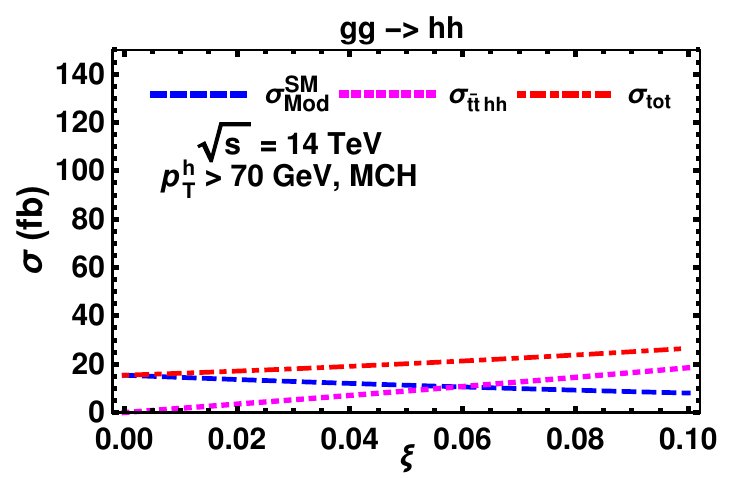} 
\includegraphics[angle=0,width=0.4\linewidth]{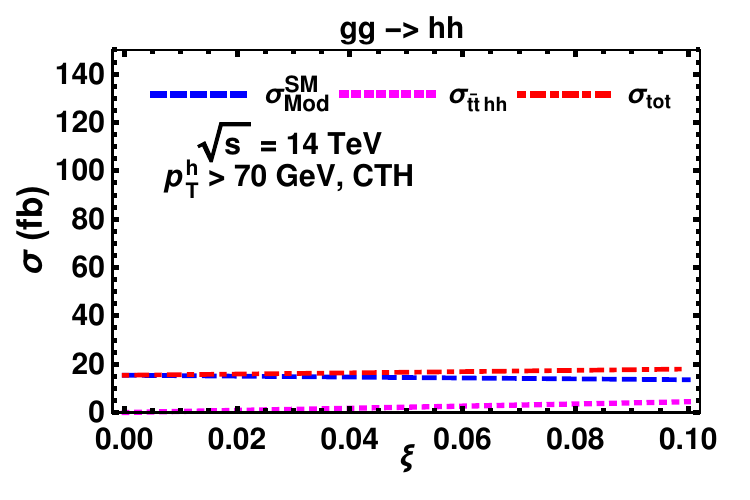}\\
\includegraphics[angle=0,width=0.4\linewidth]{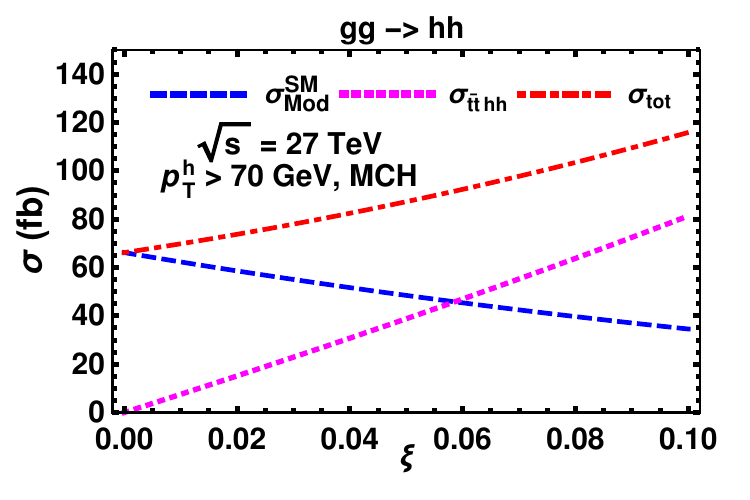} 
\includegraphics[angle=0,width=0.4\linewidth]{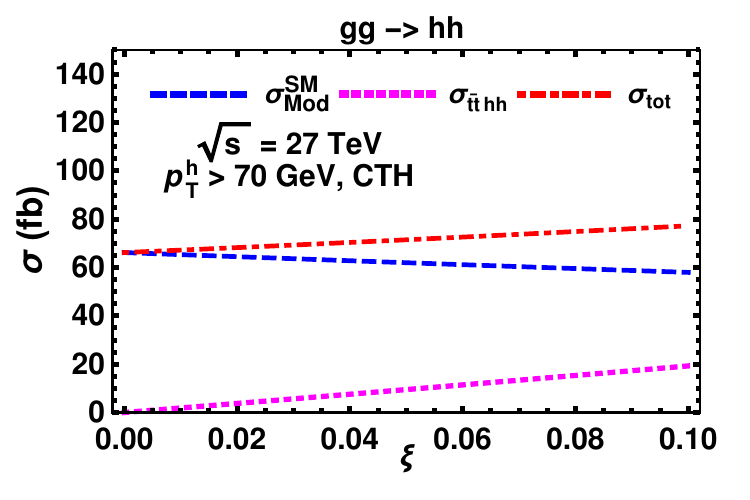}
\vspace{-.5cm} 
\caption{Variation of different contributions to the SM-like cross sections, cf. Eq.~(\ref{eqn:interfernce_HH_composite_intrmd}), as a function of $\xi$ in the MCH and CTH scenarios, at the LHC. 
The Magenta line, parametrizing the effect of $t\bar{t}hh$, crosses the blue line,  parametrizing the effect of $t\bar{t}h$ and $hhh $ couplings, around $\xi=0.06$.}  
\label{fig:cs-HH-13TeV_MCH_parts}
\end{center}
\end{figure}

Fig.~\ref{fig:cs-HH-14-27TeV} shows the variation of the total cross section with the parameter $\xi$. 
With a fixed $\xi$ value, the total cross section in the MCH model is significantly larger than that in the CTH model. 
In both models, the trilinear Higgs  coupling remains to be the same because of the universal form of the Higgs potential, however the $t\bar{t}h$ and $t\bar{t}hh$ couplings are different due to different fermion embedding
in both models. 
From Eq.~(\ref{eqn:interfernce_HH_composite_intrmd}) and Tabel~\ref{tab:coupling}, we see that the scaling of the combination $\sigma_{b,\,t\bar{t}hh}+\sigma_{t,\,t\bar{t}hh}$ in the MCH model is larger by a factor of 4 than that in the CTH model. 
 Similarly, $\sigma_{t\bar{t}hh}$ in the MCH model is larger by a factor of $16$ in comparison to the CTH model. However, this term does not contribute noticeably when $\xi$ is as small as $0.01$ because of the $\xi^2$ scaling in the cross section. However, for $\xi=0.1$, this term contributes moderately in the MCH  model. 
The above discussion explains the difference in the increase of the rates of the total cross section in the MCH model and the CTH model as $\xi$ increases. A similar conclusion also holds after applying the $p_T^h$ cut. 
Finally, in Fig.~\ref{fig:cs-HH-13TeV_MCH_parts}, we show the different contributions to the SM-like cross sections  as function of the $\xi$ parameter in the MCH and CTH models at the LHC. It shows that the $t\bar{t}hh$ coupling is important to enhance the cross sections. 
As the $\xi$ increases, although the contribution from the SM-like diagrams decreases, the total cross section increases due to the dominance of the $t\bar{t}hh$ contribution, most noticeably in the MCH model.

\subsection{Model Discrimination and $\lambda_3$ Extraction}

In this subsection, we investigate the possibility to distinguish various new physics scenarios of Higgs potentials at the $27$ TeV HE-LHC and the $100$ TeV pp collider. At the HL-LHC, due to the limited cross 
section, it is difficult to constrain the trilinear Higgs coupling $d_3$. At higher energy hadron colliders, the total cross section increases significantly, and thus the accuracy of measuring the total cross section, and the constraint on $d_3$, improves significantly.

\begin{figure}[!htb]
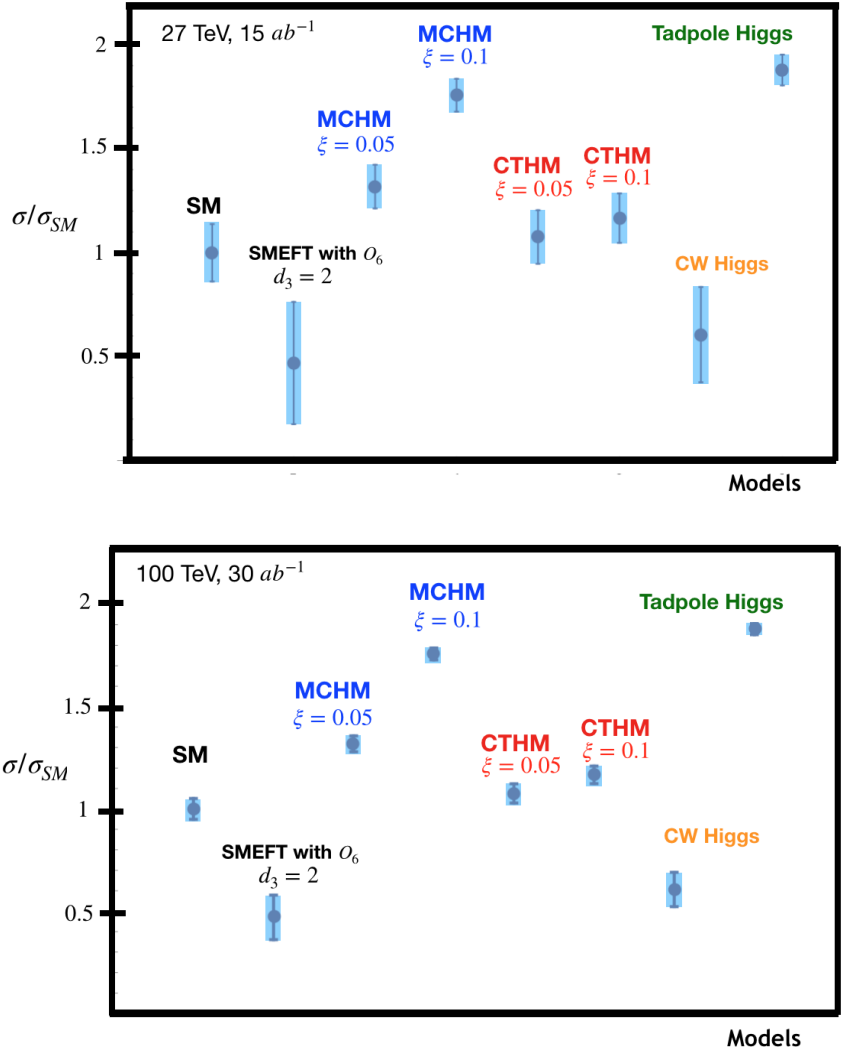

\begin{center}
\includegraphics[width=0.7\linewidth]{plots/HH_27TeV_15ab.pdf}\\
\includegraphics[width=0.7\linewidth]{plots/HH_100TeV_30ab.pdf}
\end{center}
\vspace{-0.9cm}
\caption{The cross section ratio $\sigma/\sigma_{\rm SM}$ in the double-Higgs production at the $27$ TeV HE-LHC with an integrated luminosity of $15\ \text{ab}^{-1}$ (upper), and the $100$ TeV pp collider with an integrated luminosity of $30\ \text{ab}^{-1}$ (lower) for various models. Here, we consider the case that the  SM cross section can be measured with an accuracy of $13.8\%$ and $5\%$, at the $1\sigma$ level, respectively, at the $27$ TeV HE-LHC and the $100$ TeV pp collider. The accuracy for the NP models are obtained using the rescaling procedure  described in the text. The blue bars denote the expected accuracy for a given model.
}
\label{HHproduction}
\end{figure}

It has been shown in the literature~\cite{Goncalves:2018yva} that the double Higgs boson production cross section of the SM, at the $27$ TeV HE-LHC with the integrated luminosity $15\ \text{ab}^{-1}$, can be measured with the accuracy of $13.8\%$ at the $1 \, \sigma$ level. This accuracy would be further improved at the $100$ TeV pp collider with a $30\ \text{ab}^{-1}$ integrated luminosity. Accordingly, the SM signal for the double-Higgs production can be measured with the accuracy of $5\%$ at the $1 \, \sigma$  level~\cite{Goncalves:2018yva}. We shall use this information as the benchmark point and 
perform a recast to obtain the signal significance in various NP scenarios. Using the fixed luminosity and the backgrounds from Ref.~\cite{Goncalves:2018yva}, the significance is obtained using $Z=\Phi^{-1}(1-1/2p)=\sqrt{2}\textrm{Erf}^{-1}(1-p)$~\cite{Cowan:2010js,Tanabashi:2018oca}, where $\Phi$ is the cumulative distribution of the standard Gaussian and Erf is the error function.
 In this case, the $Z$ value is 
\bea
Z=\sqrt{2\left[n_0\text{ln}\frac{n_0}{n_1}+(n_1-n_0)\right]}.
\label{eq:Zformula}	
\eea
Here, $n_0$ is defined as $n_0=n_b+n_s$, and $n_1=n_b+n_s^\prime$. 
The $n_b$ denotes the number of background events and $n_s$ denotes the number of signal events rescaled in each NP scenarios as 
\bea
	{n_s} \sim \frac{\sigma^{_{\rm SM}}_{_{\textrm{after all cuts}}}}{\sigma^{_{\rm SM}}_{_{\textrm{after PT cuts}}}} \sigma^{_{\rm NP}}_{_{\textrm{after PT cuts}}}.
\label{eq:NPrescale}	
\eea 
The $n_s^\prime$ denotes the signal event number that can be constrained at the $1 \, \sigma$ level, which can be obtained by solving Eq.~(\ref{eq:Zformula}) with $Z=1$ for a given $n_0$. 
With $n_s$ and $n_s^\prime$, the relative accuracy for each NP scenario is obtained as $|n_s-n_s^\prime|/n_s$.
As expected, the larger cross sections lead to smaller relative errors for different new physics models.

The results are shown in Fig.~\ref{HHproduction}. At the 27 TeV HE-LHC and the 100 TeV pp collider, based on the total cross sections of the double-Higgs production, it is sufficient to distinguish new physics scenarios with different Higgs potentials.  The following conclusions can be drawn:
\bit
\item For the SMEFT with non-vanishing $O_6\sim (H^\dagger H)^3$ operator, the total cross section tends to be smaller than that of the SM. Because of the tree-level vacuum stability constraint discussed in Sec.~\ref{sec:unitarity}, the Wilson coefficients of the $O_6$ operator is preferred to be positive, which renders  $d_3$ to be larger than one and yields a small cross section, as shown in Fig.~\ref{fig:cs-HH-14-27TeV}. 
For the benchmark $d_3=2$, it leads to an accuracy of being $29.4\%$ at the $27$ TeV HE-LHC, and $10.9\%$ at the $100$ TeV pp  collider, respectively.
\item  For the Nambu-Goldstone Higgs, the total cross section tends to be larger than the SM prediction because of the existence of the contact $t\bar{t}hh$ coupling. We show that for the benchmark $\xi\simeq 0.1$, different Nambu-Goldstone Higgs models, with the top quark embedded in different representation,  can also be distinguished. The relative accuracy at the $1\sigma$ level is about $10\%$ at the $27$ TeV HE-LHC, and about $5\%$ at the $100$ TeV pp collider, respectively.
\item The trilinear Higgs coupling in the Coleman-Weinberg Higgs scenario is universally predicted to be $d_3 = 5/3$. So, similar to SMEFT, models of Coleman-Weinberg Higgs also have a smaller cross section compared to the SM one.  The $1\sigma$ relative accuracy is about $23\%$ at the $27$ TeV HE-LHC, and about $4.7\%$ at the $100$ TeV pp collider, respectively.
\item The trilinear Higgs coupling in the Tadpole-induced Higgs scenario is highly suppressed. Therefore Tadpole-induced Higgs models can have a much larger cross section compared to the SM value, due to the enhanced box contribution and small interference. It turns out this scenario could be examined very well at both the $27$ TeV HE-LHC (relative accuracy of $7.4\%$ at the $1\sigma$ level) and the $100$ TeV pp collider (relative accuracy of $2.7\%$ at the $1\sigma$ level), and it can be well-discriminated from the SM scenario. 
\eit

\begin{figure}[!htb]
\vspace{-0.1cm} 
\begin{center}
\includegraphics[angle=0,width=0.31\linewidth]{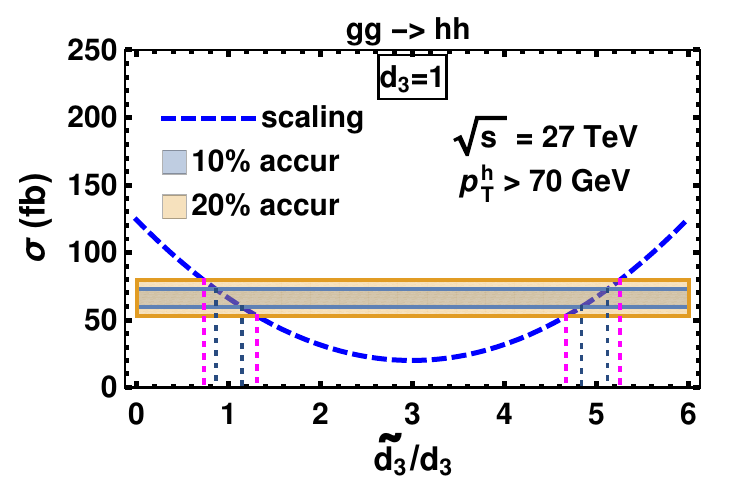}
\includegraphics[angle=0,width=0.31\linewidth]{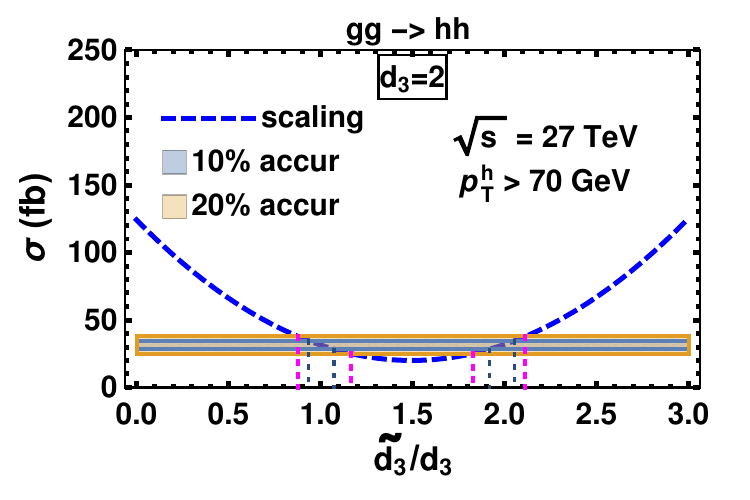}
\includegraphics[angle=0,width=0.31\linewidth]{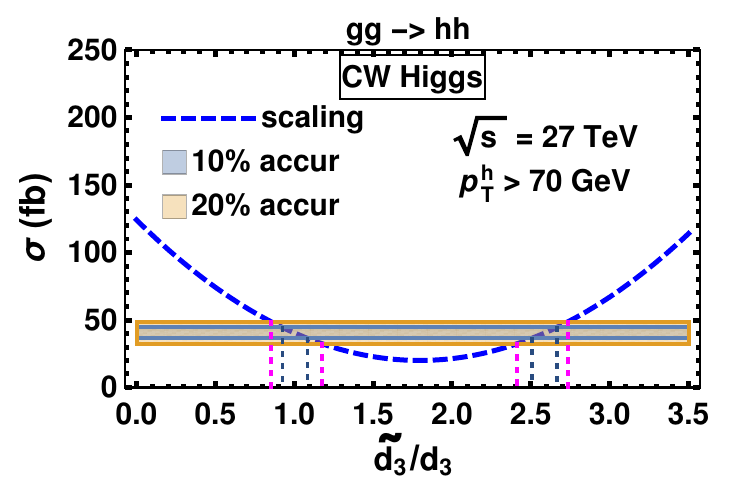}
\end{center}
\vspace{-0.9cm}
\caption{ Constraints on the scaling $\tilde{d_3}/d_3$, assuming that  the cross section can be measured up to 10\% and 20\% accuracy, respectively. Here, $\tilde{d_3}$ denotes the scaled $d_3$ value. }
\label{fig:accuracy_10-20-s3_SM-HH}
\end{figure}

\begin{figure}[!htb]
\vspace{-0.1cm} 
\begin{center}
\includegraphics[angle=0,width=0.35\linewidth]{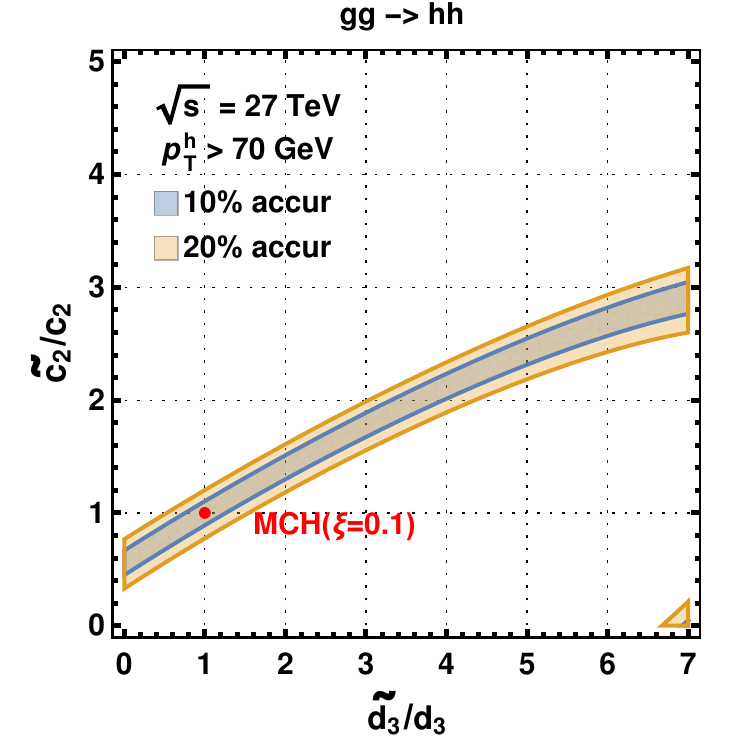}
\includegraphics[angle=0,width=0.35\linewidth]{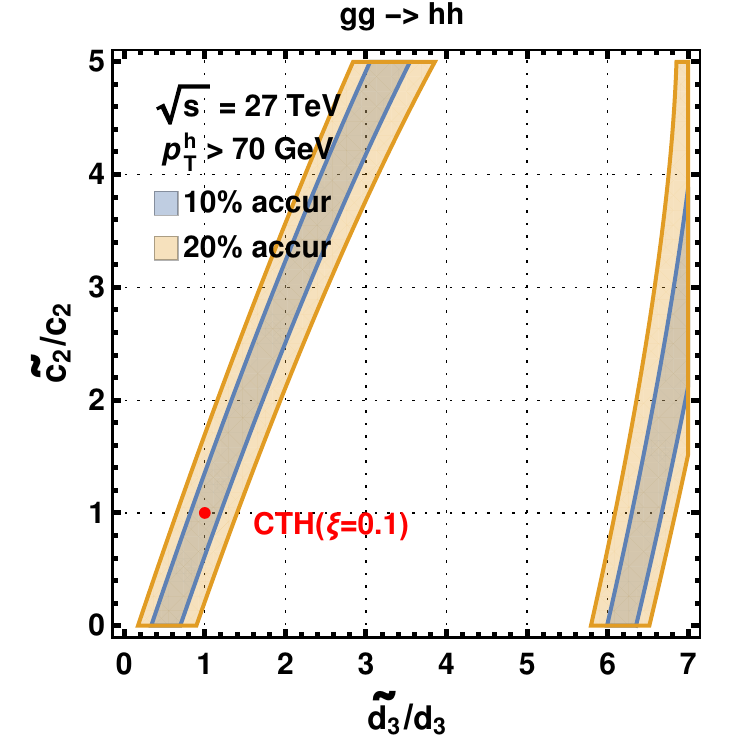}
\end{center}
\vspace{-0.9cm}
\caption{ Constraints on $\tilde{c_2}/c_2$ and $\tilde{d_3}/d_3$, assuming that  the cross section can be measured up to 10\% and 20\% accuracy, respectively, in the MCH and CTH models. Here, $\tilde{c_2}$ and $\tilde{d_3}$ denote the scaled $c_2$ and $d_3$ values, respectively.}
\label{fig:accuracy_10-20-s3_SM-HH2}
\end{figure}

With the total cross section of the double-Higgs production measured up to certain precision, we would like to extract the information on $d_3$.  
In Fig.~\ref{fig:accuracy_10-20-s3_SM-HH}, assuming the measured accuracy of the double-Higgs production cross section is $10\%$ and $20\%$ respectively, we extract the parameter range for the trilinear Higgs coupling $d_3$. We use $\tilde{d}_3$ to denote the scaled $d_3$. As shown in Fig.~\ref{fig:accuracy_10-20-s3_SM-HH}, we find that the ranges are $0.86<\tilde{d_3}/d_3<1.15 \cup 4.83<\tilde{d_3}/d_3<5.12$ ($0.73<\tilde{d_3}/d_3<1.31 \cup 4.67<\tilde{d_3}/d_3<5.25$) if the accuracy is $10\%$ ($20\%$) for $d_3 = 1$, and $0.94<\tilde{d_3}/d_3<1.07 \cup 1.92<\tilde{d_3}/d_3<2.06$ ($0.88<\tilde{d_3}/d_3<1.16 \cup 1.83<\tilde{d_3}/d_3<2.11$) if the accuracy is $10\%$ ($20\%$) for $d_3=2$, respectively.

In Fig.~\ref{fig:accuracy_10-20-s3_SM-HH2}, we show the parameter contour of the  general effective couplings $c_2$ and $d_3$ (with fixed $c_1$) that can be constrained by the double-Higgs production at the $27$ TeV HE-LHC, assuming the $1\sigma$ accuracy is $10\%$ and $20\%$, respectively. The scaling factors of the trilinear Higgs coupling and the contact $t\bar{t}hh$ coupling are denoted as the ratio $\tilde{d_3}/d_3$ and $\tilde{c_2}/c_2$, respectively. Compared to the CTH model, the constrained regions in the MCH model are tighter because the absolute value of $c_2$ in the CTH model is smaller than that in the MCH model, cf. Table.~\ref{tab:coupling}. Hence, the cross section does not vary much with the scaling of $c_2$.
Overall, we see that the $27$ TeV HE-LHC can already set strict bounds on these Higgs couplings.

\section{Triple-Higgs Production: Shape Determination}

In this section, we investigate the possibility and sensitivity to measure the quartic Higgs coupling, $d_4$,  by using the  $hhh$ production via gluon fusion, $gg\to hhh$. This process can help in a better understanding of the shape of the Higgs potential in different scenarios.

As discussed in the literature~\cite{Plehn:2005nk,Binoth:2006ym,Papaefstathiou:2015paa,Chen:2015gva,Fuks:2015hna,Fuks:2017zkg,Kilian:2017nio, Agrawal:2017cbs}, measuring the quartic Higgs coupling in the triple-Higgs production channel is not easy even at the $100$ TeV pp collider, but may be possible. This is because the signal cross section of the triple-Higgs production $pp\to hhh$ is small as compared to its SM backgrounds. Furthermore, the contribution of the quartic Higgs coupling is over-shadowed by other Higgs couplings. The quartic Higgs coupling appears in a very few diagrams which make a very small contribution to the
total cross section. According to literature~\cite{Chen:2015gva, Fuks:2015hna}, the quartic Higgs coupling is only constrained in the range of $[-20,30]$ (at the  $2\sigma$ level) by measuring the triple-Higgs boson production rate  at the $100$ TeV pp collider with a $30\ \textrm{fb}^{-1}$ data. In another approach, there have been attempts to measure the trilinear and quartic Higgs couplings indirectly using higher-order loop corrections~\cite{Bizon:2018syu,Borowka:2018pxx,Liu:2018peg}. These indirect searches put quite loose bound on the quartic Higgs coupling at future colliders, such as the double Higgs production at the future linear collider (ILC). A partial list of other related studies is included as Refs.~\cite{Kilian:2018bhs,Maltoni:2018ttu,Dicus:2016rpf}.

To further pin down the quartic Higgs coupling, it is straightforward to utilize the triple-Higgs production channel at the $100$ TeV pp collider with high luminosity run. We calculate the triple-Higgs production cross sections with general parametrization of new physics effects in different NP scenarios. Five scenarios: independent scaling of SM trilinear and quartic Higgs couplings, the SMEFT models with correlated trilinear and quartic Higgs coupling, the Nambu-Goldstone Higgs, Coleman-Weinberg Higgs and Tadpole-induced Higgs models are considered. We shall first compute and discuss cross sections and distributions in these models, then we estimate how well the quartic Higgs coupling can be measured, assuming other couplings are already determined by other experiments. It is expected that one could determine the $t\bar{t}h$ coupling, trilinear Higgs coupling, and $t\bar{t}hh$ coupling more precisely before measuring the quartic Higgs coupling.

\subsection{Cross Section and Distributions}

\begin{figure}[!htb]
\begin{center}
\includegraphics[angle=0,width=0.98\linewidth]{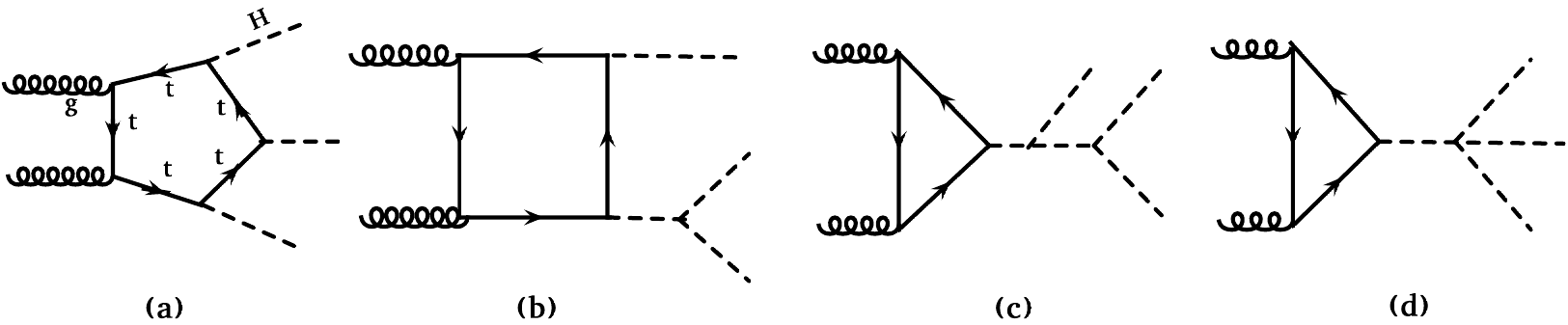}
\end{center}
\caption{Different classes of Feynman diagrams for the $gg \to hhh$ production in the SM. }  
 \label{fig:feyn-HHH-pen-bx-tr1-tr2}
\end{figure}

As shown in Fig.~\ref{fig:feyn-HHH-pen-bx-tr1-tr2}, there are several basic classes of Feynman diagrams contributing to the process $gg\to hhh$, i.e. the pentagon-class diagrams, box-class diagrams, and triangle-class diagrams. In the pentagon-class diagrams, the Higgs self-coupling does not exist and the relevant coupling is the $t\bar{t}h$ coupling. In the box-class diagrams, the trilinear Higgs coupling plays a major role. Only the triangle-class diagrams have a dependence on both the trilinear and quartic Higgs couplings. However, only a few diagrams depend on the quartic Higgs coupling~\footnote{To be specific, for each quark flavor in the loop, there are $24$ pentagon-class diagrams, $18$ box-class diagrams, and $8$ triangle-class diagrams. Out of these $50$ diagrams, only two triangle diagrams have a dependence on quartic Higgs coupling.}. Besides, the contribution of the triangle-class diagrams is comparatively small. Because of this, the process $gg\to hhh$ is only moderately sensitive to the quartic Higgs coupling. 
The cross section could change significantly with large modification of the quartic Higgs coupling and the trilinear Higgs coupling.

\begin{figure}[!htb]
\begin{center}
\includegraphics[scale=0.9]{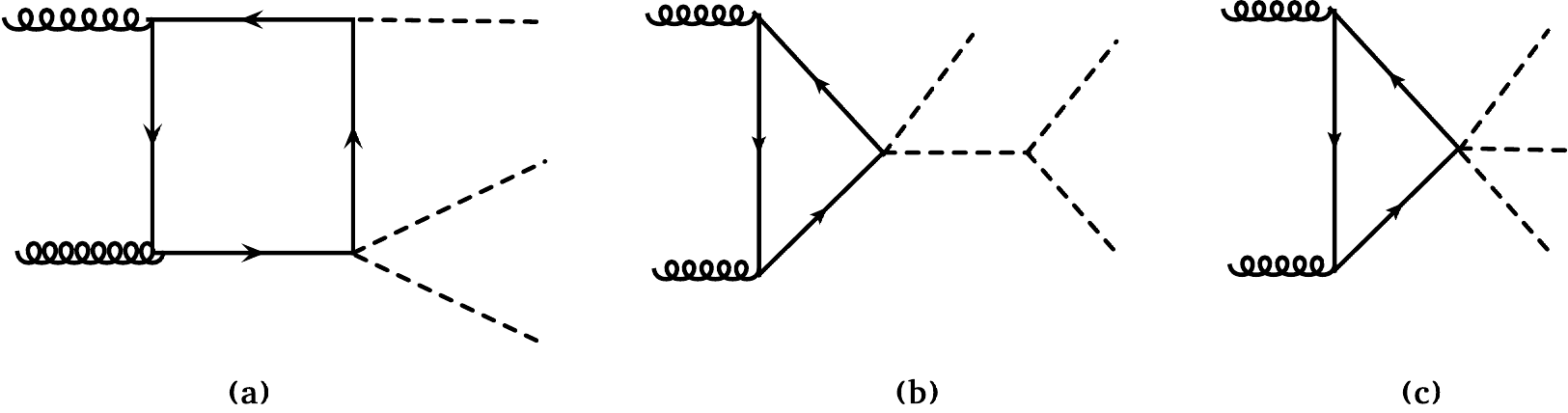}
\end{center}
\caption{New Feynman diagrams for the $gg \to hhh$ production  in the presence of the $t\bar{t}hh$ and $t\bar{t}hhh$ vertices.}
\label{fig:new-diagram_HHH}
\end{figure}

Furthermore, as shown in Fig.~\ref{fig:new-diagram_HHH}, several new diagrams would appear if additional $t\bar{t}hh$ and $t\bar{t}hhh$ couplings are non-zero. This scenario is realized explicitly, e.g. in the Nambu-Goldstone Higgs case, because of the Higgs non-linearity. In these scenarios, there are strong correlations among the $t\bar{t}h$, $t\bar{t}hh$, and $t\bar{t}hhh$ couplings.
As we will see, given the nonlinear parameter $\xi\sim 0.1$, the diagrams with $t\bar{t}hh$ and $t\bar{t}hhh$ couplings could have large contributions, which render it more difficult to extract the quartic Higgs coupling.

In the $pp\to hhh$ process, there are strong destructive interferences between different classes of diagrams. Interference between pentagon, box, and triangle diagrams plays a crucial role in dictating the cross section and various distributions. We first parameterize the contribution of each class of diagrams to the total cross section. To be specific, the total cross section is written as
\begin{eqnarray}
\sigma &=&  c_1^6 \sigma_p^{_{\rm SM}} + c_1^4d_3^2 \sigma_b^{_{\rm SM}}+ c_1^2d_3^4 \sigma_{3t}^{_{\rm SM}}+ c_1^2d_4^2 \sigma_{4t}^{_{\rm SM}} +  c_1^5d_3 \sigma_{p,b}^{_{\rm SM}}+  c_1^4d_3^2 \sigma_{p,3t}^{_{\rm SM}}+  c_1^4d_4\,\sigma_{p,4t}^{_{\rm SM}}+\nonumber\\
&& c_1^3d_3^3 \sigma_{b,3t}^{_{\rm SM}}+ c_1^3d_3d_4 \sigma_{b,4t}^{_{\rm SM}} 
+   c_1^2d_3^2d_4 \sigma_{3t,4t}^{_{\rm SM}}\nonumber\\
&& +  \left( c_1^4c_2 \sigma_{p,\,b-2t2h}+ c_1^3d_3c_2 \sigma_{b,\,b-2t2h}+ c_1^2d_3^2c_2 \sigma_{3t,\,b-2t2h} + c_1^2d_4c_2 \sigma_{4t,\,b-2t2h} + c_1^2c_2^2 \sigma_{b-2t2h} +\right.\nonumber\\
&& c_1^3c_2d_3 \sigma_{p,\,t-2t2h}+ c_1^2c_2d_3^2 \sigma_{b,\,t-2t2h} + c_1c_2d_3^3 \sigma_{3t,\,t-2t2h}+\nonumber\\ 
&&  \left. c_1c_2d_3d_4 \sigma_{4t,\,t-2t2h} + c_1c_2^2d_3 \sigma_{b-2t2h,t-2t2h} + c_2^2d_3^2 \sigma_{t-2t2h}  \right) \nonumber\\
&& + \left( c_1^3c_3 \sigma_{p,\,t-2t3h} + c_1^2d_3c_3 \sigma_{b,\,t-2t3h} + c_1d_3^2c_3 \sigma_{3t,\,t-2t3h} +\right.\nonumber\\ 
&& \left. c_1d_4c_3 \sigma_{4t,\,t-2t3h} + c_1c_2c_3 \sigma_{b-2t2h,t-2t3h} + c_2d_3c_3 \sigma_{t-2t2h,t-2t3h} + c_3^2\,\sigma_{t-2t3h} \right)\ ,
\label{eqn:interfernce_model_HHH}
\end{eqnarray}
where we separate individual Feynman diagrams, and thus can explicitly read out their dependence on various Higgs couplings.

\begin{table}[!htb]
\resizebox{\columnwidth}{!}{%
\begin{tabular}{|c|c|c||c|c|c||c|c|c|}
\hline
\multirow{2}{*}{Parts} & \multicolumn{2}{c||}{$p_T^h$} &\multirow{2}{*}{Parts} & \multicolumn{2}{c||}{$p_T^h$}&\multirow{2}{*}{Parts} & \multicolumn{2}{c||}{$p_T^h$}   \\
\cline{2-3}\cline{5-6}\cline{8-9}
&  $\rm{no\_cut}$ & $\rm{>\ 70GeV}$ & & $\rm{no\_cut}$ & $\rm{>\ 70GeV}$ & & $\rm{no\_cut}$ & $\rm{>\ 70GeV}$ \\
\hline
$\sigma_p^{SM}$        & 7777  & 3526  & $\sigma_{p,\,b-2t2h}$    & -41310  & -20509  & $\sigma_{p,\,t-2t3h}$     & -9702   & -13422  \\  
$\sigma_b^{SM}$        & 4113  & 1542  & $\sigma_{b,\,b-2t2h}$    & 39685   & 19693   & $\sigma_{b,\,t-2t3h}$     & -35207  & - 19578 \\
$\sigma_{3t}^{SM}$     & 92.2  & 26.0  & $\sigma_{3t,\,b-2t2h}$   & -3960   & -1558   & $\sigma_{3t,\,t-2t3h}$    & 5829    & 3034    \\
$\sigma_{4t}^{SM}$     & 46.57 & 22.52 & $\sigma_{4t,\,b-2t2h}$   & - 3164  & -1628   & $\sigma_{4t,\,t-2t3h}$    & 6131    & 4067    \\
$\sigma_{p,b}^{SM}$    & -8026 & -2873 & $\sigma_{b-2t2h}$        & 130729  & 85499   & $\sigma_{b-2t2h,t-2t3h}$  & -228538 & -159601 \\
$\sigma_{p,3t}^{SM}$   & 381.5 & 7.5   & $\sigma_{p,\,t-2t2h}$    & 1363    & -1719   & $\sigma_{t-2t2h,t-2t3h}$  & 148590  & 104409  \\
$\sigma_{p,4t}^{SM}$   & 133.5 & -49.5 & $\sigma_{b,\,t-2t2h}$    &  -13626 & -5906   & $\sigma_{t-2t3h}$         & 443606  & 377483  \\
$\sigma_{b,3t}^{SM}$   & -985  & -298  & $\sigma_{3t,\,t-2t2h}$   & 2412    & 976     &  &  & \\
$\sigma_{b,4t}^{SM}$   & -673.3& -266  & $\sigma_{4t,\,t-2t2h}$   & 1943    & 1011    &  &  & \\
$\sigma_{3t,4t}^{SM}$  & 121.5 & 45.0  & $\sigma_{b-2t2h,t-2t2h}$ & -66447  & -36259  &  &  & \\
                       &       &       & $\sigma_{t-2t2h}$        & 21774   & 12329 &   &  &\\
\hline                       
\end{tabular}}
\caption{ Numerical values of various SM-like cross sections, cf. Eq.~(\ref{eqn:interfernce_model_HHH}), at the 100 TeV pp collider. } \label{tab:form_factors_SM_HHH}
\label{tab:parts_HHH}
\end{table}

\begin{figure}[!htb] 
\centering
\includegraphics[angle=0,width=0.6\linewidth]{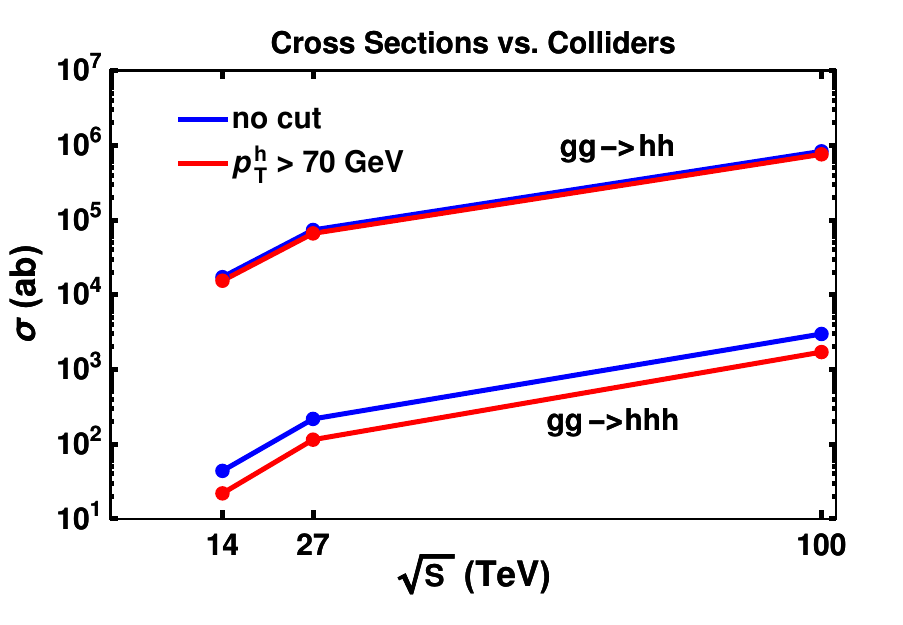}
\caption{The total cross sections of the $pp\to hh$ and $pp\to hhh$ processes for the SM at the $14$ TeV LHC, the $27$ TeV HE-LHC and the $100$ TeV pp collider, respectively. The blue lines denote the cross sections without the cut, and the red lines denote the ones with $p_T^h > 70$ GeV. Here, we do not include the QCD $K$-factors, which are known to be about $1.7$~\cite{Dawson:1998py} for $pp\to hh$ and around $2$~\cite{Maltoni:2014eza} for $pp\to hhh$, respectively.
}  
\label{fig:tot_xs_double_triple_higgs}
\end{figure}

We carry out the calculation in the way discussed in Refs.~\cite{Agrawal:1998ch,Agrawal:2012as}. 
We use FORM~\cite{Vermaseren:2000nd} to compute the trace of gamma matrices in the amplitude and to write the amplitude in terms of tensor integrals. These tensor integrals are computed using an in-house package, OVReduce~\cite{Agrawal:1998ch}, which implements the Oldenborgh-Vermaseren~\cite{vanOldenborgh:1989wn} technique of tensor integral reduction. Scalar integrals are computed using the package OneLOop~\cite{vanHameren:2010cp}. 
We use the leading-order CTEQ parton distribution functions, CT14LLO~\cite{Dulat:2015mca}, and set the renormalization (and factorization) scale to be the invariant mass of the hard-scattering process $\sqrt{\hat{s}}$. 
The numerical value of each individual SM-like cross section is calculated and summarized in Table~\ref{tab:parts_HHH}. Here we do not include the higher-order QCD correction, which may 
lead to a $K$-factor (the ratio of the next-to-leading to the leading-order cross section) of about $2$~\cite{Maltoni:2014eza}.  Due to the extremely small cross section of this process\footnote{To be specific, the total cross section is about $44$ ab at the $14$ TeV LHC, and is only about $218$ ab at the $27$ TeV HE-LHC, respectively.} at the HL-LHC and HE-LHC and the large QCD backgrounds at the same time, we directly present results at the $100$ TeV pp collider. Basic $p_T$ cuts are also implemented for each Higgs boson in the Higgs final state. At the 100 TeV collider, the SM cross-sections without cuts and with $p_T > 70 $ GeV cut are 2987 ab and 1710 ab, respectively. We summarize the total cross sections of the double- and triple-Higgs productions for the SM in Fig.~\ref{fig:tot_xs_double_triple_higgs} at the $14$ TeV LHC, the $27$ TeV HE-LHC and the $100$ TeV pp collider.\\

At the 100 TeV collider, the total cross-sections for the Tadpole-induced Higgs model and Coleman-Weinberg model without any cut are 7796 ab and 1272 ab, while with $ p_T^h > 70\ \rm{GeV}$ cut these are 3579 ab and 836 ab, respectively. For the benchmark value $\xi=0.05$ in the MCH and CTH models, the cross-sections without any cut  are  5033 ab and 3479 ab, while with $ p_T^h > 70\ \rm{GeV}$ cut these are 3302 ab and 2057 ab, respectively.

\begin{figure}[!htb] 
\centering
\includegraphics[angle=0,width=0.45\linewidth]{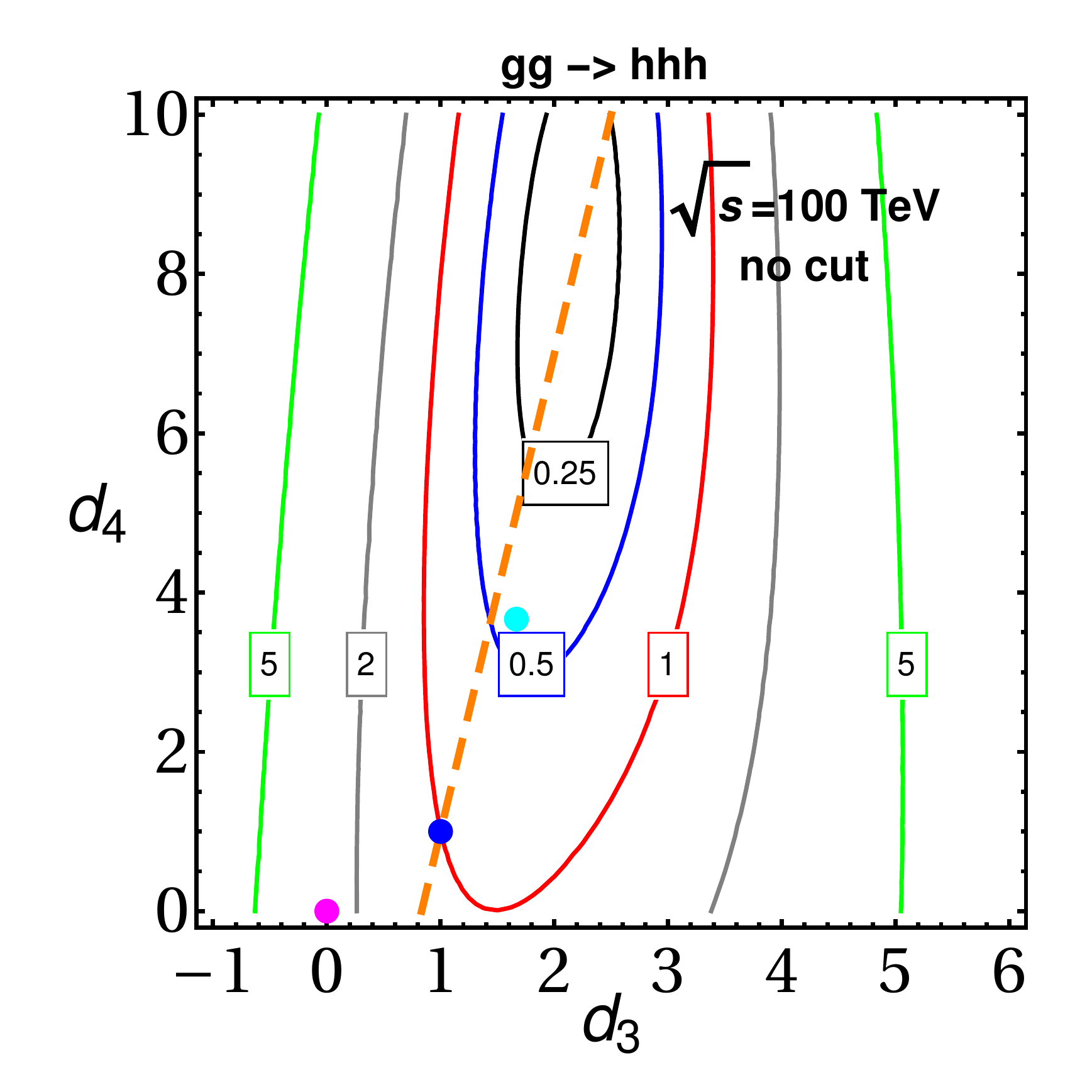}
\includegraphics[angle=0,width=0.45\linewidth]{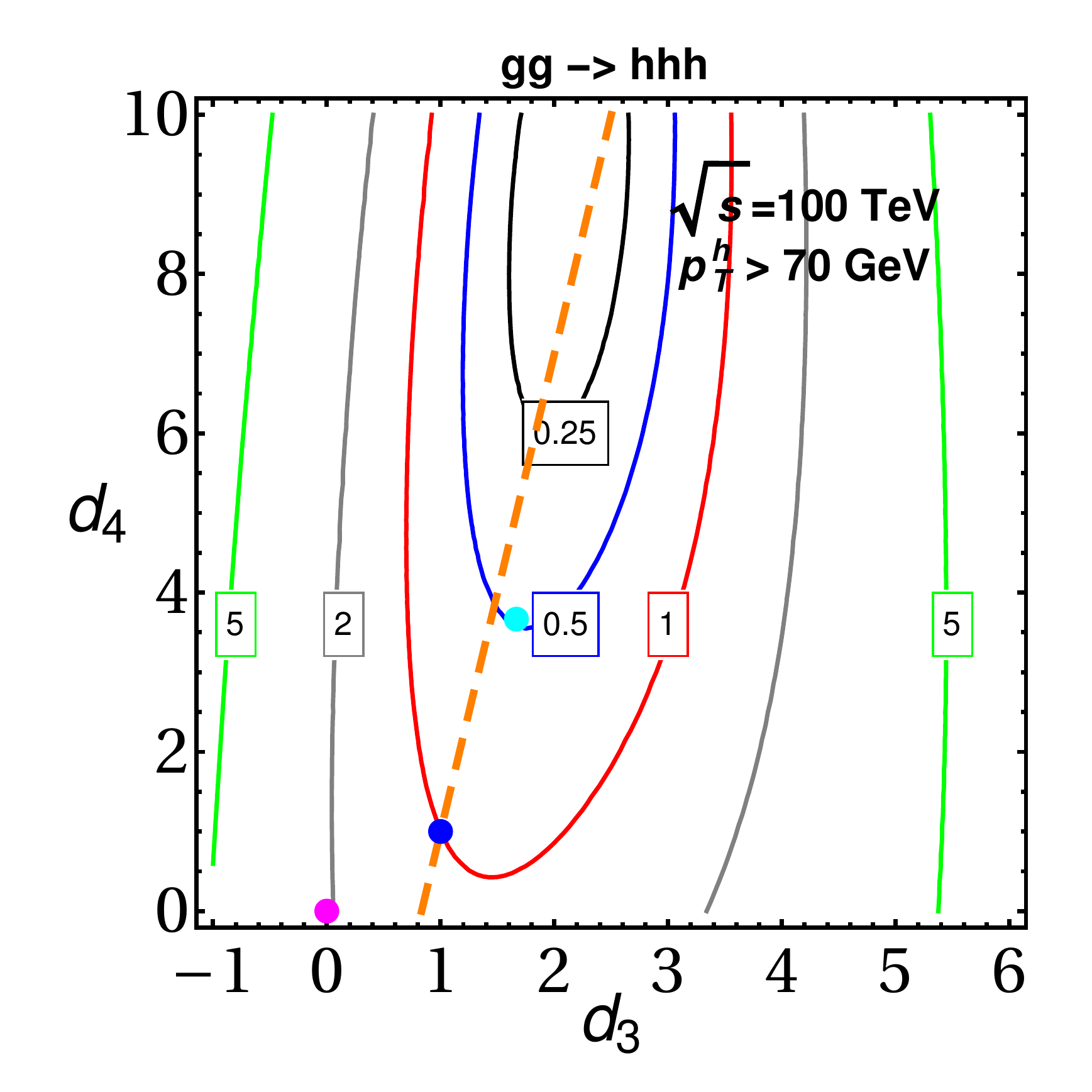}
\caption{The cross section ratio $\sigma/\sigma_{SM}$, as a function of the scaling of the trilinear and quartic Higgs couplings with various cuts. At the 100 TeV pp collider, the SM cross section without any cut and with $p_T > 70 $ GeV cut are 2987 ab and 1710 ab, respectively. The blue, cyan, and magenta dots denote the SM, CW Higgs and Tadpole-induced Higgs scenarios, respectively. The orange dashed line denotes the SMEFT (with non-vanishing $O_6$) for $d_3$ in the range of [5/6,2.5].
}  
\label{fig:lam_3-lam_4_HHH}
\end{figure}

\begin{figure}[!htb] 
\centering
\includegraphics[angle=0,width=0.50\linewidth]{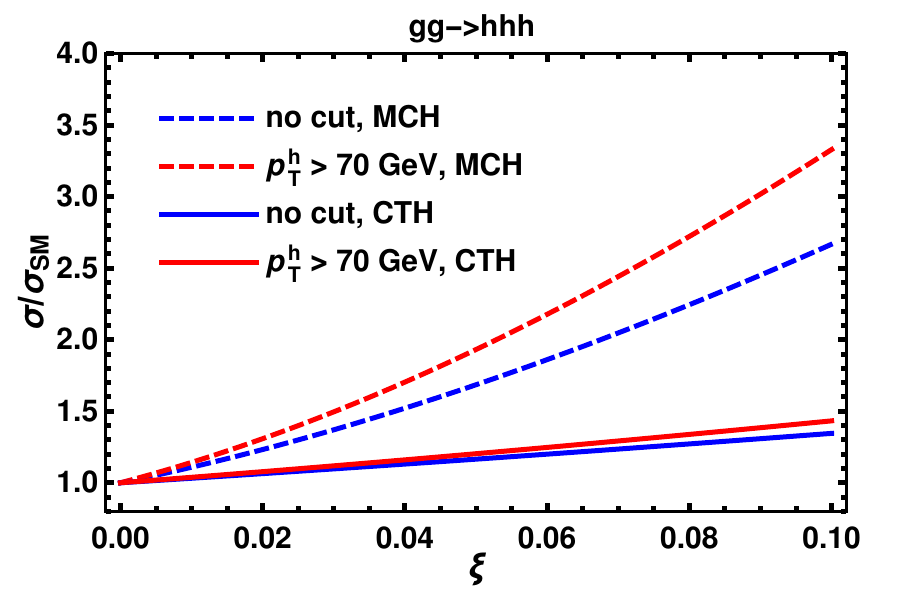}
\caption{The cross section ratio $\sigma/\sigma_{SM}$, as a function of the parameter $\xi$ in the MCH and CTH Models, at the 100 TeV pp collider. }  
\label{xi_CTH_MCH}
\end{figure}

Based on these numerical values, we display the cross sections in the $(d_3, d_4)$ parameter contour in Fig.~\ref{fig:lam_3-lam_4_HHH} and the $\xi$ dependence in Fig.~\ref{xi_CTH_MCH}, for different NP scenarios, without and with including the contact $t\bar{t}hh$ and $t\bar{t}hhh$ couplings. Fig.~\ref{fig:lam_3-lam_4_HHH} shows the total cross section $\sigma$ as a function of the trilinear and quartic Higgs couplings, i.e. $d_3$ and $d_4$.
We see that there is a significant increase in the cross section for the negative value of the $d_3$ because the largest negative interference between the  box and pentagon diagrams $\sigma_{p,b}^{_{\rm SM}}$, either vanishes or becomes positive. 
There is only a marginal increase in the cross section for the negative value of the $d_4$. 
In this figure, we also mark the SM scenario, the Coleman-Weinberg Higgs scenario, the Tadpole-induced Higgs scenario by blue, cyan, and magenta dots, respectively. The orange line denotes the SMEFT with nonzero $O_6\sim (H^\dagger H)^3$ operator in the linear expansion as in Eq.~(\ref{eq:wilsoncoef1}) and Eq.~(\ref{eq:wilsoncoef2}). However, since the Nambu-Goldstone boson models contain additional $t\bar{t}hh$ and  $t\bar{t}hhh$ couplings, and the $t\bar{t}h$ coupling is different from the SM one, they cannot be directly compared with in this figure.
Instead, the result of Nambu-Goldstone Higgs scenario is presented in Fig.~\ref{xi_CTH_MCH}, which only depends on the nonlinearity parameter $\xi$. To be concrete, we consider two specific models, i.e. the MCH and CTH models, in Fig.~\ref{xi_CTH_MCH}. 
Compared to the MCH, the cross section of the CTH remains close to the SM prediction for the given $\xi$.

\begin{figure}[!htb] 
\begin{center}
\includegraphics[angle=0,width=0.29\linewidth, height=0.29\linewidth]{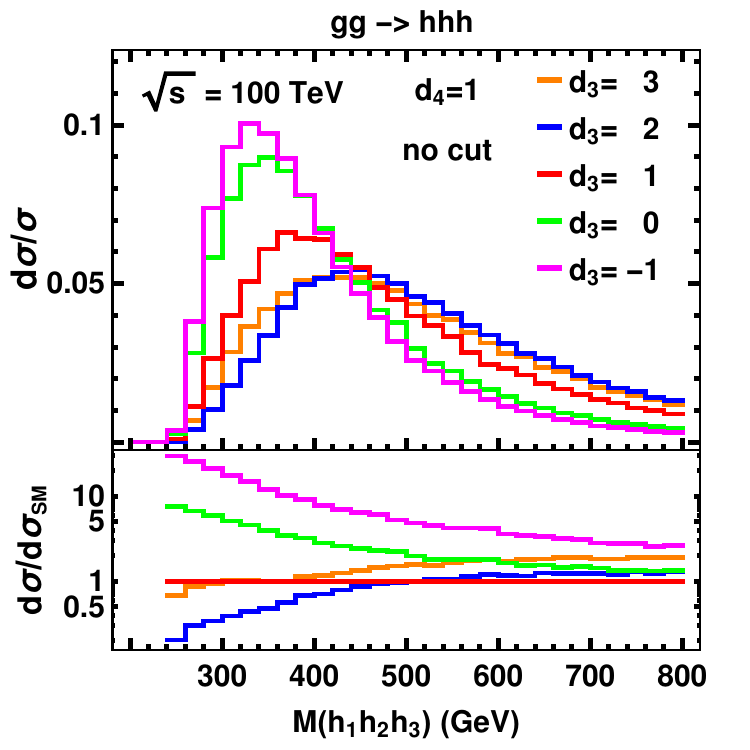}
\includegraphics[angle=0,width=0.29\linewidth,, height=0.29\linewidth]{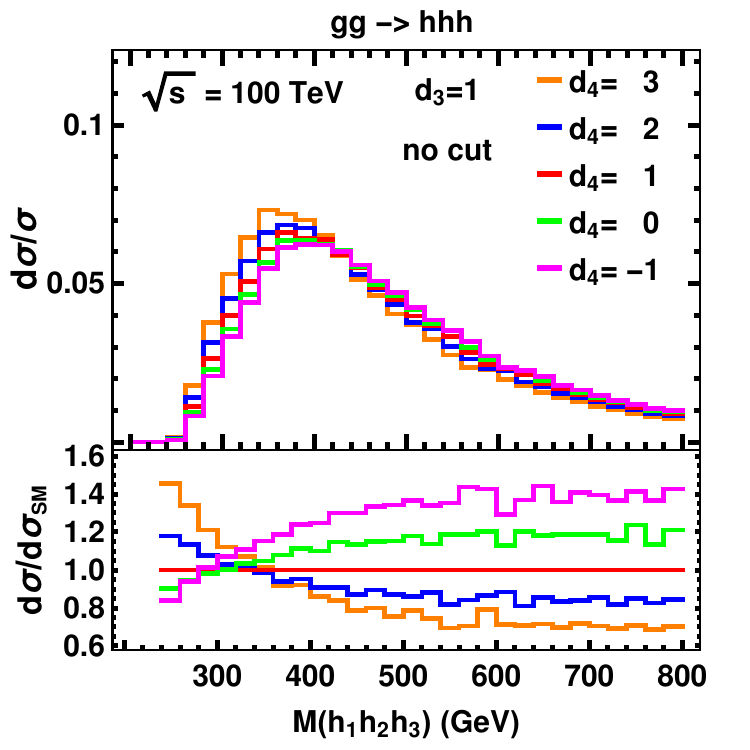}
\includegraphics[angle=0,width=0.4\linewidth, height=0.29\linewidth]{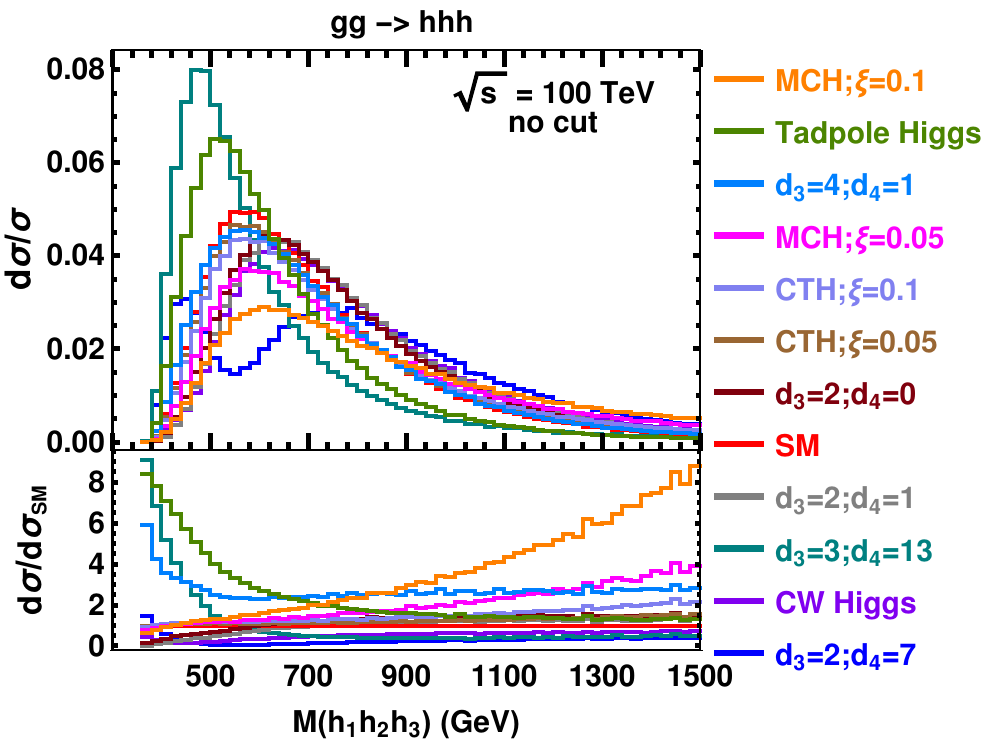}
\end{center}
\caption{{Distributions with partonic center-of-mass energy $M(hhh)$ for $hhh$ production via gluon-gluon fusion with different benchmark values of $d_3$ and $d_4$ at the 100 TeV pp collider. No cut on $p_T$ of Higgs bosons has been imposed.}}
\label{fig:dist-lam3-HHH-M}
\end{figure}

To complete discussion of this subsection, we present several basic differential distributions. In Fig.~\ref{fig:dist-lam3-HHH-M}, we show the invariant mass $M(hhh)$ distribution for various $d_3$ and $d_4$ values, and the normalized plots to examine the modification of the shape of the distributions. We observe the quite distinct behavior near the threshold of triple-Higgs boson production for different values of the $d_3$ and $d_4$ couplings. 
In the case of $d_3$, there is a larger increase in the cross section near the threshold for its negative and zero value, while decrease for positive values of $d_3$. The behavior is the opposite in the case of $d_4$.
Most of the increase is for smaller values of the invariant mass of the triple-Higgs system, up to about $700$ GeV, and it is near the threshold where the triangle diagram with quartic Higgs coupling is important.

\subsection{Interference Effects}
In this subsection, we investigate the interference patterns for the triple-Higgs production $pp\to hhh$ process, for better understanding variation of the total cross section and distributions in the different Higgs scenarios.

\subsubsection{Interference without $t\bar{t}hh$ or $t\bar{t}hhh$}

\begin{figure}[!htp] 
\begin{center}
\includegraphics[angle=0,width=0.47\linewidth]{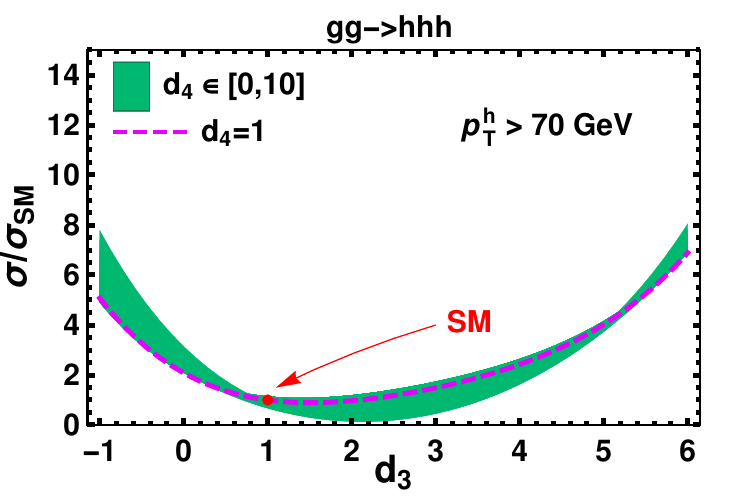}
\includegraphics[angle=0,width=0.47\linewidth]{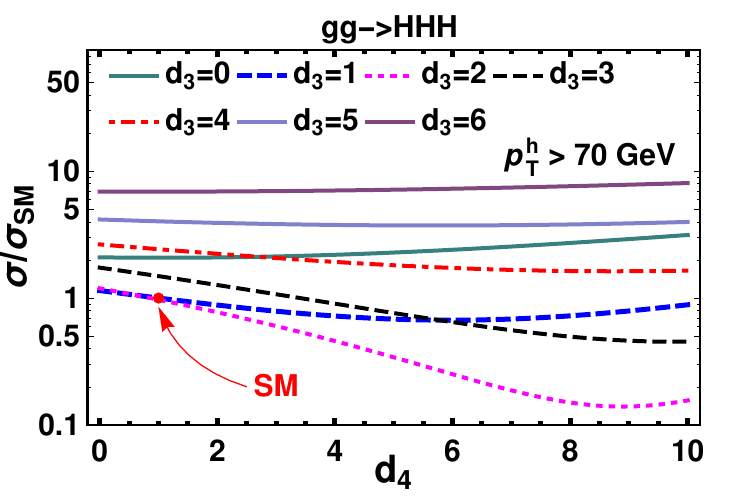} 
\end{center}
\vspace{-0.9cm}
\caption{ Variation of the ratio of the cross section  $\sigma/\sigma_{SM}$ with respect to $d_3$ and $d_4$ at the 100 TeV pp collider. In the left, we show a band for varying $d_4$ in the range of [0,10]. In the right, variation  with $d_4$ for various fixed $d_3$ values is shown. The standard model cross section without any cut and with $p_T^h > 70$ GeV are 2987 ab and 1710 ab, respectively.}  
\label{fig:lam_3-lam_4_HHH_band}
\end{figure}

Let us first consider the scenarios without the $t\bar{t}hh$ and $t\bar{t}hhh$ couplings. There are 10 relevant terms in the total cross section, as shown in Eq.~(\ref{eqn:interfernce_model_HHH}). The first four terms are always positive, 
and the rest of the six terms are interference terms and can be either positive or negative. 
As shown in the left panel of Fig.~\ref{fig:lam_3-lam_4_HHH_band}, the cross section first decreases and then increases within the range $-1<d_3<6$ as the $d_3$ increases. 
In addition, we show the variation of cross section, as the green band in Fig.~\ref{fig:lam_3-lam_4_HHH_band}, with the quartic Higgs coupling $d_4$ varying within $0<d_4<10$.
In the right panel of Fig.~\ref{fig:lam_3-lam_4_HHH_band}, we explicitly see the variation of $\sigma/\sigma_{\rm{SM}}$ as function of $d_4$, with $d_3$ fixed. 
Although it is theoretically less plausible to have a large $d_3$ ($d_3 \le 6$, as constrained by vacuum stability), we still include this possibility here. In that case, the cross section only moderately varies with the $d_4$ values. In this case, there is large degeneracy in the $d_4$ determination when $d_3$ is around 5 to 6.

\subsubsection{Interference with $t\bar{t}hh$ and $t\bar{t}hhh$ }

In this subsection, we discuss NP scenarios in which the $t\bar{t}hh$ and $t\bar{t}hhh$ couplings are non-vanishing, e.g. the Nambu-Goldstone Higgs scenario, and investigate the interference terms involving these couplings in details. 
In this scenario, all the Higgs couplings are only related to the single parameter $\xi$ due to the Higgs non-linearity. 

\begin{figure}[!htb] 
\begin{center}
\includegraphics[angle=0,width=0.4\linewidth]{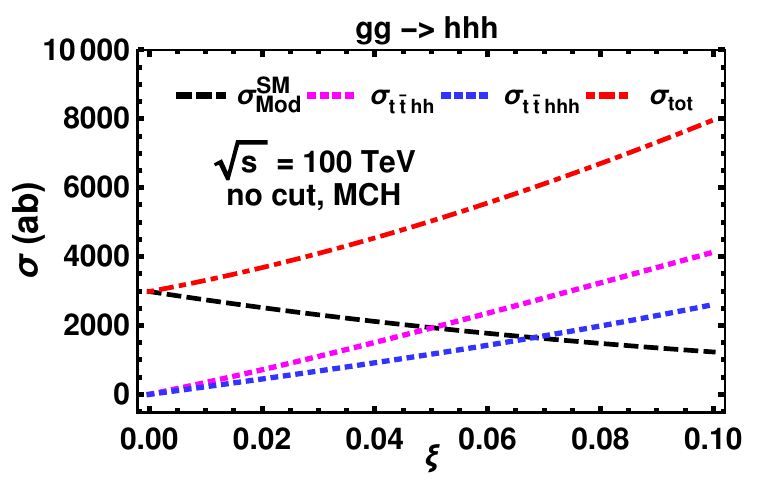}
\includegraphics[angle=0,width=0.4\linewidth]{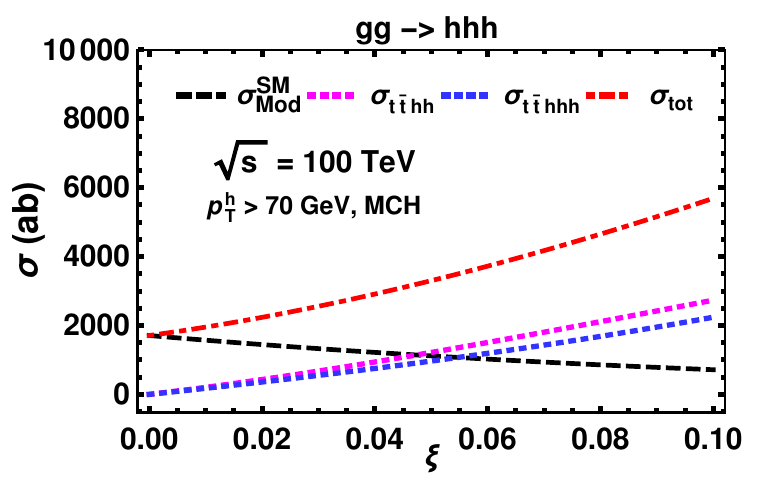}\\
\includegraphics[angle=0,width=0.4\linewidth]{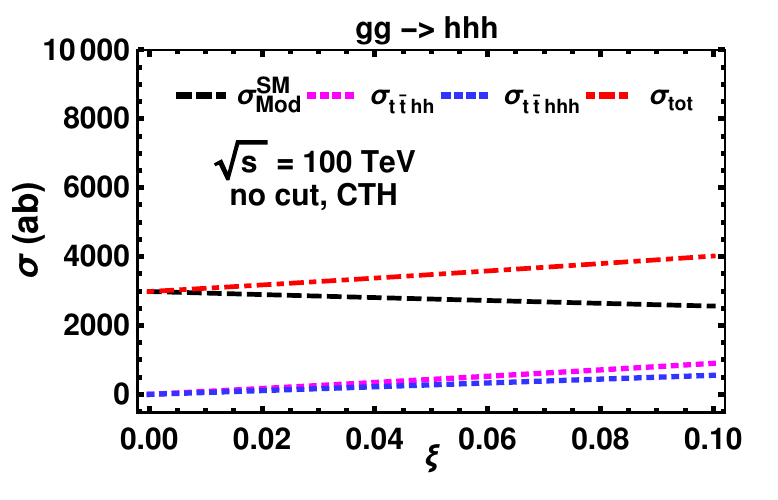}
\includegraphics[angle=0,width=0.4\linewidth]{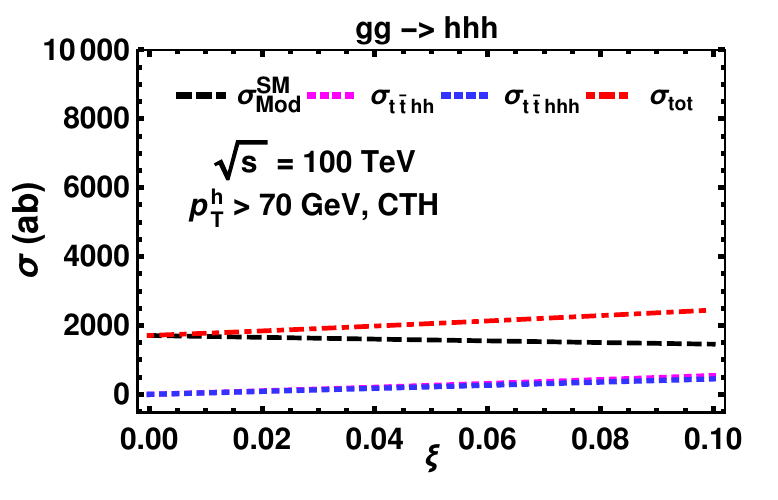}\\
\end{center}
\caption{The cross section [in ab] as a function of the parameter $\xi$ in the MCH and CTH models. The magenta line shows the effect of $t\bar{t}hh$ coupling. In the MCH model, it exceeds the "SM-like" effect ($\sigma_{Mod}^{SM}$) around the value $\xi=0.05$. The blue line shows the effect of $t\bar{t}hhh$ coupling, which includes the interference effect between  $t\bar{t}hhh$ and $t\bar{t}hh$ couplings (this interference is destructive for the shown range of $\xi$).}
\label{CTH-MCH-tthh-effect}
\end{figure}

In Fig.~\ref{CTH-MCH-tthh-effect}, we show the interference effect of the $t\bar{t}hh$ and $t\bar{t}hhh$ couplings in two specific NG Higgs models, i.e. the MCH and CTH models. 
As expected, in the case of CTH model, the contribution of these couplings remains very small, except at large value of $\xi$, where it is also not that significant. However, in the case of MCH model, both the $t\bar{t}hh$ and $t\bar{t}hhh$ couplings play important role. At a larger value of $\xi$, the significant increase in the cross section is induced by these couplings. As $\xi$ increases, the contribution ($\sigma_{Mod}^{SM}$) of SM-like diagrams decreases due to the smaller $t {\bar t}h$, $d_3$ and $d_4$ couplings, but the contribution of diagrams with $t\bar{t}hh$ and $t\bar{t}hhh$ couplings increases.

\begin{figure}[H] 
\begin{center}
\includegraphics[angle=0,width=0.47\linewidth]{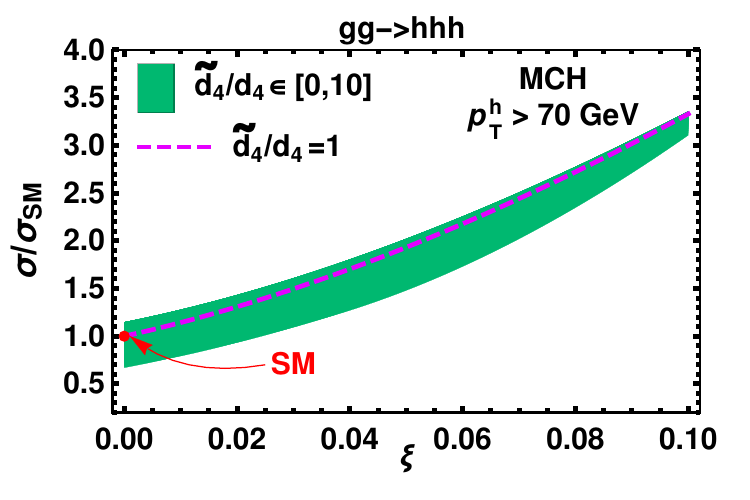}
\includegraphics[angle=0,width=0.47\linewidth]{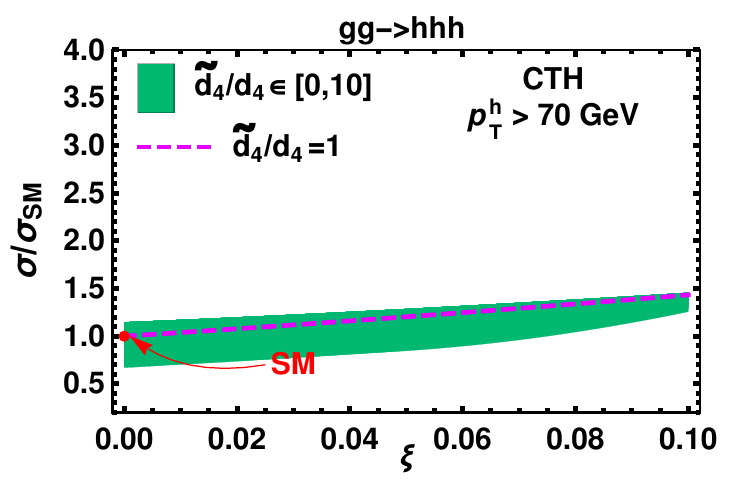}
\end{center}
\vspace{-0.9cm}
\caption{ Variation of the ratio of the cross section to the SM, with respect to $\xi$ and $\tilde{d_4}/{d_4}$, at the 100 TeV pp collider. The band is obtained by varying  $\tilde{d_4}/{d_4}$ in the range of [0,10] for the MCH and CTH models, respectively. 
}  
\label{fig:s_4-band_HHH_model}
\end{figure}

In Fig.~\ref{fig:s_4-band_HHH_model}, the ratios of the cross sections in the  MCH and CTH models with respect to the SM value are shown, and the ratios depend on the single parameter $\xi$.
The green band shows variation of the ratios due to the scaling of the quartic Higgs coupling, denoted by $\tilde{d_4}/d_4$, and the dashed line is for $d_4=1$. 
We see the variation due to the scaling of the quartic Higgs coupling decreases for the larger values of the parameter $\xi$.

\subsection{Shape Determination and $\lambda_4$ Extraction}

Here in this subsection we investigate how to measure the quartic Higgs coupling at the $100$ TeV pp collider to discriminate  various NP scenarios. 
Similar to the study of the double-Higgs production process in the above section, we do not perform any detailed collider analysis, 
but  only utilize the existing collider simulations as the benchmark point and perform recast to obtain the signal significance in various scenarios.
There have already been several signal-to-background studies
to observe the $ p p \to h h h$ process at the 100 TeV collider. To reduce the
backgrounds, the most promising signature has one of the Higgs bosons decaying into the rare decay channel of two photons, while
the other two Higgs bosons each decay into a pair of bottom jets~\cite{Chen:2015gva, Fuks:2015hna}.
Different studies tag different number of bottom jets. Tagging 
more bottom jets reduces the signal events, but the backgrounds
reduce very significantly. So it is not surprising that the
studies where the four bottom jets are tagged perform better than
where two bottom jets are tagged. It has been shown that
if two or three bottom jets are tagged, then the largest
background is due to the production of $\gamma \gamma b {\bar b} jj$. However, if four bottom jets are tagged, then the largest
background is due to the production of $h (\to \gamma \gamma) b {\bar b} b {\bar b}$~\cite{Fuks:2015hna}. This is because of the small mistagging
rate for the light jets.

\begin{table}
 \begin{center}
  \begin{tabular}{|c|c|c|c|c|c|}
   \hline
  With $2$ b-tagged jets & SM & SMEFT & NG Higgs & CW Higgs & Tadpole Higgs  \\
   \hline
  luminosity ($\rm{ab}^{-1}$) & $1.8\times 10^4$ & $5.1\times 10^5$ & $1.6\times 10^3$ & $7.5\times 10^4$ & $4.1\times 10^3$ \\
  \hline
  \end{tabular}
 \end{center}

 \begin{center}
  \begin{tabular}{|c|c|c|c|c|c|}
   \hline
  With $3$ b-tagged jets & SM & SMEFT & NG Higgs & CW Higgs & Tadpole Higgs  \\
   \hline
  luminosity ($\rm{ab}^{-1}$)  &  $1198$ &  $33819$ &  $111$ &  $4976$ &  $277$\\
  \hline
  \end{tabular}
 \end{center}

 \begin{center}
  \begin{tabular}{|c|c|c|c|c|c|}
   \hline
  With $4$ b-tagged jets & SM & SMEFT & NG Higgs & CW Higgs & Tadpole Higgs  \\
   \hline
  luminosity ($\rm{ab}^{-1}$)  &  $51$ &  $873$ &  $8.5$ &  $163$ &  $16.7$\\
  \hline
  \end{tabular}
 \end{center}
   
	\caption{The integrated luminosity required for the $5 \, \sigma$ observation of the process $pp \to hhh$ in the SM and other new physics scenarios. Here, we take $d_3=2$  (and $d_4=7$, cf. Eqs.~(\ref{eq:wilsoncoef1}) and~(\ref{eq:wilsoncoef2})) for the SMEFT, and $\xi=0.1$ for the MCH of NG Higgs scenario. These numbers are obtained without including the QCD $K$-factor, which is known up to NNLO~\cite{Maltoni:2014eza,deFlorian:2016sit,Spira:2016zna} and is about a factor of 2 for the SM. The $K$-factors for $hhh$ production with scaled SM Higgs couplings have also been evaluated in~\cite{deFlorian:2016sit}. By including this 
  	$K$-factor, we expect the required luminosity would be slightly reduced for discovering these NP scenarios.}
  \label{tab:lum}
\end{table}

\begin{figure}[!htb]
\begin{center}
\includegraphics[width=0.7\linewidth]{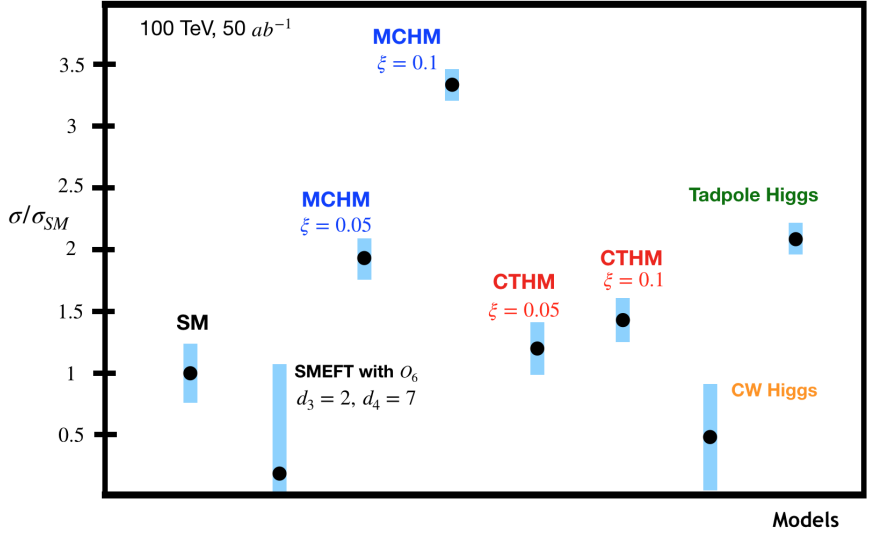}
\end{center}
\vspace{-0.5cm}
\caption{The cross section ratio $\sigma/\sigma_{\rm SM}$ in the triple-Higgs production at the $100$ TeV pp collider, with an  integrated luminosity of $50\ \text{ab}^{-1}$, for various models. Here, we consider the case that the  SM cross section can be measured with an accuracy of $30\%$ at the $1\sigma$ level. The accuracy for the NP models are obtained using the rescaling procedure  described in the text. 
The blue bars denote the expected accuracy for a given model. 	
}
\label{HHHproduction}
\end{figure}

In Table~\ref{tab:lum}, we have summarized the required luminosity for a $5 \, \sigma$ discovery by rescaling the signal cross section, cf. Eq.~(\ref{eq:NPrescale}). The results are presented for two, three or four bottom jet tagging scenarios.
In order to measure the SM cross section of the 
$pp\to hhh$ process to be within $30\%$ accuracy at the $1 \, \sigma$ level, the needed integrated luminosity is around $50\ \rm{ab}^{-1}$ when we adopt the four bottom jet tagging scenario. The corresponding accuracies for other scenarios are obtained through the same procedure as in the section 5.
%
With an increasing luminosity of the 100 TeV pp collider, it is still challenging to extract the quartic Higgs coupling due to the large contribution from the $tthh$ and $tthhh$ couplings to the total cross sections, although the cross sections in the MCH and CTH models are relatively larger than the one in the SM. 
Furthermore, it would be difficult to extract the quartic Higgs coupling in the Tadpole-induced Higgs scenario, with $d_4\simeq 0$, though the required integrated luminosity for observing the triple-Higgs production process in this scenario is smaller than that in the SM. 
With an integrated luminosity of $50\ \text{ab}^{-1}$,  
the expected accuracy for measuring the cross section of the triple-Higgs production process for a given model is shown in 
Fig.~\ref{HHHproduction}. 
To be specific, the SMEFT with $d_3=2$ and $d_4=7$ can be measured with an accuracy of $86\%$ at the $1\sigma$  level, while $16\%$ ($21\%$) for MCH with the parameter $\xi=0.1$ ($\xi=0.05$), $24\%$ ($28.9\%$) for CTH with the parameter $\xi=0.1$ ($\xi=0.05$), $45\%$ for Coleman-Weinberg Higgs scenario, and $20\%$ for the Tadpole-induced Higgs scenario, respectively. 
Since the expected uncertainties on the measurement of the triple-Higgs production cross section in these scenarios are larger than those of the double-Higgs production, cf. Fig.~\ref{HHproduction}, 
it would be relatively easier to discriminate these models via precision measurement of the double-Higgs production.

\begin{figure}[!htb] 
\begin{center}
\includegraphics[angle=0,width=0.4\linewidth, height=0.3\linewidth]{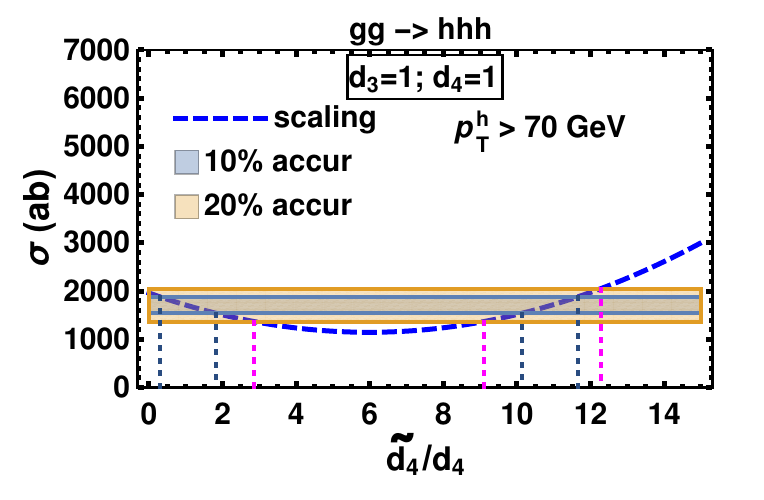}
\includegraphics[angle=0,width=0.4\linewidth, height=0.3\linewidth]{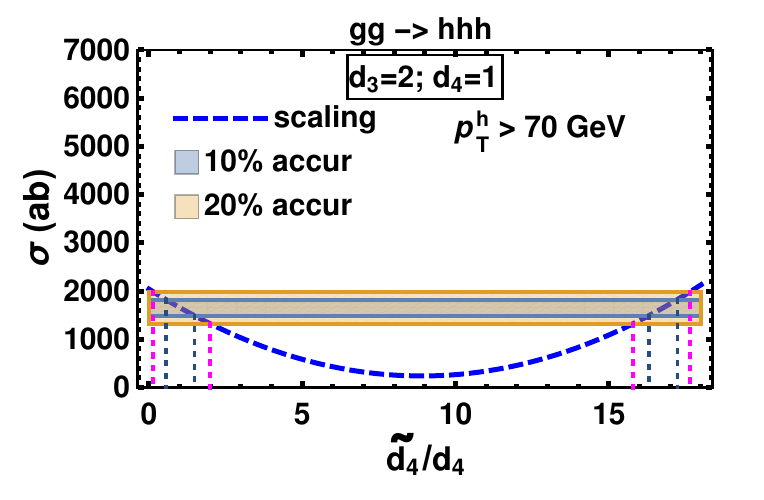}\\
\includegraphics[angle=0,width=0.4\linewidth, height=0.3\linewidth]{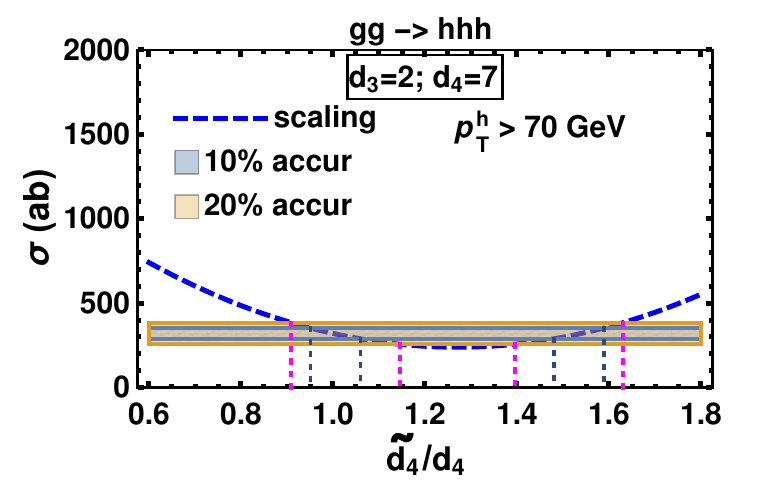}
\includegraphics[angle=0,width=0.4\linewidth, height=0.3\linewidth]{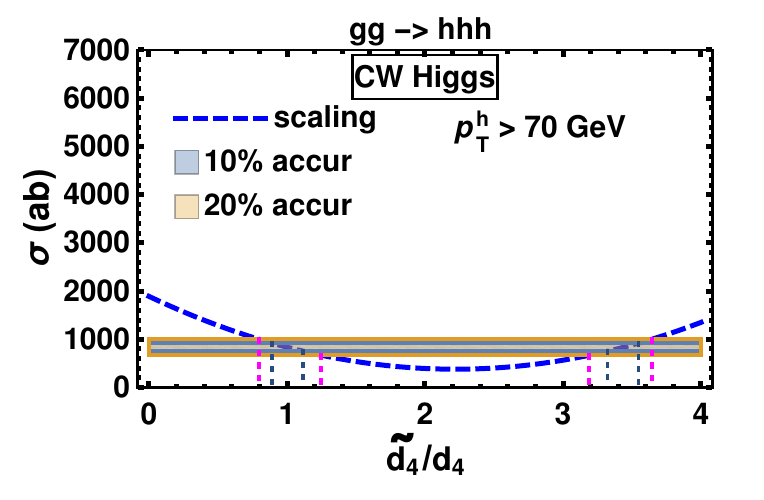}
\end{center}
\vspace{-0.9cm}
\caption{ Constraints on $\tilde{d_4}/{d_4}$ in various new physics scenarios, when the cross section can be measured up to 10\% and 20\% accuracy, respectively. The parameter $\tilde{d_4}/{d_4}$ scales the quartic Higgs coupling in a given model.} 
\label{fig:accuracy_10-20-s4_MCH-HHH}
\end{figure}

More importantly, our goal is to extract out the range of the quartic Higgs coupling from the triple-Higgs production process.
In Fig.~\ref{fig:accuracy_10-20-s4_MCH-HHH}, we show the variation of the cross section for the triple-Higgs production as a function of the scaling factor $\tilde{d_4}/d_4$, denoted by the dashed line. 
In these plots, we present the bands for the $10\%$ and $20\%$ accuracies on measuring the $pp\to hhh$ cross section in each scenario. We consider the SM ($d_3=1,d_4=1$), and take the SMEFT with  ($d_3=2,d_4=1$) and ($d_3=2,d_4=7$), and the Coleman-Weinberg Higgs case with ($d_3=5/3,d_4=11/3$), respectively, as the benchmark scenarios. 

\begin{figure}[H]
\begin{center}
\includegraphics[angle=0,width=0.35\linewidth]{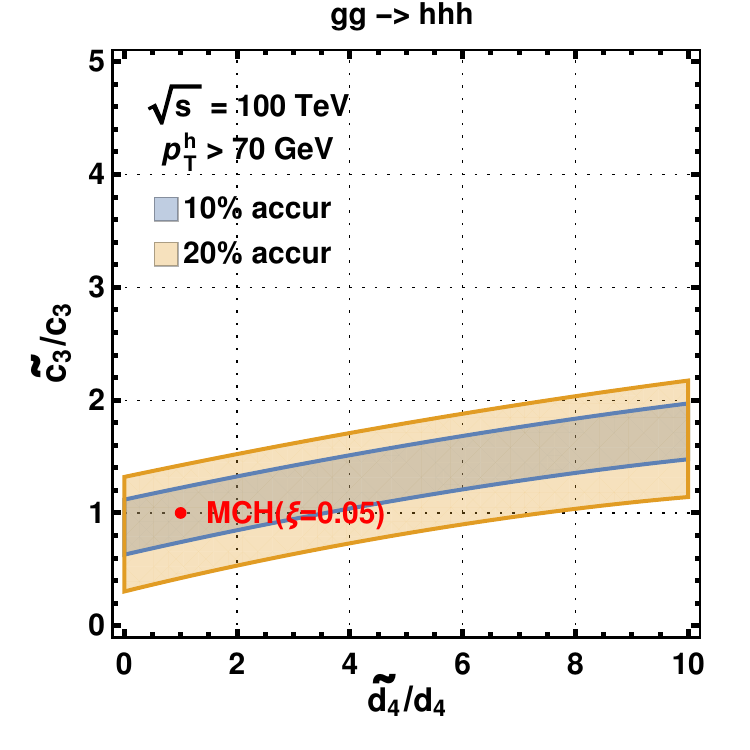}
\includegraphics[angle=0,width=0.35\linewidth]{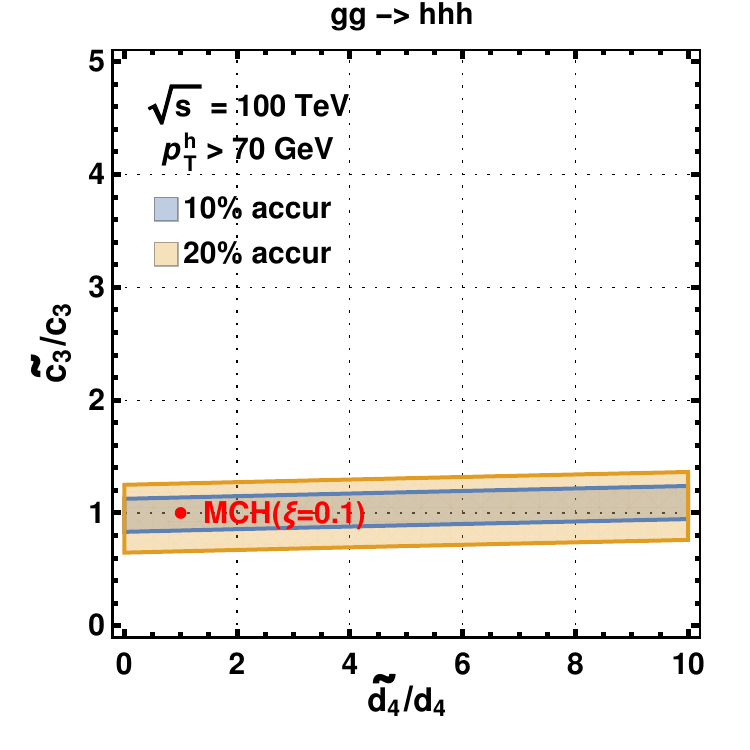}\\
\includegraphics[angle=0,width=0.35\linewidth]{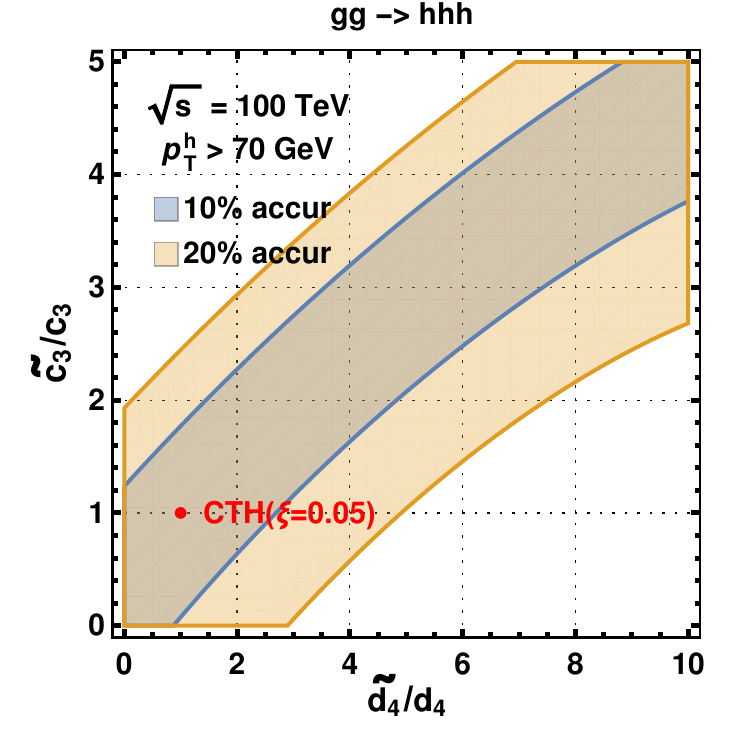}
\includegraphics[angle=0,width=0.35\linewidth]{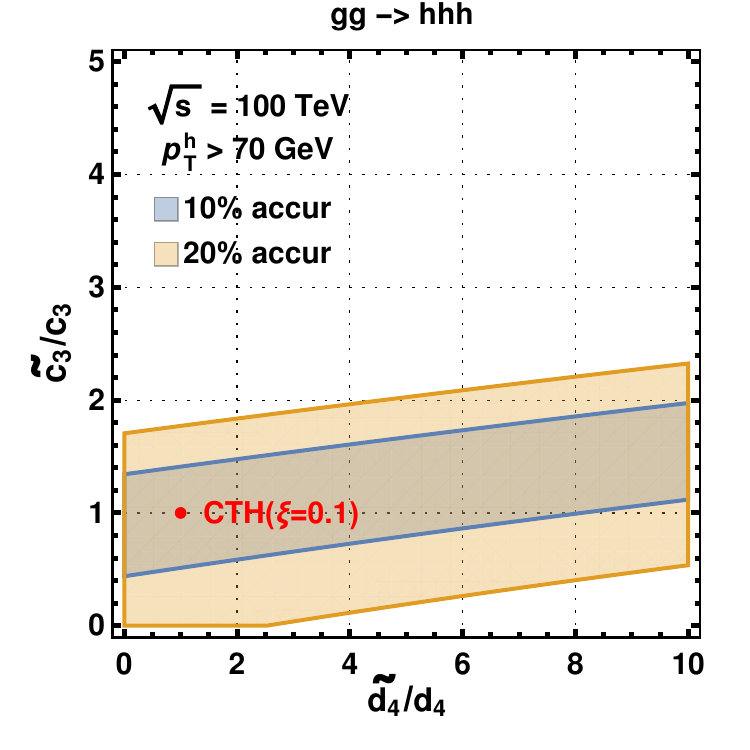}
\end{center}
\vspace{-0.9cm}
\caption{ Constraints on $\tilde{c_3}/{c_3}$ and $\tilde{d_4}/{d_4}$, assuming that  the cross section can be measured up to 10\% and 20\% accuracy, respectively, in the MCH and CTH models.
}
\label{fig:accuracy_10-20-sc3sd4_SM-HHH}
\end{figure}

Fig.~\ref{fig:accuracy_10-20-s4_MCH-HHH} shows how well the scaling factor $\tilde{d_4}/d_4$, and hence the quartic Higgs coupling, can be measured for a given benchmark scenario. 
In case there are non-vanishing contact $t\bar{t}hh$ and $t\bar{t}hhh$ couplings, e.g., in the Nambu-Goldstone Higgs scenario, the situation is slightly different. We focus on the MCH and CTH models with the nonlinear parameter $\xi=0.05$ and $\xi=0.1$, respectively. Assuming the $pp \to hhh$ cross section could be measured with the accuracy of $10\%$ and $20\%$, we show the corresponding contours in Fig.~\ref{fig:accuracy_10-20-sc3sd4_SM-HHH}, in which $\tilde{c_3}/c_3$ and $\tilde{d_4}/d_4$ are respectively the scaling factors for the $t\bar{t}hhh$ coupling and the quartic Higgs coupling, when the other couplings are fixed in the given models. We do not include the result for the Tadpole-induced Higgs scenario ($d_3\simeq 0, d_4\simeq 0$) because it would be  difficult to pin down the quartic Higgs coupling in this scenario due to its tiny value. 

Fig.~\ref{fig:accuracy_10-20-s4_MCH-HHH} shows that for the (SM) case of  $d_3=d_4=1$, the scaling factor is constrained to be within the range of $0.3<\tilde{d_4}/d_4<1.82 \cup 10.13<\tilde{d_4}/d_4<11.66$ ($0<\tilde{d_4}/d_4<2.85 \cup 9.10<\tilde{d_4}/d_4<12.28$), if the accuracy 
of the cross section measurement can be reached at the $10\%$ ($20\%$) level. 
For the NP scenarios, we give a brief summary as follows.
\bit
\item For the SMEFT scenario, we note that the bound on the quartic Higgs coupling will be generally quite loose, unless the  cross sections can be measured with better than $10\%$ accuracy. 
However, as shown in Fig.~\ref{fig:lam_3-lam_4_HHH_band}, the bounds on the quartic Higgs coupling $d_4$ could be tight when the trilinear coupling $d_3\simeq 2-3$, in which case the triple-Higgs production cross section shows sizable variation as the $d_4$ value changes.
\item For the Coleman-Weinberg Higgs scenario, the bound on the quartic Higgs coupling  $d_4$ is relatively tight, as the trilinear Higgs coupling is $5/3$.
\item For the Nambu-Goldstone Higgs scenario, the scaling factor $\tilde{c_3}/c_3$ could be constrained to be within the order of $10$, but $\tilde{d_4}/d_4$ could only be constrained to the order of much larger than $10$.
\item For the Tadpole-induced Higgs scenario, because the trilinear Higgs coupling $d_3$ could be highly suppressed, the dependence on the quartic Higgs coupling $d_4$ is very weak.
This renders the precision determination of $d_4$ to be very difficult in this scenario. 
\eit

\section{Conclusion}
\label{sec:conclusion}
The nature of the Higgs boson is still mysterious, for its potential is not well understood yet. In this paper, we consider several theoretically compelling new physics scenarios, in which the Higgs self-couplings can be quite different from the SM prediction. To be specific, we have considered the elementary Higgs, Nambu-Goldstone Higgs, Coleman-Weinberg Higgs, and Tadpole-induced Higgs scenarios, with the trilinear and quartic Higgs couplings being either smaller or larger than the SM ones. Trilinear Higgs coupling is enhanced in the elementary Higgs scenario (with the preferred positive coefficient $c_6$ for the effective operator $(H^\dagger H)^3$) and Coleman-Weinberg Higgs scenario, while it is reduced in the Nambu-Goldstone Higgs scenario and Tadpole-induced Higgs scenario. The same pattern also holds for the quartic Higgs coupling. 
In the Nambu-Goldstone Higgs scenario, 
we have also considered the Higgs nonlinear effect and explored the relations among the $t\bar{t}h$, $t\bar{t}hh$ and $t\bar{t}hhh$ couplings.

In general, both the SMEFT and the Higgs EFT can be used to describe the Higgs boson's nature and parameterize Higgs interactions, depending on whether the SM gauge symmetry is linearly or nonlinearly realized. The SMEFT is defined in the unbroken phase of the electroweak symmetry, while the Higgs EFT is defined in the broken phase.
Comparing these two EFT frameworks, only the Higgs EFT can exhibit the non-decoupling feature of new physics, this renders the Higgs EFT more general than the SMEFT. Among the new physics scenarios of different Higgs potentials, the SMEFT can only describe the elementary Higgs and the Nambu-Goldstone Higgs, but the Higgs EFT can describe all the scenarios, including the Coleman-Weinberg Higgs and the Tadpole-induced Higgs scenarios.

In this work, we study how well the trilinear and quartic couplings of the Higgs boson in various new physics scenarios can be measured at the 14 TeV HL-LHC, 27 TeV HE-LHC and the future 100 TeV pp collider. 
First, we have investigated the theoretical  
constraints on the Higgs self-couplings using the partial wave unitarity and tree-level vacuum stability analyses. It turns out that the partial wave unitarity bound is not very tight for the $2\to 2$ scatterings, assuming that the Higgs couplings to the longitudinal electroweak gauge bosons $hW_LW_L, hhW_LW_L$ are the same as those in the SM. The tree-level vacuum stability prefers the trilinear Higgs couplings to be within $0<d_3<3$, while the quartic Higgs coupling can be $10$ times larger than the SM value.

Given the unique patterns of the Higgs self-couplings predicted by various new physics scenarios, we explore the possibility of discriminating various new physics scenarios through the process of double-Higgs production $pp\to hh$ at the $27$ TeV HE-LHC and the $100$ TeV pp collider. We have studied in detail the total cross sections and various differential distributions, including the effects from distinct interference patterns, in each NP scenario. The values of the cross sections are typically smaller, compared to the SM value, for the elementary Higgs, and the Coleman-Weinberg Higgs cases, while they are larger for the Nambu-Goldstone Higgs and Tadpole-induced Higgs cases. With larger cross sections, the corresponding uncertainties in the determination of the Higgs self-couplings are reduced. Thus, one can distinguish different new physics scenarios at the $27$ TeV HE-LHC, given the SM is expected to be measured with the accuracy of $14\%$ at the $1\sigma$  level. The discrimination power is further enhanced at the $100$ TeV pp collider. For completeness, we have also extracted the possible range of the allowed trilinear Higgs coupling values for several new physics scenarios, assuming the cross section is measured with $10\%$ and $20\%$ accuracy, respectively. These are shown in Figs.~\ref{HHproduction}, \ref{fig:accuracy_10-20-s3_SM-HH}, and \ref{fig:accuracy_10-20-s3_SM-HH2}.

To fully pin down the quartic Higgs coupling and thus the shape of the Higgs potential in various scenarios, we also need to investigate the triple-Higgs production process $pp\to hhh$ at future colliders. However, due to the small rate of the signal event with respect to its backgrounds, it might be quite challenging to determine the quartic Higgs coupling. It appears, on the basis of current studies,
that one needs about $50\ \rm{ab}^{-1}$, cf. Table.~\ref{tab:lum}, at the $100$ TeV pp collider, to discover this process and precisely measure the quartic Higgs coupling in the case of the SM.
The integrated luminosity required for the $5 \sigma$ observation of the triple-Higgs production process in new physics scenarios is also shown in Table~\ref{tab:lum}. However, using some special techniques, including machine learning, one might be able to bring this luminosity to within the proposed luminosity for the 100 TeV collider. After investigating the interference patterns of the $pp\to hhh$ process, we find the dependence of the cross section on the quartic Higgs coupling is moderate because other couplings obscure the extraction of the quartic coupling. Thus, even when the total cross section can be relatively well measured with $10\%$ and $20\%$ accuracy, it is still not easy to measure the quartic Higgs coupling, cf. Figs.~\ref{HHHproduction}, \ref{fig:accuracy_10-20-s4_MCH-HHH}, and \ref{fig:accuracy_10-20-sc3sd4_SM-HHH}. 
Hence, more effort and better techniques are called for a better understanding of the Higgs potential.

\section*{Acknowledgements}
J.-H.Yu is supported by the National Science Foundation of China under Grants No. 11875003 and the Chinese Academy of Sciences (CAS) Hundred-Talent Program.
C.-P. Yuan is supported by the U.S. National Science Foundation under Grant No. PHY-1719914. C.-P. Yuan is also grateful for the support from the Wu-Ki Tung endowed chair in particle physics.
L.-X. Xu is supported in part by the National Science Foundation of China under Grants No. 11635001, 11875072.

\bibliography{HiggsPotential}

\begin{thebibliography}{100}

\bibitem{Sirunyan:2018two}
Albert~M Sirunyan et~al.
\newblock {Combination of searches for Higgs boson pair production in
  proton-proton collisions at $\sqrt{s} = $ 13 TeV}.
\newblock {\em Phys. Rev. Lett.}, 122(12):121803, 2019.

\bibitem{Aaboud:2018sfw}
Morad Aaboud et~al.
\newblock {Search for resonant and non-resonant Higgs boson pair production in
  the ${b\bar{b}\tau^+\tau^-}$ decay channel in $pp$ collisions at
  $\sqrt{s}=13$ TeV with the ATLAS detector}.
\newblock {\em Phys. Rev. Lett.}, 121(19):191801, 2018.
\newblock [Erratum: Phys. Rev. Lett.122,no.8,089901(2019)].

\bibitem{Aaboud:2018knk}
Morad Aaboud et~al.
\newblock {Search for pair production of Higgs bosons in the $b\bar{b}b\bar{b}$
  final state using proton-proton collisions at $\sqrt{s} = 13$ TeV with the
  ATLAS detector}.
\newblock {\em JHEP}, 01:030, 2019.

\bibitem{Aaboud:2018ksn}
Morad Aaboud et~al.
\newblock {Search for Higgs boson pair production in the $WW^{(*)}WW^{(*)}$
  decay channel using ATLAS data recorded at $\sqrt{s}=13$ TeV}.
\newblock {\em Submitted to: JHEP}, 2018.

\bibitem{Aaboud:2018zhh}
Morad Aaboud et~al.
\newblock {Search for Higgs boson pair production in the $b\bar{b}WW^{*}$ decay
  mode at $\sqrt{s}=13$ TeV with the ATLAS detector}.
\newblock {\em JHEP}, 04:092, 2019.

\bibitem{Aaboud:2018ewm}
Morad Aaboud et~al.
\newblock {Search for Higgs boson pair production in the $\gamma\gamma WW^{*}$
  channel using $pp$ collision data recorded at $\sqrt{s} = 13$ TeV with the
  ATLAS detector}.
\newblock {\em Eur. Phys. J.}, C78(12):1007, 2018.

\bibitem{CMS:2018ccd}
{Prospects for HH measurements at the HL-LHC}.
\newblock {\em CMS-PAS-FTR-18-019}, 2018.

\bibitem{ATLAS:2018ch1}
{Measurement prospects of the pair production and self-coupling of the Higgs
  boson with the ATLAS experiment at the HL-LHC}.
\newblock {\em ATL-PHYS-PUB-2018-053}, (ATL-PHYS-PUB-2018-053), Dec 2018.

\bibitem{Kaplan:1983fs}
David~B. Kaplan and Howard Georgi.
\newblock {SU(2) x U(1) Breaking by Vacuum Misalignment}.
\newblock {\em Phys. Lett.}, 136B:183--186, 1984.

\bibitem{Kaplan:1983sm}
David~B. Kaplan, Howard Georgi, and Savas Dimopoulos.
\newblock {Composite Higgs Scalars}.
\newblock {\em Phys. Lett.}, 136B:187--190, 1984.

\bibitem{Contino:2010rs}
Roberto Contino.
\newblock {The Higgs as a Composite Nambu-Goldstone Boson}.
\newblock In {\em {Physics of the large and the small, TASI 09, proceedings of
  the Theoretical Advanced Study Institute in Elementary Particle Physics,
  Boulder, Colorado, USA, 1-26 June 2009}}, pages 235--306, 2011.

\bibitem{Panico:2015jxa}
Giuliano Panico and Andrea Wulzer.
\newblock {The Composite Nambu-Goldstone Higgs}.
\newblock {\em Lect. Notes Phys.}, 913:pp.1--316, 2016.

\bibitem{Bellazzini:2014yua}
Brando Bellazzini, Csaba Csaki, and Javi Serra.
\newblock {Composite Higgses}.
\newblock {\em Eur. Phys. J.}, C74(5):2766, 2014.

\bibitem{Hill:2014mqa}
Christopher~T. Hill.
\newblock {Is the Higgs Boson Associated with Coleman-Weinberg Dynamical
  Symmetry Breaking?}
\newblock {\em Phys. Rev.}, D89(7):073003, 2014.

\bibitem{Helmboldt:2016mpi}
Alexander~J. Helmboldt, Pascal Humbert, Manfred Lindner, and Juri Smirnov.
\newblock {Minimal conformal extensions of the Higgs sector}.
\newblock {\em JHEP}, 07:113, 2017.

\bibitem{Hashino:2015nxa}
Katsuya Hashino, Shinya Kanemura, and Yuta Orikasa.
\newblock {Discriminative phenomenological features of scale invariant models
  for electroweak symmetry breaking}.
\newblock {\em Phys. Lett.}, B752:217--220, 2016.

\bibitem{Galloway:2013dma}
Jamison Galloway, Markus~A. Luty, Yuhsin Tsai, and Yue Zhao.
\newblock {Induced Electroweak Symmetry Breaking and Supersymmetric
  Naturalness}.
\newblock {\em Phys. Rev.}, D89(7):075003, 2014.

\bibitem{Chang:2014ida}
Spencer Chang, Jamison Galloway, Markus Luty, Ennio Salvioni, and Yuhsin Tsai.
\newblock {Phenomenology of Induced Electroweak Symmetry Breaking}.
\newblock {\em JHEP}, 03:017, 2015.

\bibitem{Buchmuller:1985jz}
W.~Buchmuller and D.~Wyler.
\newblock {Effective Lagrangian Analysis of New Interactions and Flavor
  Conservation}.
\newblock {\em Nucl. Phys.}, B268:621--653, 1986.

\bibitem{Grzadkowski:2010es}
B.~Grzadkowski, M.~Iskrzynski, M.~Misiak, and J.~Rosiek.
\newblock {Dimension-Six Terms in the Standard Model Lagrangian}.
\newblock {\em JHEP}, 10:085, 2010.

\bibitem{Giudice:2007fh}
G.~F. Giudice, C.~Grojean, A.~Pomarol, and R.~Rattazzi.
\newblock {The Strongly-Interacting Light Higgs}.
\newblock {\em JHEP}, 06:045, 2007.

\bibitem{Li:2019ghf}
Hao-Lin Li, Ling-Xiao Xu, Jiang-Hao Yu, and Shou-Hua Zhu.
\newblock {EFTs meet Higgs Nonlinearity, Compositeness and (Neutral)
  Naturalness}.
\newblock 2019.

\bibitem{Appelquist:1980vg}
Thomas Appelquist and Claude~W. Bernard.
\newblock {Strongly Interacting Higgs Bosons}.
\newblock {\em Phys. Rev.}, D22:200, 1980.

\bibitem{Longhitano:1980iz}
Anthony~C. Longhitano.
\newblock {Heavy Higgs Bosons in the Weinberg-Salam Model}.
\newblock {\em Phys. Rev.}, D22:1166, 1980.

\bibitem{Koulovassilopoulos:1993pw}
Vassilis Koulovassilopoulos and R.~Sekhar Chivukula.
\newblock {The Phenomenology of a nonstandard Higgs boson in W(L) W(L)
  scattering}.
\newblock {\em Phys. Rev.}, D50:3218--3234, 1994.

\bibitem{Contino:2010mh}
Roberto Contino, Christophe Grojean, Mauro Moretti, Fulvio Piccinini, and
  Riccardo Rattazzi.
\newblock {Strong Double Higgs Production at the LHC}.
\newblock {\em JHEP}, 05:089, 2010.

\bibitem{Alonso:2012px}
R.~Alonso, M.~B. Gavela, L.~Merlo, S.~Rigolin, and J.~Yepes.
\newblock {The Effective Chiral Lagrangian for a Light Dynamical "Higgs
  Particle"}.
\newblock {\em Phys. Lett.}, B722:330--335, 2013.
\newblock [Erratum: Phys. Lett.B726,926(2013)].

\bibitem{Buchalla:2013rka}
Gerhard Buchalla, Oscar Cata, and Claudius Krause.
\newblock {Complete Electroweak Chiral Lagrangian with a Light Higgs at NLO}.
\newblock {\em Nucl. Phys.}, B880:552--573, 2014.
\newblock [Erratum: Nucl. Phys.B913,475(2016)].

\bibitem{Buchalla:2013eza}
Gerhard Buchalla, Oscar Cata, and Claudius Krause.
\newblock {On the Power Counting in Effective Field Theories}.
\newblock {\em Phys. Lett.}, B731:80--86, 2014.

\bibitem{Glover:1987nx}
E.~W.~Nigel Glover and J.~J. van~der Bij.
\newblock {HIGGS BOSON PAIR PRODUCTION VIA GLUON FUSION}.
\newblock {\em Nucl. Phys.}, B309:282--294, 1988.

\bibitem{Baur:2003gp}
U.~Baur, T.~Plehn, and David~L. Rainwater.
\newblock {Probing the Higgs selfcoupling at hadron colliders using rare
  decays}.
\newblock {\em Phys. Rev.}, D69:053004, 2004.

\bibitem{Baur:2003gpa}
U.~Baur, T.~Plehn, and David~L. Rainwater.
\newblock {Examining the Higgs boson potential at lepton and hadron colliders:
  A Comparative analysis}.
\newblock {\em Phys. Rev.}, D68:033001, 2003.

\bibitem{Dolan:2012rv}
Matthew~J. Dolan, Christoph Englert, and Michael Spannowsky.
\newblock {Higgs self-coupling measurements at the LHC}.
\newblock {\em JHEP}, 10:112, 2012.

\bibitem{Baglio:2012np}
J.~Baglio, A.~Djouadi, R.~Grober, M.~M. Muhlleitner, J.~Quevillon, and
  M.~Spira.
\newblock {The measurement of the Higgs self-coupling at the LHC: theoretical
  status}.
\newblock {\em JHEP}, 04:151, 2013.

\bibitem{Papaefstathiou:2012qe}
Andreas Papaefstathiou, Li~Lin Yang, and Jose Zurita.
\newblock {Higgs boson pair production at the LHC in the $b \bar{b} W^+ W^-$
  channel}.
\newblock {\em Phys. Rev.}, D87(1):011301, 2013.

\bibitem{Goertz:2013kp}
Florian Goertz, Andreas Papaefstathiou, Li~Lin Yang, and Jose Zurita.
\newblock {Higgs Boson self-coupling measurements using ratios of cross
  sections}.
\newblock {\em JHEP}, 06:016, 2013.

\bibitem{Barger:2013jfa}
Vernon Barger, Lisa~L. Everett, C.~B. Jackson, and Gabe Shaughnessy.
\newblock {Higgs-Pair Production and Measurement of the Triscalar Coupling at
  LHC(8,14)}.
\newblock {\em Phys. Lett.}, B728:433--436, 2014.

\bibitem{Barr:2013tda}
Alan~J. Barr, Matthew~J. Dolan, Christoph Englert, and Michael Spannowsky.
\newblock {Di-Higgs final states augMT2ed -- selecting $hh$ events at the high
  luminosity LHC}.
\newblock {\em Phys. Lett.}, B728:308--313, 2014.

\bibitem{Li:2015yia}
Qiang Li, Zhao Li, Qi-Shu Yan, and Xiaoran Zhao.
\newblock {Probe Higgs boson pair production via the 3?2j+$\not{E}$ mode}.
\newblock {\em Phys. Rev.}, D92(1):014015, 2015.

\bibitem{deLima:2014dta}
Danilo~Enoque Ferreira~de Lima, Andreas Papaefstathiou, and Michael Spannowsky.
\newblock {Standard model Higgs boson pair production in the ( $ b\overline{b}
  $ )( $ b\overline{b} $ ) final state}.
\newblock {\em JHEP}, 08:030, 2014.

\bibitem{Alves:2017ued}
Alexandre Alves, Tathagata Ghosh, and Kuver Sinha.
\newblock {Can We Discover Double Higgs Production at the LHC?}
\newblock {\em Phys. Rev.}, D96(3):035022, 2017.

\bibitem{Adhikary:2017jtu}
Amit Adhikary, Shankha Banerjee, Rahool~Kumar Barman, Biplob Bhattacherjee, and
  Saurabh Niyogi.
\newblock {Revisiting the non-resonant Higgs pair production at the HL-LHC}.
\newblock {\em JHEP}, 07:116, 2018.

\bibitem{Goncalves:2018yva}
Dorival Goncalves, Tao Han, Felix Kling, Tilman Plehn, and Michihisa Takeuchi.
\newblock {Higgs boson pair production at future hadron colliders: From
  kinematics to dynamics}.
\newblock {\em Phys. Rev.}, D97(11):113004, 2018.

\bibitem{Borowka:2018pxx}
Sophia Borowka, Claude Duhr, Fabio Maltoni, Davide Pagani, Ambresh Shivaji, and
  Xiaoran Zhao.
\newblock {Probing the scalar potential via double Higgs boson production at
  hadron colliders}.
\newblock {\em JHEP}, 04:016, 2019.

\bibitem{Homiller:2018dgu}
Samuel Homiller and Patrick Meade.
\newblock {Measurement of the Triple Higgs Coupling at a HE-LHC}.
\newblock {\em JHEP}, 03:055, 2019.

\bibitem{Kim:2018cxf}
Jeong~Han Kim, Kyoungchul Kong, Konstantin~T. Matchev, and Myeonghun Park.
\newblock {Probing the Triple Higgs Self-Interaction at the Large Hadron
  Collider}.
\newblock {\em Phys. Rev. Lett.}, 122(9):091801, 2019.

\bibitem{Heinrich:2019bkc}
G.~Heinrich, S.~P. Jones, M.~Kerner, G.~Luisoni, and L.~Scyboz.
\newblock {Probing the trilinear Higgs boson coupling in di-Higgs production at
  NLO QCD including parton shower effects}.
\newblock {\em JHEP}, 06:066, 2019.

\bibitem{Cepeda:2019klc}
M.~Cepeda et~al.
\newblock {Report from Working Group 2}.
\newblock {\em CERN Yellow Rep. Monogr.}, 7:221--584, 2019.

\bibitem{Kim:2019wns}
Jeong~Han Kim, Minho Kim, Kyoungchul Kong, Konstantin~T. Matchev, and Myeonghun
  Park.
\newblock {Portraying Double Higgs at the Large Hadron Collider}.
\newblock 2019.

\bibitem{Kanemura:2008ub}
Shinya Kanemura and Koji Tsumura.
\newblock {Effects of the anomalous Higgs couplings on the Higgs boson
  production at the Large Hadron Collider}.
\newblock {\em Eur. Phys. J.}, C63:11--21, 2009.

\bibitem{Contino:2012xk}
Roberto Contino, Margherita Ghezzi, Mauro Moretti, Giuliano Panico, Fulvio
  Piccinini, and Andrea Wulzer.
\newblock {Anomalous Couplings in Double Higgs Production}.
\newblock {\em JHEP}, 08:154, 2012.

\bibitem{Li:2019uyy}
Gang Li, Ling-Xiao Xu, Bin Yan, and C.~P. Yuan.
\newblock {Resolving the degeneracy in top quark Yukawa coupling with Higgs
  pair production}.
\newblock 2019.

\bibitem{Cao:2015oaa}
Qing-Hong Cao, Bin Yan, Dong-Ming Zhang, and Hao Zhang.
\newblock {Resolving the Degeneracy in Single Higgs Production with Higgs Pair
  Production}.
\newblock {\em Phys. Lett.}, B752:285--290, 2016.

\bibitem{Cao:2016zob}
Qing-Hong Cao, Gang Li, Bin Yan, Dong-Ming Zhang, and Hao Zhang.
\newblock {Double Higgs production at the 14 TeV LHC and a 100 TeV $pp$
  collider}.
\newblock {\em Phys. Rev.}, D96(9):095031, 2017.

\bibitem{Chen:2014xra}
Chuan-Ren Chen and Ian Low.
\newblock {Double take on new physics in double Higgs boson production}.
\newblock {\em Phys. Rev.}, D90(1):013018, 2014.

\bibitem{Kim:2018uty}
Jeong~Han Kim, Yasuhito Sakaki, and Minho Son.
\newblock {Combined analysis of double Higgs production via gluon fusion at the
  HL-LHC in the effective field theory approach}.
\newblock {\em Phys. Rev.}, D98(1):015016, 2018.

\bibitem{Huang:2017nnw}
Peisi Huang, Aniket Joglekar, Min Li, and Carlos E.~M. Wagner.
\newblock {Corrections to di-Higgs boson production with light stops and
  modified Higgs couplings}.
\newblock {\em Phys. Rev.}, D97(7):075001, 2018.

\bibitem{Basler:2018dac}
Philipp Basler, Sally Dawson, Christoph Englert, and Margarete Muhlleitner.
\newblock {Showcasing HH production: Benchmarks for the LHC and HL-LHC}.
\newblock {\em Phys. Rev.}, D99(5):055048, 2019.

\bibitem{Babu:2018uik}
K.~S. Babu and Sudip Jana.
\newblock {Enhanced Di-Higgs Production in the Two Higgs Doublet Model}.
\newblock {\em JHEP}, 02:193, 2019.

\bibitem{Azatov:2015oxa}
Aleksandr Azatov, Roberto Contino, Giuliano Panico, and Minho Son.
\newblock {Effective field theory analysis of double Higgs boson production via
  gluon fusion}.
\newblock {\em Phys. Rev.}, D92(3):035001, 2015.

\bibitem{Dawson:2015oha}
S.~Dawson, A.~Ismail, and Ian Low.
\newblock {What?s in the loop? The anatomy of double Higgs production}.
\newblock {\em Phys. Rev.}, D91(11):115008, 2015.

\bibitem{Grober:2010yv}
R.~Grober and M.~Muhlleitner.
\newblock {Composite Higgs Boson Pair Production at the LHC}.
\newblock {\em JHEP}, 06:020, 2011.

\bibitem{Gillioz:2012se}
M.~Gillioz, R.~Grober, C.~Grojean, M.~Muhlleitner, and E.~Salvioni.
\newblock {Higgs Low-Energy Theorem (and its corrections) in Composite Models}.
\newblock {\em JHEP}, 10:004, 2012.

\bibitem{Grober:2016wmf}
Ramona Grober, Margarete Muhlleitner, and Michael Spira.
\newblock {Signs of Composite Higgs Pair Production at Next-to-Leading Order}.
\newblock {\em JHEP}, 06:080, 2016.

\bibitem{Millet:2019oqe}
Francois Millet, L.~Tavian, U.~Cardella, O.~Amstutz, P.~Selva, and A.~Kuendig.
\newblock {Preliminary Conceptual design of FCC-hh cryoplants: Linde
  evaluation}.
\newblock {\em IOP Conf. Ser. Mater. Sci. Eng.}, 502(1):012131, 2019.

\bibitem{CEPC-SPPCStudyGroup:2015csa}
Muhammd Ahmad et~al.
\newblock {CEPC-SPPC Preliminary Conceptual Design Report. 1. Physics and
  Detector}.
\newblock 2015.

\bibitem{Plehn:2005nk}
Tilman Plehn and Michael Rauch.
\newblock {The quartic higgs coupling at hadron colliders}.
\newblock {\em Phys. Rev.}, D72:053008, 2005.

\bibitem{Chen:2015gva}
Chien-Yi Chen, Qi-Shu Yan, Xiaoran Zhao, Yi-Ming Zhong, and Zhijie Zhao.
\newblock {Probing triple-Higgs productions via 4b2? decay channel at a 100 TeV
  hadron collider}.
\newblock {\em Phys. Rev.}, D93(1):013007, 2016.

\bibitem{Fuks:2015hna}
Benjamin Fuks, Jeong~Han Kim, and Seung~J. Lee.
\newblock {Probing Higgs self-interactions in proton-proton collisions at a
  center-of-mass energy of 100 TeV}.
\newblock {\em Phys. Rev.}, D93(3):035026, 2016.

\bibitem{Papaefstathiou:2015paa}
Andreas Papaefstathiou and Kazuki Sakurai.
\newblock {Triple Higgs boson production at a 100 TeV proton-proton collider}.
\newblock {\em JHEP}, 02:006, 2016.

\bibitem{Fuks:2017zkg}
Benjamin Fuks, Jeong~Han Kim, and Seung~J. Lee.
\newblock {Scrutinizing the Higgs quartic coupling at a future 100 TeV
  proton?proton collider with taus and b-jets}.
\newblock {\em Phys. Lett.}, B771:354--358, 2017.

\bibitem{Kilian:2017nio}
Wolfgang Kilian, Sichun Sun, Qi-Shu Yan, Xiaoran Zhao, and Zhijie Zhao.
\newblock {New Physics in multi-Higgs boson final states}.
\newblock {\em JHEP}, 06:145, 2017.

\bibitem{Falkowski:2015wza}
Adam Falkowski, Benjamin Fuks, Kentarou Mawatari, Ken Mimasu, Francesco Riva,
  and Vernica Sanz.
\newblock {Rosetta: an operator basis translator for Standard Model effective
  field theory}.
\newblock {\em Eur. Phys. J.}, C75(12):583, 2015.

\bibitem{Contino:2013kra}
Roberto Contino, Margherita Ghezzi, Christophe Grojean, Margarete Muhlleitner,
  and Michael Spira.
\newblock {Effective Lagrangian for a light Higgs-like scalar}.
\newblock {\em JHEP}, 07:035, 2013.

\bibitem{Goertz:2014qta}
Florian Goertz, Andreas Papaefstathiou, Li~Lin Yang, and Jose Zurita.
\newblock {Higgs boson pair production in the D=6 extension of the SM}.
\newblock {\em JHEP}, 04:167, 2015.

\bibitem{Corbett:2017ieo}
Tyler Corbett, Aniket Joglekar, Hao-Lin Li, and Jiang-Hao Yu.
\newblock {Exploring Extended Scalar Sectors with Di-Higgs Signals: A Higgs EFT
  Perspective}.
\newblock {\em JHEP}, 05:061, 2018.

\bibitem{Dawson:2017vgm}
Sally Dawson and Christopher~W. Murphy.
\newblock {Standard Model EFT and Extended Scalar Sectors}.
\newblock {\em Phys. Rev.}, D96(1):015041, 2017.

\bibitem{Belusca-Maito:2016dqe}
Hermes Belusca-Maito, Adam Falkowski, Duarte Fontes, Jorge~C. Romao, and
  Joao~P. Silva.
\newblock {Higgs EFT for 2HDM and beyond}.
\newblock {\em Eur. Phys. J.}, C77(3):176, 2017.

\bibitem{Coleman:1969sm}
Sidney~R. Coleman, J.~Wess, and Bruno Zumino.
\newblock {Structure of phenomenological Lagrangians. 1.}
\newblock {\em Phys. Rev.}, 177:2239--2247, 1969.

\bibitem{Callan:1969sn}
Curtis~G. Callan, Jr., Sidney~R. Coleman, J.~Wess, and Bruno Zumino.
\newblock {Structure of phenomenological Lagrangians. 2.}
\newblock {\em Phys. Rev.}, 177:2247--2250, 1969.

\bibitem{Yu:2016bku}
Jiang-Hao Yu.
\newblock {Radiative-$\mathbb Z_2$-breaking twin Higgs model}.
\newblock {\em Phys. Rev.}, D94(11):111704, 2016.

\bibitem{Yu:2016swa}
Jiang-Hao Yu.
\newblock {A tale of twin Higgs: natural twin two Higgs doublet models}.
\newblock {\em JHEP}, 12:143, 2016.

\bibitem{Xu:2018ofw}
Ling-Xiao Xu, Jiang-Hao Yu, and Shou-Hua Zhu.
\newblock {Minimal Neutral Naturalness Model}.
\newblock 2018.

\bibitem{Xu:2019xuo}
Ling-Xiao Xu, Jiang-Hao Yu, and Shou-hua Zhu.
\newblock {Holographic Completion of Minimal Neutral Naturalness Model and
  Deconstruction}.
\newblock 2019.

\bibitem{Cao:2018cms}
Qing-Hong Cao, Ling-Xiao Xu, Bin Yan, and Shou-Hua Zhu.
\newblock {Signature of pseudo Nambu?Goldstone Higgs boson in its decay}.
\newblock {\em Phys. Lett.}, B789:233--237, 2019.

\bibitem{Alonso:2016btr}
Rodrigo Alonso, Elizabeth~E. Jenkins, and Aneesh~V. Manohar.
\newblock {Sigma Models with Negative Curvature}.
\newblock {\em Phys. Lett.}, B756:358--364, 2016.

\bibitem{Agashe:2004rs}
Kaustubh Agashe, Roberto Contino, and Alex Pomarol.
\newblock {The Minimal composite Higgs model}.
\newblock {\em Nucl. Phys.}, B719:165--187, 2005.

\bibitem{Contino:2006qr}
Roberto Contino, Leandro Da~Rold, and Alex Pomarol.
\newblock {Light custodians in natural composite Higgs models}.
\newblock {\em Phys. Rev.}, D75:055014, 2007.

\bibitem{Geller:2014kta}
Michael Geller and Ofri Telem.
\newblock {Holographic Twin Higgs Model}.
\newblock {\em Phys. Rev. Lett.}, 114:191801, 2015.

\bibitem{Barbieri:2015lqa}
Riccardo Barbieri, Davide Greco, Riccardo Rattazzi, and Andrea Wulzer.
\newblock {The Composite Twin Higgs scenario}.
\newblock {\em JHEP}, 08:161, 2015.

\bibitem{Low:2015nqa}
Matthew Low, Andrea Tesi, and Lian-Tao Wang.
\newblock {Twin Higgs mechanism and a composite Higgs boson}.
\newblock {\em Phys. Rev.}, D91:095012, 2015.

\bibitem{Coleman:1973jx}
Sidney~R. Coleman and Erick~J. Weinberg.
\newblock {Radiative Corrections as the Origin of Spontaneous Symmetry
  Breaking}.
\newblock {\em Phys. Rev.}, D7:1888--1910, 1973.

\bibitem{Gildener:1976ih}
Eldad Gildener and Steven Weinberg.
\newblock {Symmetry Breaking and Scalar Bosons}.
\newblock {\em Phys. Rev.}, D13:3333, 1976.

\bibitem{Simmons:1988fu}
Elizabeth~H. Simmons.
\newblock {Phenomenology of a Technicolor Model With Heavy Scalar Doublet}.
\newblock {\em Nucl. Phys.}, B312:253--268, 1989.

\bibitem{Carone:1993xc}
Christopher~D. Carone and Howard Georgi.
\newblock {Technicolor with a massless scalar doublet}.
\newblock {\em Phys. Rev.}, D49:1427--1436, 1994.

\bibitem{Susskind:1978ms}
Leonard Susskind.
\newblock {Dynamics of Spontaneous Symmetry Breaking in the Weinberg-Salam
  Theory}.
\newblock {\em Phys. Rev.}, D20:2619--2625, 1979.

\bibitem{Weinberg:1975gm}
Steven Weinberg.
\newblock {Implications of Dynamical Symmetry Breaking}.
\newblock {\em Phys. Rev.}, D13:974--996, 1976.
\newblock [Addendum: Phys. Rev.D19,1277(1979)].

\bibitem{Lee:1977eg}
Benjamin~W. Lee, C.~Quigg, and H.~B. Thacker.
\newblock {Weak Interactions at Very High-Energies: The Role of the Higgs Boson
  Mass}.
\newblock {\em Phys. Rev.}, D16:1519, 1977.

\bibitem{Chanowitz:1978mv}
Michael~S. Chanowitz, M.~A. Furman, and I.~Hinchliffe.
\newblock {Weak Interactions of Ultraheavy Fermions. 2.}
\newblock {\em Nucl. Phys.}, B153:402--430, 1979.

\bibitem{DiLuzio:2017tfn}
Luca Di~Luzio, Ramona Grober, and Michael Spannowsky.
\newblock {Maxi-sizing the trilinear Higgs self-coupling: how large could it
  be?}
\newblock {\em Eur. Phys. J.}, C77(11):788, 2017.

\bibitem{Contino:2011np}
Roberto Contino, David Marzocca, Duccio Pappadopulo, and Riccardo Rattazzi.
\newblock {On the effect of resonances in composite Higgs phenomenology}.
\newblock {\em JHEP}, 10:081, 2011.

\bibitem{Falkowski:2019tft}
Adam Falkowski and Riccardo Rattazzi.
\newblock {Which EFT}.
\newblock 2019.

\bibitem{Chang:2019vez}
Spencer Chang and Markus~A. Luty.
\newblock {The Higgs Trilinear Coupling and the Scale of New Physics}.
\newblock 2019.

\bibitem{Agrawal:1998ch}
Pankaj Agrawal and Glenn Ladinsky.
\newblock {Production of two photons and a jet through gluon fusion}.
\newblock {\em Phys. Rev.}, D63:117504, 2001.

\bibitem{Agrawal:2012as}
Pankaj Agrawal and Ambresh Shivaji.
\newblock {Production of $\gamma Z g$ and associated processes via gluon fusion
  at hadron colliders}.
\newblock {\em JHEP}, 01:071, 2013.

\bibitem{Dulat:2015mca}
Sayipjamal Dulat, Tie-Jiun Hou, Jun Gao, Marco Guzzi, Joey Huston, Pavel
  Nadolsky, Jon Pumplin, Carl Schmidt, Daniel Stump, and C.~P. Yuan.
\newblock {New parton distribution functions from a global analysis of quantum
  chromodynamics}.
\newblock {\em Phys. Rev.}, D93(3):033006, 2016.

\bibitem{Grober:2015cwa}
Ramona Grober, Margarete Muhlleitner, Michael Spira, and Juraj Streicher.
\newblock {NLO QCD Corrections to Higgs Pair Production including Dimension-6
  Operators}.
\newblock {\em JHEP}, 09:092, 2015.

\bibitem{deFlorian:2017qfk}
Daniel de~Florian, Ignacio Fabre, and Javier Mazzitelli.
\newblock {Higgs boson pair production at NNLO in QCD including dimension 6
  operators}.
\newblock {\em JHEP}, 10:215, 2017.

\bibitem{Grober:2017gut}
R.~Grober, M.~Muhlleitner, and M.~Spira.
\newblock {Higgs Pair Production at NLO QCD for CP-violating Higgs Sectors}.
\newblock {\em Nucl. Phys.}, B925:1--27, 2017.

\bibitem{Buchalla:2018yce}
G.~Buchalla, M.~Capozi, A.~Celis, G.~Heinrich, and L.~Scyboz.
\newblock {Higgs boson pair production in non-linear Effective Field Theory
  with full $m_t$-dependence at NLO QCD}.
\newblock {\em JHEP}, 09:057, 2018.

\bibitem{Dawson:1998py}
S.~Dawson, S.~Dittmaier, and M.~Spira.
\newblock {Neutral Higgs boson pair production at hadron colliders: QCD
  corrections}.
\newblock {\em Phys. Rev.}, D58:115012, 1998.

\bibitem{Grigo:2013rya}
Jonathan Grigo, Jens Hoff, Kirill Melnikov, and Matthias Steinhauser.
\newblock {On the Higgs boson pair production at the LHC}.
\newblock {\em Nucl. Phys.}, B875:1--17, 2013.

\bibitem{Borowka:2016ehy}
S.~Borowka, N.~Greiner, G.~Heinrich, S.~P. Jones, M.~Kerner, J.~Schlenk,
  U.~Schubert, and T.~Zirke.
\newblock {Higgs Boson Pair Production in Gluon Fusion at Next-to-Leading Order
  with Full Top-Quark Mass Dependence}.
\newblock {\em Phys. Rev. Lett.}, 117(1):012001, 2016.
\newblock [Erratum: Phys. Rev. Lett.117,no.7,079901(2016)].

\bibitem{Borowka:2016ypz}
S.~Borowka, N.~Greiner, G.~Heinrich, S.~P. Jones, M.~Kerner, J.~Schlenk, and
  T.~Zirke.
\newblock {Full top quark mass dependence in Higgs boson pair production at
  NLO}.
\newblock {\em JHEP}, 10:107, 2016.

\bibitem{deFlorian:2013jea}
Daniel de~Florian and Javier Mazzitelli.
\newblock {Higgs Boson Pair Production at Next-to-Next-to-Leading Order in
  QCD}.
\newblock {\em Phys. Rev. Lett.}, 111:201801, 2013.

\bibitem{Grazzini:2018bsd}
Massimiliano Grazzini, Gudrun Heinrich, Stephen Jones, Stefan Kallweit,
  Matthias Kerner, Jonas~M. Lindert, and Javier Mazzitelli.
\newblock {Higgs boson pair production at NNLO with top quark mass effects}.
\newblock {\em JHEP}, 05:059, 2018.

\bibitem{Cowan:2010js}
Glen Cowan, Kyle Cranmer, Eilam Gross, and Ofer Vitells.
\newblock {Asymptotic formulae for likelihood-based tests of new physics}.
\newblock {\em Eur. Phys. J.}, C71:1554, 2011.
\newblock [Erratum: Eur. Phys. J.C73,2501(2013)].

\bibitem{Tanabashi:2018oca}
M.~Tanabashi et~al.
\newblock {Review of Particle Physics}.
\newblock {\em Phys. Rev.}, D98(3):030001, 2018.

\bibitem{Binoth:2006ym}
T.~Binoth, S.~Karg, N.~Kauer, and R.~Ruckl.
\newblock {Multi-Higgs boson production in the Standard Model and beyond}.
\newblock {\em Phys. Rev.}, D74:113008, 2006.

\bibitem{Agrawal:2017cbs}
Pankaj Agrawal, Debashis Saha, and Ambresh Shivaji.
\newblock {Production of $HHH$ and $HHV(V=\gamma,Z)$ at the hadron colliders}.
\newblock {\em Phys. Rev.}, D97(3):036006, 2018.

\bibitem{Bizon:2018syu}
Wojciech Bizon, Ulrich Haisch, and Luca Rottoli.
\newblock {Constraints on the quartic Higgs self-coupling from double-Higgs
  production at future hadron colliders}.
\newblock 2018.

\bibitem{Liu:2018peg}
Tao Liu, Kun-Feng Lyu, Jing Ren, and Hua~Xing Zhu.
\newblock {Probing the quartic Higgs boson self-interaction}.
\newblock {\em Phys. Rev.}, D98(9):093004, 2018.

\bibitem{Kilian:2018bhs}
Wolfgang Kilian, Sichun Sun, Qi-Shu Yan, Xiaoran Zhao, and Zhijie Zhao.
\newblock {Multi-Higgs Production and Unitarity in Vector-Boson Fusion at
  Future Hadron Colliders}.
\newblock 2018.

\bibitem{Maltoni:2018ttu}
Fabio Maltoni, Davide Pagani, and Xiaoran Zhao.
\newblock {Constraining the Higgs self-couplings at e$^{+}$e$^{?}$ colliders}.
\newblock {\em JHEP}, 07:087, 2018.

\bibitem{Dicus:2016rpf}
Duane~A. Dicus, Chung Kao, and Wayne~W. Repko.
\newblock {Self Coupling of the Higgs boson in the processes
  $p\,p\,\rightarrow\,ZHHH+X$ and $p\,p\,\rightarrow\,WHHH+X$}.
\newblock {\em Phys. Rev.}, D93(11):113003, 2016.

\bibitem{Vermaseren:2000nd}
J.~A.~M. Vermaseren.
\newblock {New features of FORM}.
\newblock 2000.

\bibitem{vanOldenborgh:1989wn}
G.~J. van Oldenborgh and J.~A.~M. Vermaseren.
\newblock {New Algorithms for One Loop Integrals}.
\newblock {\em Z. Phys.}, C46:425--438, 1990.

\bibitem{vanHameren:2010cp}
A.~van Hameren.
\newblock {OneLOop: For the evaluation of one-loop scalar functions}.
\newblock {\em Comput. Phys. Commun.}, 182:2427--2438, 2011.

\bibitem{Maltoni:2014eza}
F.~Maltoni, E.~Vryonidou, and M.~Zaro.
\newblock {Top-quark mass effects in double and triple Higgs production in
  gluon-gluon fusion at NLO}.
\newblock {\em JHEP}, 11:079, 2014.

\bibitem{deFlorian:2016sit}
Daniel de~Florian and Javier Mazzitelli.
\newblock {Two-loop corrections to the triple Higgs boson production cross
  section}.
\newblock {\em JHEP}, 02:107, 2017.

\bibitem{Spira:2016zna}
Michael Spira.
\newblock {Effective Multi-Higgs Couplings to Gluons}.
\newblock {\em JHEP}, 10:026, 2016.

\end{thebibliography}
\bibliographystyle{unsrt}

\end{document}